\documentclass{elsarticle}
\usepackage{graphicx}

\usepackage{amsmath}
\usepackage{rotating}  
\usepackage{psfrag,graphicx} 
\usepackage{subfigure}

\begin{document}
\begin{frontmatter}


\title{The High-Acceptance Dielectron Spectrometer HADES}

\author[8]{G.~Agakishiev}
\author[2]{C.~Agodi}
\author[16]{H.~Alvarez-Pol}
\author[19]{E.~Atkin}
\author[4]{E.~Badura}
\author[3,d]{A.~Balanda}
\author[9]{A.~Bassi}
\author[9]{R.~Bassini}
\author[2,a]{G.~Bellia}
\author[16]{D.~Belver}

% ...
\address[2]{Istituto Nazionale di Fisica Nucleare - Laboratori Nazionali del Sud, 95125~Catania, Italy}
\address[3]{Smoluchowski Institute of Physics, Jagiellonian University of Krak\'{o}w, 30-059~Krak\'{o}w, Poland}
\address[4]{GSI Helmholtzzentrum f\"{u}r Schwerionenforschung, 64291~Darmstadt, Germany}
% ...
\address[8]{II. Physikalisches Institut, Justus-Liebig-Universit\"{a}t Gie$\beta$ en, 35392~Gie$\beta$en, Germany}
\address[9]{Istituto Nazionale di Fisica Nucleare, Sezione di Milano, 20133~Milano, Italy}
% ...
\address[16]{Departamento de F\'{\i}sica de Part\'{\i}culas, Universidad de Santiago de Compostela, 15706~Santiago de Compostela, Spain}
% ...
\address[19]{Moscow Engineering Physics Institute (State University), 115409~Moscow, Russia}
% ...
\address[a]{Also at Dipartimento di Fisica e Astronomia, Universit\`{a} di Catania, 95125~Catania, Italy}
% ...
\address[d]{Also at Panstwowa Wyzsza Szkola Zawodowa , 33-300~Nowy Sacz, Poland}

\begin{abstract}
HADES is a versatile magnetic spectrometer aimed at studying dielectron production in pion, proton and heavy-ion induced collisions.
Its main features include a ring imaging gas Cherenkov detector for electron-hadron
discrimination, a tracking system consisting of a set of 6 superconducting coils producing a
toroidal field
and drift chambers and a multiplicity and electron trigger array
for additional electron-hadron discrimination and event characterization.
A two-stage trigger system enhances events
containing electrons. The physics program is focused on the
investigation of hadron properties in nuclei and in the
hot and dense hadronic matter.
The detector system is characterized by an
85~\% azimuthal coverage over a polar angle interval from $18^{\circ}$
to $85^{\circ}$, a single electron efficiency of 50~\%
and a vector meson mass resolution of 2.5~\%. Identification of pions, kaons
and protons is achieved combining time-of-flight and energy loss measurements
over a large momentum range.
This paper describes the main features
and the performance of the detector system.
\end{abstract}

\end{frontmatter}


%

\section{Introduction}
\label{sec:introduction}

\subsection{Physics motivation}
A central topic of contemporary hadron physics is the investigation
of had\-ronic matter. Theoretical models based on non-perturbative
Quantum Chromo-Dynamics indicate that the properties of hadrons are
modified, if the particles are
embedded in a strongly interacting medium (for a theory overview see \cite{rapp_wambach}).

The High-Acceptance DiElectron Spectrometer (HADES) in operation at the
GSI Helmholtzzentrum f\"ur Schwerionenforschung
has been specifically designed to study medium modifications of
the light vector mesons $\rho,\omega,\phi$ \cite{schicker_nim}.
Experimentally, these probes are well suited for two reasons.
The vector mesons are short-lived with lifetimes comparable to the
duration of the compression phase of relativistic heavy-ion reactions
in the 1 to 2~AGeV regime of the heavy-ion synchrotron SIS18.
Equally important is their electromagnetic decay branch into e$^+$e$^-$ pairs.
This channel is not subject to strong final state interaction
and thus provides an undistorted signal of the matter phase.
The goal of the HADES experiments is to measure the spectral properties of the vector
mesons  such as their in-medium masses and widths.

The HADES heavy-ion program is focused on incident kinetic energies from 1 to 2~AGeV.
Above about 0.7~AGeV these nucleus-nucleus reactions
become increasingly complex as  new particles - predominantly
mesons - are produced which induce secondary reactions \cite{metag}.
Some of these elementary reactions are not well known and need to
be explored as well.
While relativistic heavy-ion collisions produce hadronic matter at
a few times normal nuclear matter density and elevated temperature,
pion or proton induced reactions embed vector mesons into
normal nuclear matter.
A dedicated physics program including heavy ions,
deuteron, proton and pion beams has been proposed for the HADES
detector \cite{piotr_erice,juergen}.

Dilepton decays of vector mesons at SIS energies are rare
events and their observation presents a challenge for the detector design.
Thus, HADES has been equipped with a hadron-blind ring imaging
Cherenkov counter, a tracking system and a multiplicity and electron trigger array.
A two-stage trigger system selects events containing
electron candidates in real time. With its much larger solid angle and improved
resolution, HADES continues  and has the capability to complete the physics program which was pioneered by the
DLS spectrometer at the BEVALAC \cite{dls}.

\begin{figure}[htb]
\center
\includegraphics[width=10cm,height=10cm]{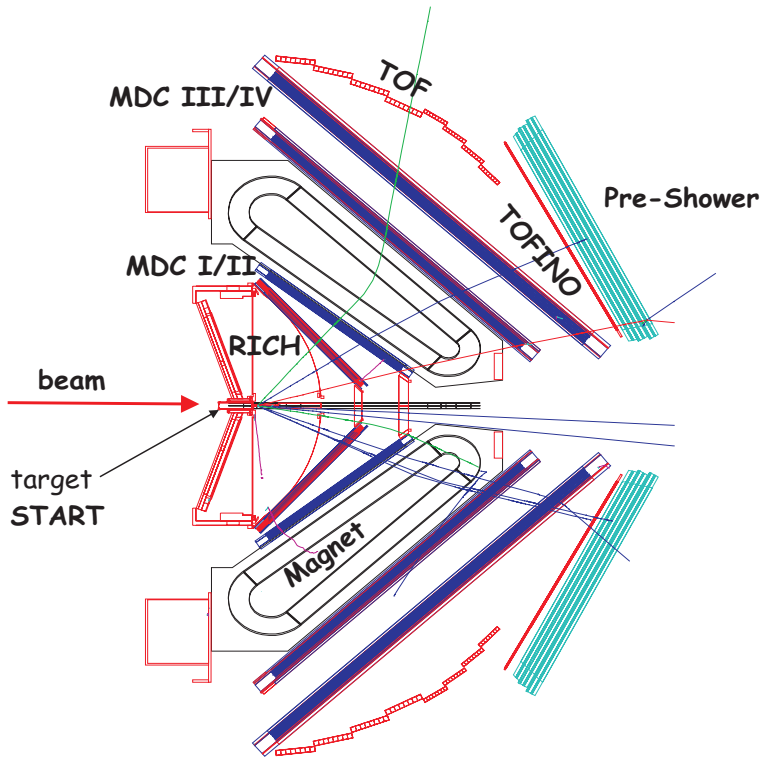}
\caption[]{Schematic layout of the HADES detector.
 A RICH detector with
gaseous radiator, carbon fiber mirror and UV photon detector
with solid CsI photocathode is used for electron identification. Two
sets of Mini-Drift Chambers (MDCs) with 4 modules per sector are
placed in front and behind the toroidal magnetic field to measure particle
momenta. A time of flight wall (TOF/TOFINO) accompanied by a
Pre-Shower detector at forward angles is used for additional
electron identification and trigger purposes. The target is
placed at half radius off the centre of the mirror. For reaction time
measurement, a START detector is located in front of the target.
A few particle tracks are depicted too.}
\label{hades}
\end{figure}

\subsection{Detector overview}
HADES features six
identical sectors defined by the superconducting coils producing the toroidal geometry magnetic field.
The spectrometer has 85~\% azimuthal acceptance and covers polar angles between
$\theta=18^\circ$ and $\theta=85^\circ$.
The angular and momentum
acceptance has been optimized for the detection of dielectron decays
of hadrons produced in the SIS energy regime.
A section of the detector in the vertical plane containing the
beam axis is shown in fig.~\ref{hades}.

Momentum reconstruction is carried out by measuring the deflection
angle of the particle trajectories derived from the 4 hit positions in the
planes of the Mini-Drift Chambers (MDC) located before and
after the magnetic field region.
Electron identification is performed with the hadron-blind gas Ring
Imaging Cherenkov detector (RICH)
together with the Multiplicity and Electron Trigger Array (META) consisting
of time-of-flight scintillator walls (TOF/TOFINO) and
electromagnetic shower detectors (Pre-Shower).
A powerful two-stage trigger system is employed to select events
within a predefined charged particle multiplicity interval (first-level
trigger LVL1), as well as electron candidates (second-level trigger LVL2).

In the following, a detailed description of the main
spectrometer components is given: magnet (sect.~\ref{Chapter_magnet}), RICH
(sect.~\ref{Chapter_RICH}), tracking system (sect.~\ref{Chapter_MDC}),
META (sects.~\ref{Chapter_tof} and \ref{Chapter_shower})
and beam detectors
(sect.~\ref{Chapter_start}). The detector description is followed by a
discussion of the data acquisition and trigger system
(sect.~\ref{Chapter_daq}). The  data analysis framework
and the detector performance are
discussed in sect.~\ref{Chapter_simana}.


\section{Major spectrometer components}
\subsection{Magnet}
\label{Chapter_magnet}

\subsubsection{Basic design considerations}

The purpose of the magnet is to provide a transverse kick to charged
particles in order to obtain their momenta with sufficient
resolution being of the order of $\sigma_p/p$\,=\,1.5\,-\,2~\% for electrons.
On the other hand, electron
identification with the RICH detector requires a nearly field free
region around the target. Furthermore, a large momentum range of $p$
= 0.1 - 2~GeV/c should be accepted simultaneously within a
large solid angle ($\theta=18^{\circ}-85^{\circ}$, as close as possible to full azimuthal
coverage). Simulations of reactions in the SIS18 energy regime have shown that
these requirements call for a non-focusing spectrometer
with a transverse momentum kick $p_k$ of about 0.05 to 0.1~GeV/c,
where $p_k$ is the momentum difference between the incoming
and outgoing momentum vectors in the plane perpendicular to the
field.  The $p_k$ is proportional to the product of magnetic
field strength $B$ and path length $L$. Assuming
a magnetic field path length of $L \simeq 0.4$ m, in order to keep the
spectrometer compact, the respective magnetic field strength stays
below $B$ = 0.9~T.

For such a design, the required momentum resolution can be obtained
only by keeping multiple scattering in the region of large magnetic
field as small as possible (i.e. allowing no detector material in this
region).  For high momentum electrons ($p\sim$\,1~GeV/c), $p_k$\,=\,0.1
GeV/c also puts constraints on the position resolution of the
particle detectors (MDCs) in front and behind the field region. For example, at
$p$\,=\,1~GeV/c and $\theta$\,=\,$20^{\circ}$, the deflection angle $\Delta\theta_k$  amounts to
$5.7^{\circ}$ for $p_k=0.1$~GeV/c.  A simple model calculation
assuming two sets of two detectors each spaced by $d=0.3$~m shows that
for this case a position resolution of better than 150~$\mu$m is
required to keep the corresponding contribution to the momentum resolution below
~1~\%.

\subsubsection{Field geometry}

The toroidal field geometry provides a field free region around the
target and inside the active volume of the RICH. Since the shadow of
the coils can be aligned with the detector frames, no additional
loss of solid angle is caused by the coils. Although the field
strength is rather low, superconducting coils are necessary in order
to obtain a compact coil construction. An additional advantage is the
low operating cost.

\subsubsection{Superconducting coils}

\begin{figure}[htb]
\subfigure{\includegraphics[width=0.385\textwidth]{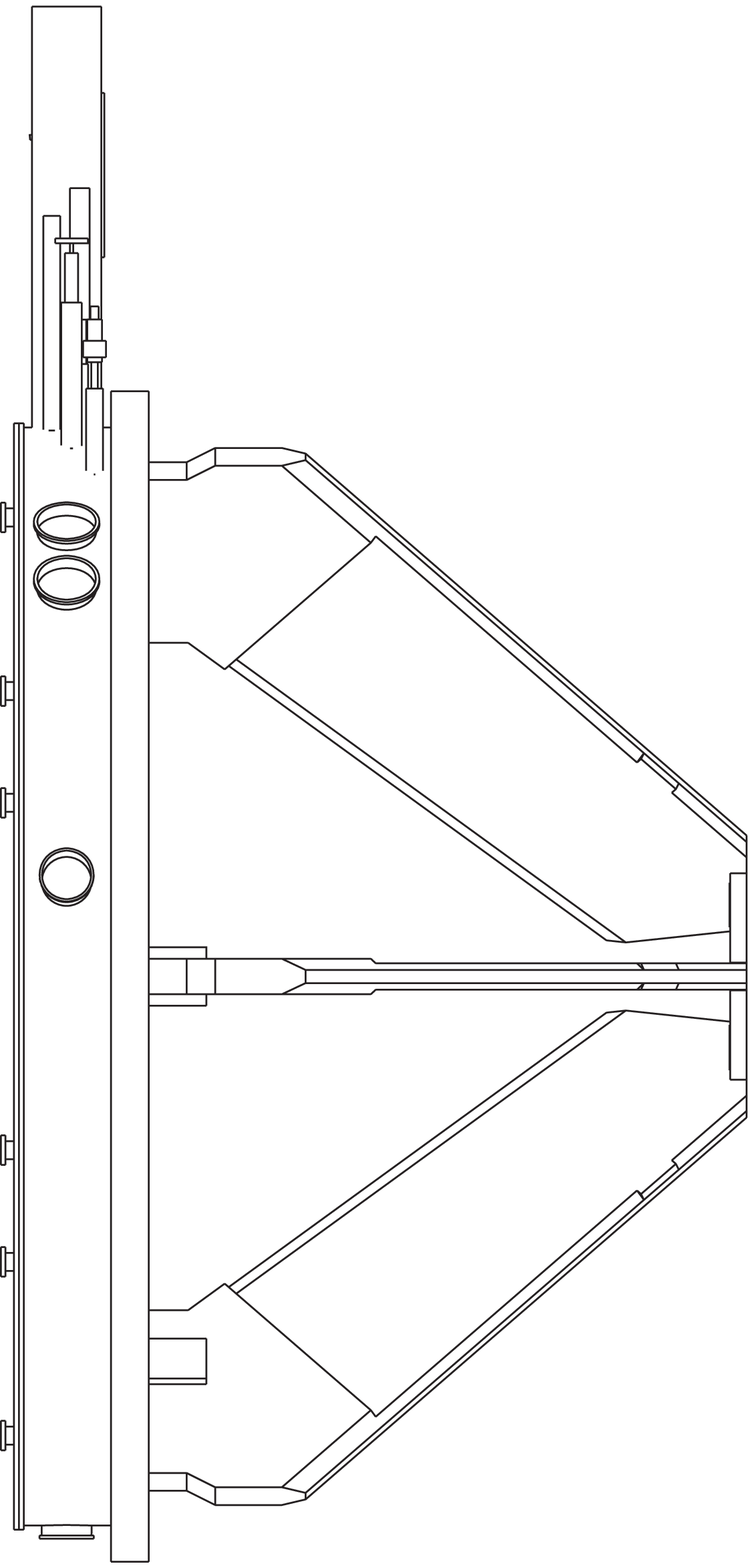}}
\subfigure{\includegraphics[width=0.532\textwidth]{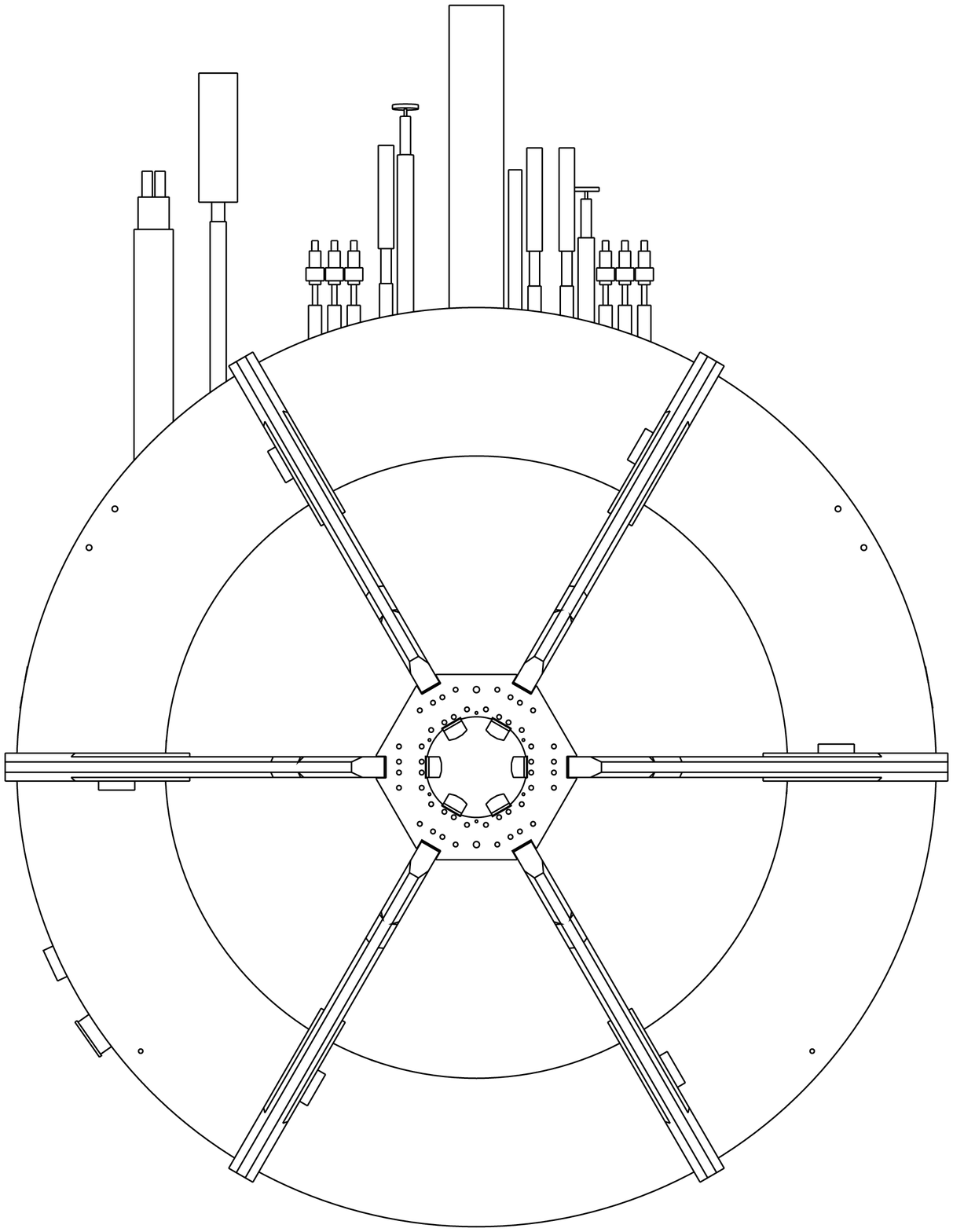}}
\caption[]{Left: Side view of the
superconducting HADES magnet. The outer diameter of the support ring
amounts to 3.56~m. Right: Back view} \label{Magnet}
\end{figure}

The system consists of 6 coils surrounding the beam axis.
Each coil is separately contained in its individual vacuum chamber.
The latter ones are connected to a support ring located upstream of the
target. Figure~\ref{Magnet} shows a side and a back view of the
magnet including the support structure of the coil cases. A
hexagonal plate, with a hole for the beam pipe, connects the back end of
the six coil cases. Through this plate and the support ring, the magnetic forces
acting on the coil cases of about $4.9 \cdot 10^4$~N per coil are
compensated. The ring upstream of the target supports the electrical
connections between the coils as well as the Helium and Nitrogen cooling
lines.  No support structure is needed in the region where the drift
chambers are located.  Furthermore, no material is placed in a
$7^{\circ}$ (starting from the target) cone around the beam axis.

Each coil consists of 2 non-parallel long straight sections connected
by two arcs. The magnetomotive force of a coil amounts to 485000 Ampere-turns.
Each coil has 140 turns, thus 3464 A have to be fed through the
current leads. The angles of the entrance and exit sections of
$40^{\circ}$ and $45^{\circ}$ were chosen to minimize the azimuthal
deflection of particles over the whole range of polar angles.
Due to the V-shape of the coil a small net
focusing - or defocusing, depending on the particle charge -
with respect to the azimuthal angles is obtained. The shape
and orientation of the coil result in a stronger $p_k$ at small
polar angles (see table~\ref{table_kick}). For beam energies of
1~-~2~AGeV, the transverse momentum kick $p_k$ provided by the field
follows roughly the kinematical variation of the particle momenta
with polar angle.

As explained in sect.~\ref{Chapter_kickplane} below, the particle
momentum $p$ can be calculated from the relation
\begin{equation}\label{momentum}
 p={\frac{1}{2}} {\frac{p_{k0}}{{\sin(\Delta \theta_k/2)}}} + p_{k1} + 2 p_{k2} \sin(\Delta \theta_k/2),
\end{equation}
where $p_{k0}$ represents the leading term (see table~\ref{table_kick}).
The coefficients $p_{k1}$ and $p_{k2}$ are correction terms accounting for the
variation of the track length through the field and depend on the sign of the charge.
All coefficients depend strongly on $\theta$ and $\phi$.

\begin{table}
\begin{center}
\begin{tabular}{|l|l|l|l|l|l|r} \hline
$\theta$ & $20^\circ$ & $30^\circ$ & $40^\circ$ & $60^\circ$ & $80^\circ$ \\ \hline
 $p_{k0}$ [MeV/c] at $\phi=0^\circ$ & 109 & 89 & 73 & 55 & 41 \\ \hline
 $p_{k0}$ [MeV/c] at $\phi=15^\circ$ & 123 & 94 & 76 & 61 & 53 \\ \hline
 $p_{k0}$ [MeV/c] at $\phi=25^\circ$ & -- & 99 & 82 & 73 & 85 \\ \hline
\end{tabular}
\vspace*{.5cm} \caption[]{Transverse momentum kick $p_{k0}$ as a
function of the polar $\theta$ and azimuthal $\phi$ angles
($\phi=0$ corresponds to midplane between adjacent coils) for electrons in the limit of large
momenta.} \label{table_kick}
\end{center}
\end{table}

\begin{figure}[h]
\centering
 \mbox{
 \subfigure{\includegraphics[angle=-90,width=0.5\textwidth]{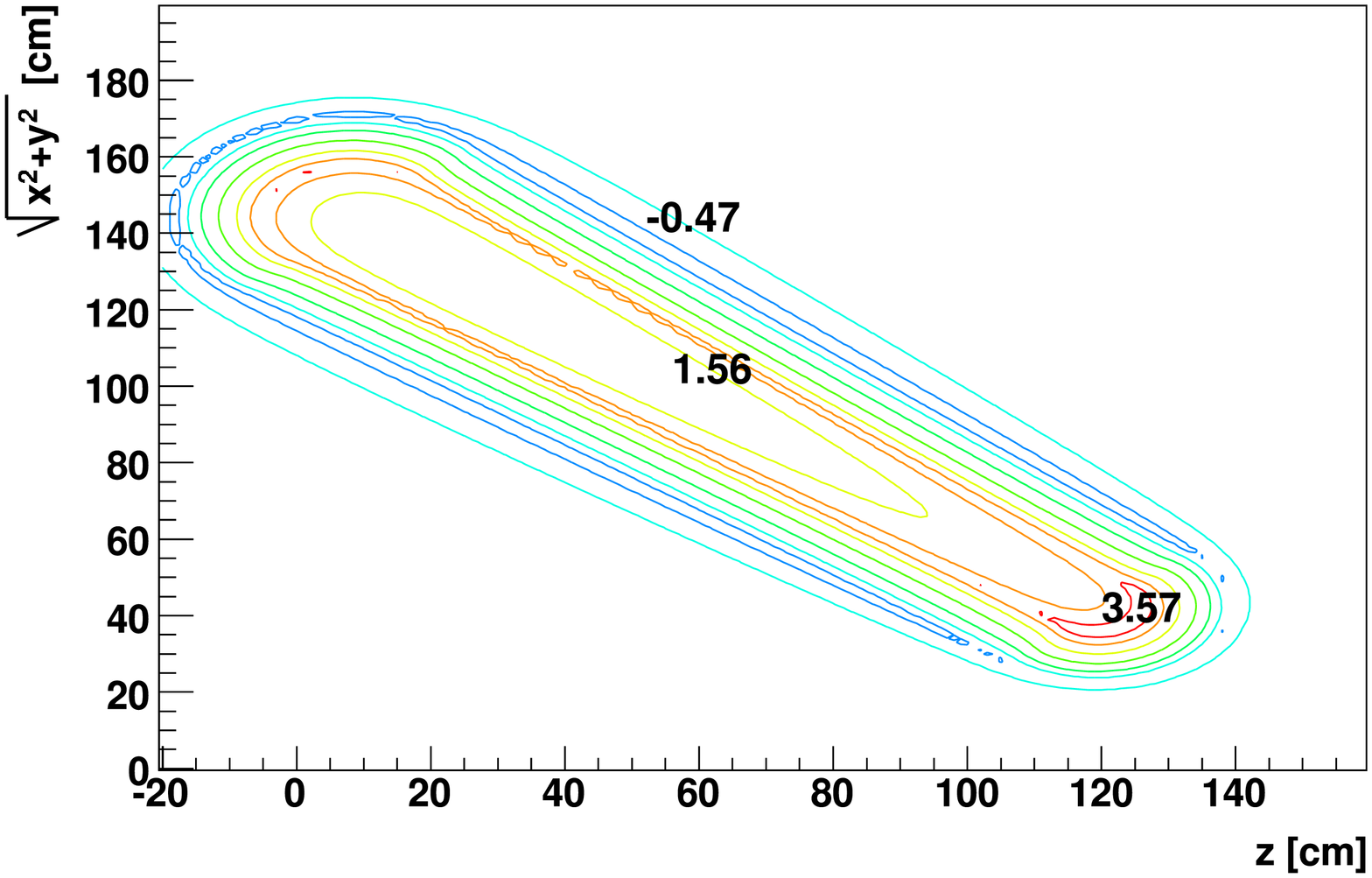}}
 \subfigure{\includegraphics[angle=-90,width=0.5\textwidth]{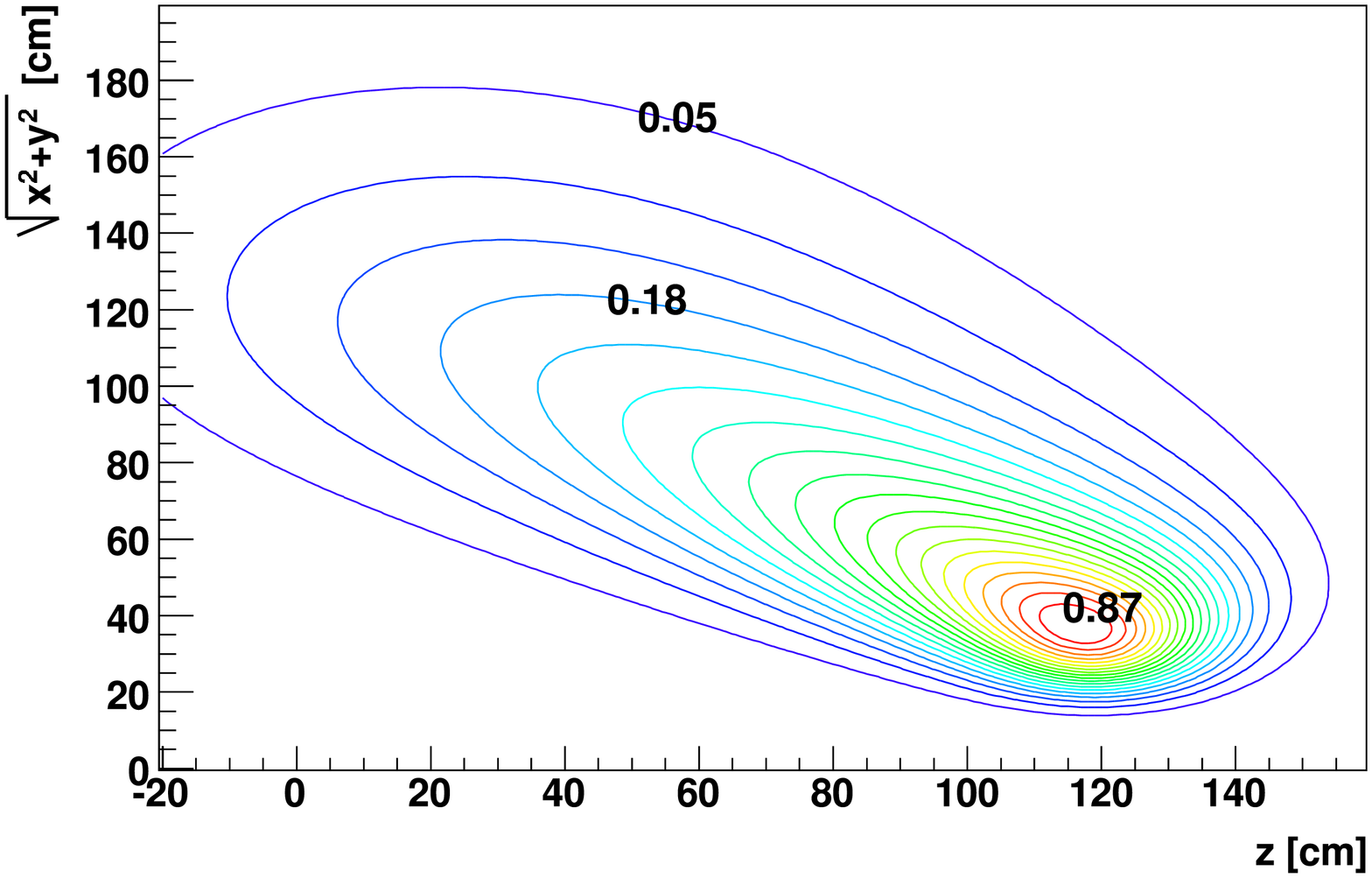}}}
  \caption{Left: Magnetic field maps, $B$ as function of z (along beam axis) and r (per. to the beam axis), at $\phi = 30^{\circ}$ (coil position). The step size for the contour lines amounts to 0.24~T. Right: Field map at $0^\circ$ (midplane between coils). The step
size for the contour lines amounts to 0.046~T.}
\label{field}
\end{figure}

The field maps exhibited in fig.~\ref{field} show the strong
inhomogeneity of the field as a function of both polar and azimuthal
angles. The maximum field is obtained at the forward arc of the coil
and amounts to 3.6~T at the sector edge ($\phi = 30^{\circ}$).
The field of each sector was mapped using Hall probes and
a dedicated optical positioning system. After correcting for the earth
magnetic field, the measured field values agree with the ones calculated by TOSCA
\cite{Tosca}
within better than 1~\%. This shows that we have a full control of the coil
geometry. The agreement with the integrated field
(straight line through the field region) is better than 0.2~\%.

\subsubsection{Cryo plant}

The coils are surrounded by a liquid Nitrogen cooled shield at 85~K.
The flow through this shield amounts to 2~g/s including all shielding
components. The remaining heat load of each coil amounts to 2~W.
Together with the heat load on all other components, the total load
amounts to 20~W excluding the current leads. The current leads are
cooled with He gas, starting at 4.7~K and warming up to about 270~K.
The heat load depends nearly quadratically on the current with a
maximum load of 80~W (corresponding to 0.7~g/s) at full field. Thus,
at full field, the cryo plant (TCF20, \cite{Lindekryo}) has to provide a cooling power
of 100~W, quite close to its 110~W limit. All heat
loads refer to an equivalent cooling power at 4.7~K. In order to
avoid gas bubbles inside the thin He pipes cooling the coils, single
phase He at 4.7~K and 0.29~MPa is used. Above 0.23~MPa (critical point), Helium
remains in the gas phase even at low temperatures, with density above the liquid
phase at pressures below 0.23~MPa. It is afterwards liquefied by
expanding to 0.13~MPa, providing thermal stability via heat exchangers,
connecting thermally the cooling pipes with the liquid reservoir.


\clearpage
\subsection{The RICH}
\label{Chapter_RICH}

\subsubsection{Overview}

\begin{figure}[ht]
\begin{minipage}[]{46mm}
\vspace{-1mm} The radiator gas perfluorobutan (C$_4$F$_{10}$) offers
high transmission down to $\lambda$\,=\,145 \,{\rm nm} and a
suitable Cherenkov threshold (Lorentz factor $\gamma_{thresh}$\,=\,18)
to suppress radiation from muons and hadrons in the given momentum
regime. It surrounds the target station in an essentially field free
region and is confined by a thin Carbon fiber shell at forward angles
(thickness = 0.4 mm), by the photon detector CaF$_2$ entrance window and
by thin Mylar foils on the beam path. The Cherenkov light is radiated from
straight particle trajectories with effective path lengths varying
from 36 cm at $\theta=20^{\circ}$ to 65 cm at $\theta=80^{\circ}$.
\end{minipage}
\hspace*{5mm}
\begin{minipage}[]{67mm}
\vspace*{-5mm}
\begin{center}
\includegraphics[width=67mm]{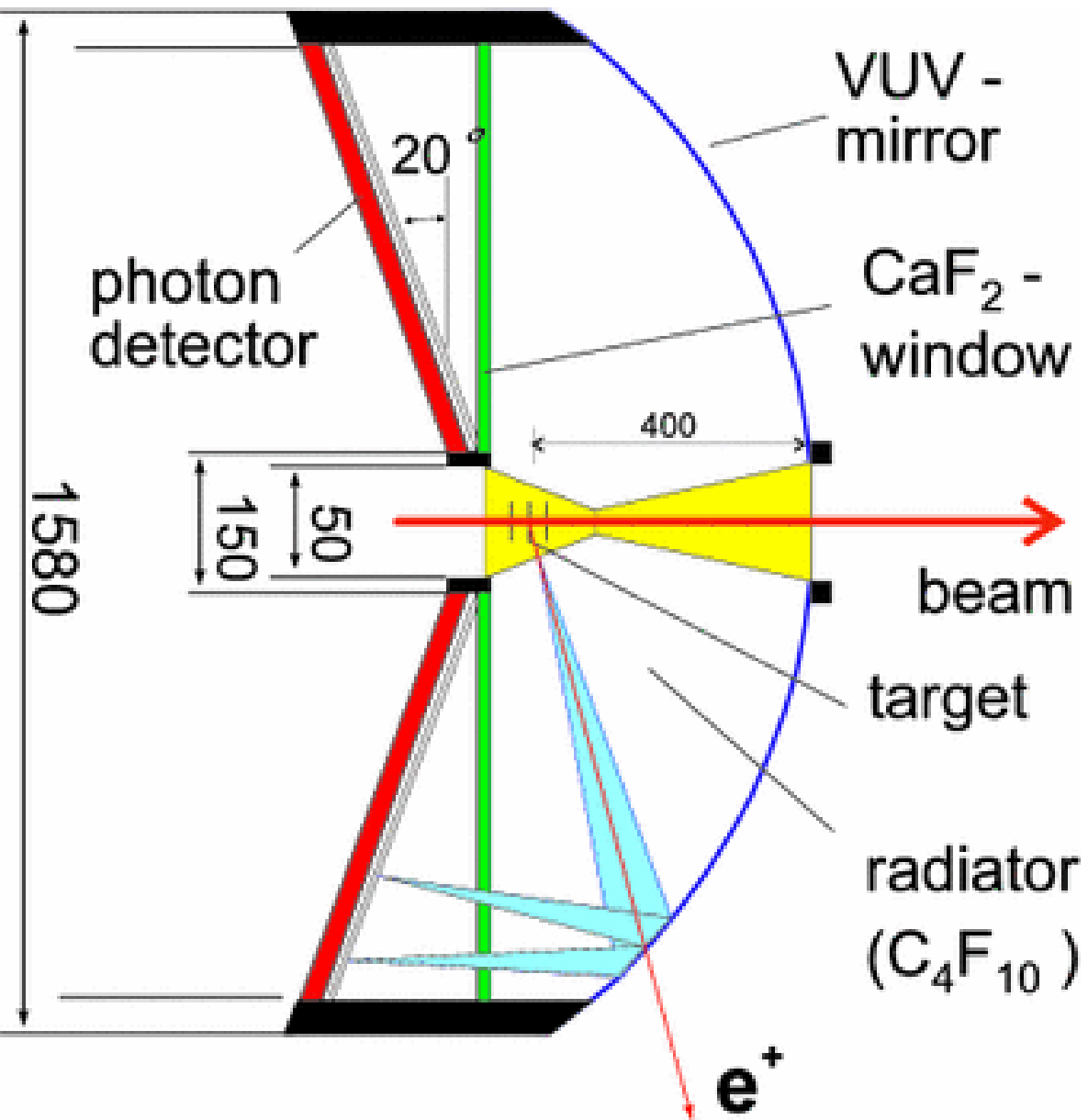}
\end{center}
\caption[]{Schematic layout of the RICH, consisting of a Carbon shell mirror, a CaF$_2$ window and a photon detector. All distances are in millimeter.} \label{rich}
\end{minipage}
\end{figure}

The Ring Imaging Cherenkov (RICH) detector constitutes the innermost
part of the spectrometer and is designed to identify
relativistic e$^\pm$ with momenta 0.1~GeV/c $\le p \le$ 1.5~GeV/c.
The layout, shown in fig.~\ref{rich}, is governed by the
limited space between target and tracking detectors and by the need
for a low material budget along the particle trajectories to
minimize external pair conversion and multiple scattering. The
photon detector is placed upstream of the target to spatially
decouple the registration of the Cherenkov light from charged
particle tracks emitted from the target. The choice of a gaseous
photon detector with a photosensitive CSI cathode restricts the sensitivity to the far vacuum
ultra violet (VUV)  wavelength region.

The photons are reflected by a low mass spherical mirror
(curvature radius $R$ = 872
mm) onto the photosensitive CsI cathodes of six Multi Wire
Proportional Chambers (MWPC) operated with CH$_{4}$ and equipped with
individual pad readout. The optical geometry is chosen such that the
photons are focused to rings of almost constant diameter across the
whole detector plane. The measured ring center positions are used to
disentangle lepton and hadron tracks in high-multiplicity central
heavy-ion collisions.

In the following sections we summarize the most important aspects of
the photon detector, of the mirror and of the window and present results from
in-beam measurements. More detailed information on the various RICH
components can be found in refs.
\cite{zeitel99,frieseCsI99,kasten99,friese03,pegasus}.

\subsubsection{Photon detector}
\label{phdet}

\begin{figure}[ht]
\begin{center}
\includegraphics[width=90mm]{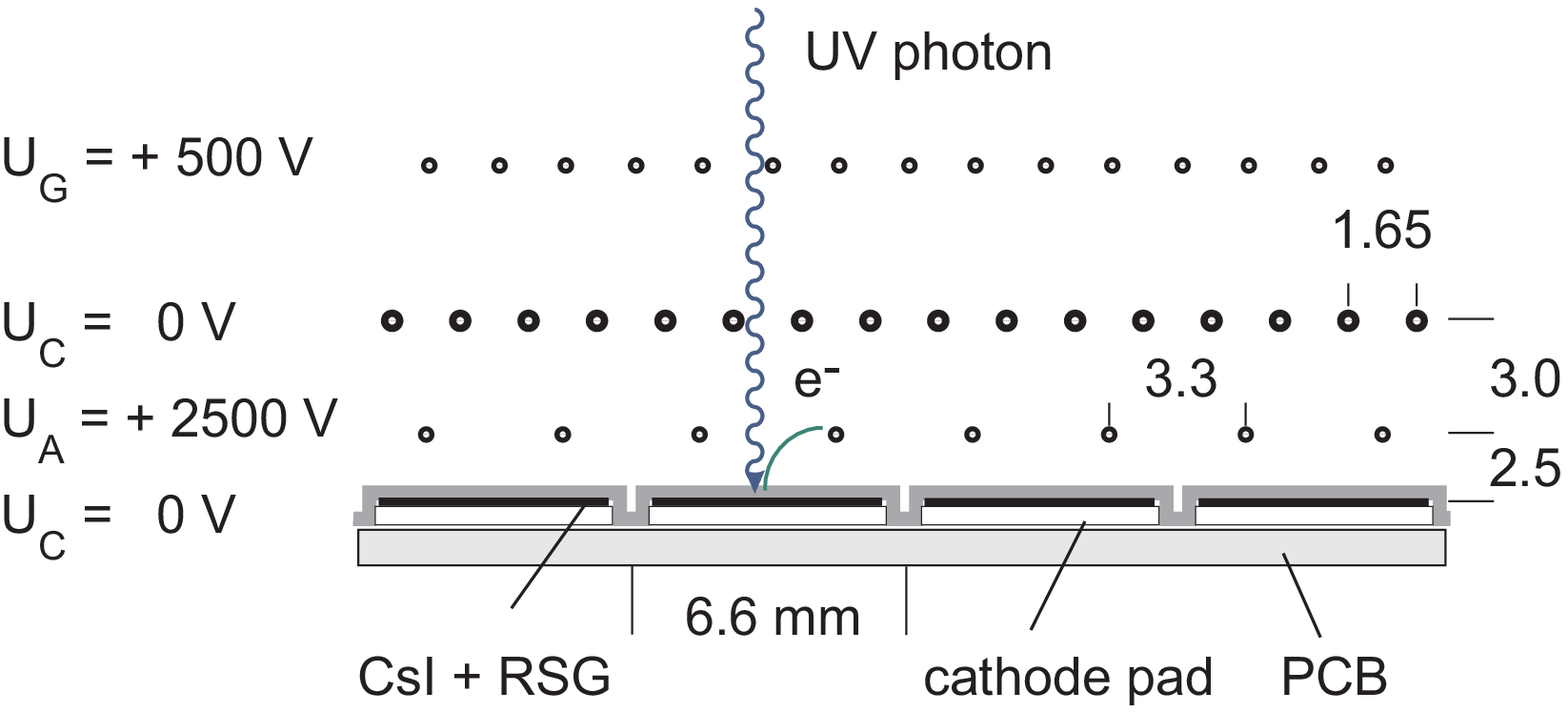}
\end{center}
\caption[]{Schematic cross section of the MWPC part of the photon detector.
The different wire planes comprise a gate, a cathode and an anode.
All distances are given in millimeter.
The asymmetric field configuration increases the charge fraction induced onto the pads to about 70~\%. The entrance window is not shown.}
\label{padcathode}
\end{figure}

The RICH photon detector is assembled around the beam pipe from six
modules of trapezoidal shape (area $\simeq 0.25$\,m$^2$) such that the
sensitive planes form a hexagonal pyramid and approximately match
the curved focal plane of the mirror. Each module consists of a thin
gap ($d$ = 5.5 mm) MWPC with
asymmetric field configuration and pad cathode readout (see
fig.~\ref{padcathode}). The anode and cathode planes are built from
$d_A=20~\mu$m and $d_C=50~\mu$m thick Gold-plated tungsten wires,
respectively. An auxiliary gate (at $U_G$\,= +500 V) anode separates the amplification
region from freely propagating electrons produced by ionisation
processes in the passive gas volume between the MWPC and the CaF$-2$ entrance
window. All wires are oriented in radial direction. The photosensitive
cathode plane is segmented into 4712 pads on a Printed
Circuit Board (PCB) with Gold-plated conductive layers. The pads are
individually coated with Resin Stabilized Graphite (RSG) and a
reflective CsI layer \cite{frieseCsI99} acting as photon-electron
converter. Operated with pure Methane at atmospheric pressure, the
chambers run with anode voltages of 2450\,-\,2550 V corresponding to
visible gas gains of $(3-9)\cdot 10^4$.

The mismatch of pad plane and mirror focal surface leads to slightly
varying image shapes and sizes across the whole sensitive area. With
increasing polar angle of e$^\pm$ tracks the Cherenkov images turn
from rings to ellipses. To first order, the eccentricity of the ring
images is compensated by a variation of the pad size along the
radial direction, {\emph i.e.} the direction of the wires. With a constant
pad width of $l_x = 6.6$\, mm perpendicular to the wires, the pad
length varies between $l_y = 7.0$\,mm and $l_y = 4.5$\,mm leading to
rings of almost constant radius ($\simeq 3.8$\,pads) for all track
angles. This facilitates the on-line ring search via simple pattern
matrix algorithms implemented in hardware and allows to provide
second-level trigger decisions for events with e$^\pm$ tracks
\cite{preformance-RICH-j-lehnert}.

The 28272 cathode pads are connected to Preprocessing Front-end
Modules (PFM) mounted on the rear side of the cathode PC-board. Each
PFM provides 64 charge-integrating amplifiers based on the GASSIPLEX
ASIC \cite{gassiplex}, ADC, zero suppression and event data memory.
Using a mixed parallel and daisy-chained mode, 75 PFMs per detector
module are cascaded and connected to two VME-based Readout
Controller modules (RC). These provide the interface to the ring
processing unit \cite{lehn99} for on-line ring search and the central
data acquisition. For details of the RICH readout electronics see
also ref.~\cite{kasten99}.

\subsubsection{VUV mirror, CaF$_2$  window and gases}

A high e$^\pm$ identification efficiency requires a sufficiently
large number of detected photons per Cherenkov ring ($N^{\gamma}_{det} \ge
10$). The short radiator length (i.e. the small number of radiated
photons) and the work function of the solar blind CsI photocathode
($\lambda_{max} = 220$\,nm) enforce a spectral sensitivity in the
VUV as large as possible. With the lower wavelength limit given by
the transmission cutoffs of the gases ($\lambda_{min} \simeq
145$\,nm), this requirement can only be met by a mirror substrate of
very low surface roughness with a MgF$_2$ protected aluminum coating
and a CaF$_2$ entrance window for the photon detector.

The mirror design goal was a low mass substrate with a thickness $d$
 such as $d/X_0 \le 1$~\%, comparable to the ones of the target and
radiator materials. Simultaneously, the surface and optical imaging
quality should guarantee for almost uniform rings with minimum
distortions of ring shapes across the whole detector plane. This
would allow both on-line e$^\pm$ recognition with a single ring
finding algorithm implemented in hardware and the extraction of hit
points from high resolution ring center analysis to be used as an
additional information for lepton tracking. These requirements
lead to a spherical mirror with the properties as listed in
table \ref{tab1}.
\begin{table}[h]
\begin{center}
\begin{tabular}{|l|r|}
\hline
Parameter & Value\\
\hline
Outer diameter $D$  & 1440 mm   \\
Radius of curvature $R_i$ & 872 mm  \\
Substrate thickness $d$   &  $\le2$ mm    \\
Radiation length $X_0$   & $\ge 20$ cm  \\
Reflectivity $R$ ($\lambda = 150 $nm) & $\ge 70$ \% \\
Surface roughness $\sigma$ (rms)  & $\le 3$ nm \\
Surface slope error SSE$_{80}$   &  $\le 1$ mrad \\
Areal density $\rho d$  & $< 0.3$ g/cm$^2$ \\
\hline
\end{tabular}\\[2pt]
\caption[]{Geometrical and optical parameters of the HADES RICH
mirror.}
\label{tab1}
\end{center}
\end{table}

\begin{figure}[ht]
\begin{minipage}[t]{52mm}
\begin{center}
\includegraphics[width=52mm]{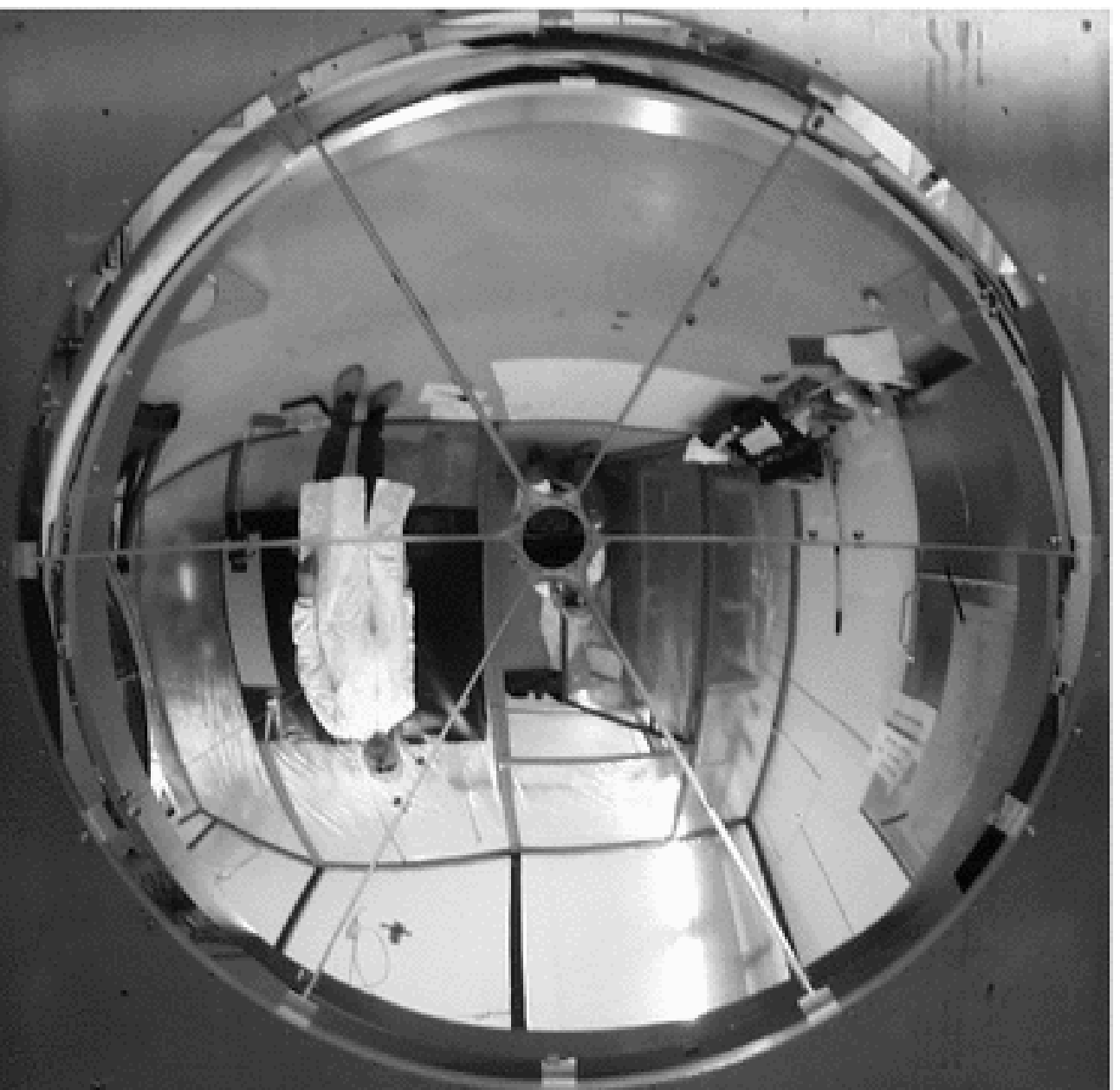}
\end{center}
\end{minipage}
\hfill
\hspace*{.5cm}
\begin{minipage}[t]{62mm}
\begin{center}
\includegraphics[width=62mm]{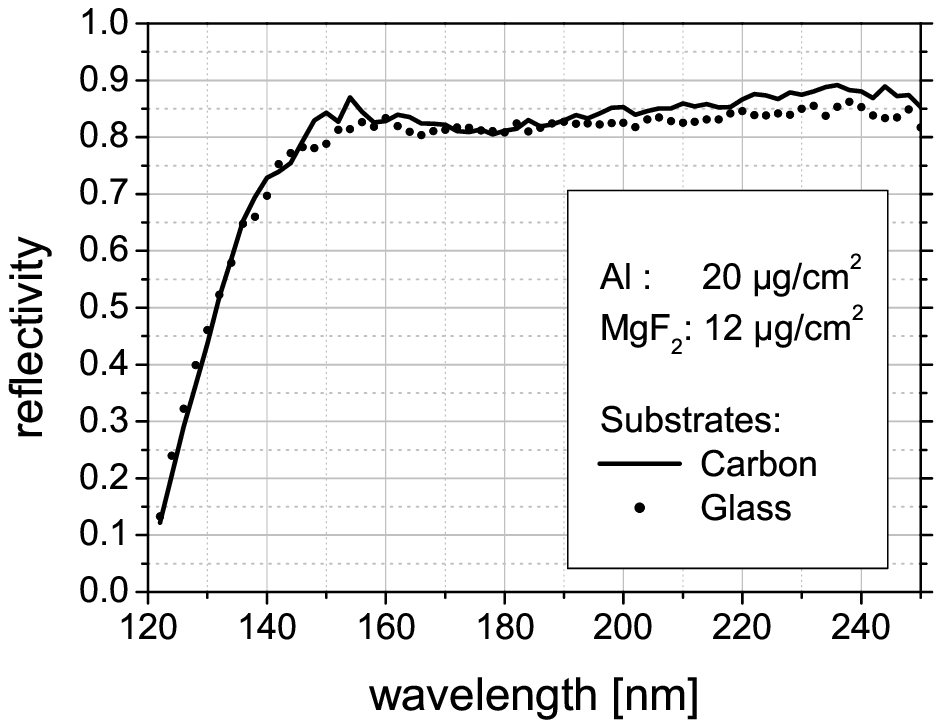}
\end{center}
\end{minipage}
\caption[]{Left: Photo of the RICH mirror. Right:
Reflectivities of glassy Carbon samples in comparison to
float glass samples.} \label{mirror}
\end{figure}

Since a self-supporting mirror shell made from a single piece was
ruled out, we have developed and constructed, together with
DSS\footnote{DSS~Dornier~Sat.~Syst., D-81663~M\"unchen,~Germany.}, a
mirror with a segmented substrate shell and a six-fold radial support
structure with spokes matched to the coil cases of the
superconducting magnet. As substrate material we have
chosen glassy Carbon, an isotropic and homogeneous material produced
by pyrolytic conversion of Phenol type resins \cite{HTW84} and
commercially available as Sigradur$^{\textrm\textregistered}$
from the company HTW\footnote{HTW GmbH, D-86672 Thierhaupten,
Germany.}. After production, the eighteen substrate panels (3 for
each sector) have been individually grinded to 2\,mm thickness,
polished to the required surface roughness and finally machined to
the desired shape. The panels were coated with a 20\,$\mu$g/cm$^2$
Aluminium reflective layer followed by a 12\,$\mu$g/cm$^2$
MgF$_2$ protection layer. The deposition was performed
in the Ultra-High-Vacuum box coater \cite{komor99} installed at
Technische Universit\"at M\"unchen
following the procedure described in~\cite{komor02}.

The optical quality of the panel surface was measured in terms of
surface slope errors SSE$_{80}$ and was found to change from
typically 0.5 mrad to 0.8 mrad. The surface roughness and the
achievable VUV reflectivity were evaluated with various methods from
witness samples obtained during the panel cutting process. Both, the
visible surface roughness and the VUV reflectivity show good
agreement with those from float glass samples and are consistent
with the assumption of a micro roughness of $\sigma \simeq 2 -
3\,$nm. Measured reflectivities are shown in the right panel of
fig.~\ref{mirror} and exhibit constant values around $R$\,=\,80~\%
down to 150~nm. Figure~\ref{mirror} shows the mirror
after assembly under clean room conditions and prior to integration
into the RICH system. For details of the mirror design, material
parameters, production techniques, and optical quality mesurements
the reader is referred to ref.~\cite{friese03}.

\begin{figure}[h]
\begin{minipage}[t]{52mm}
\begin{center}
\includegraphics[width=52mm]{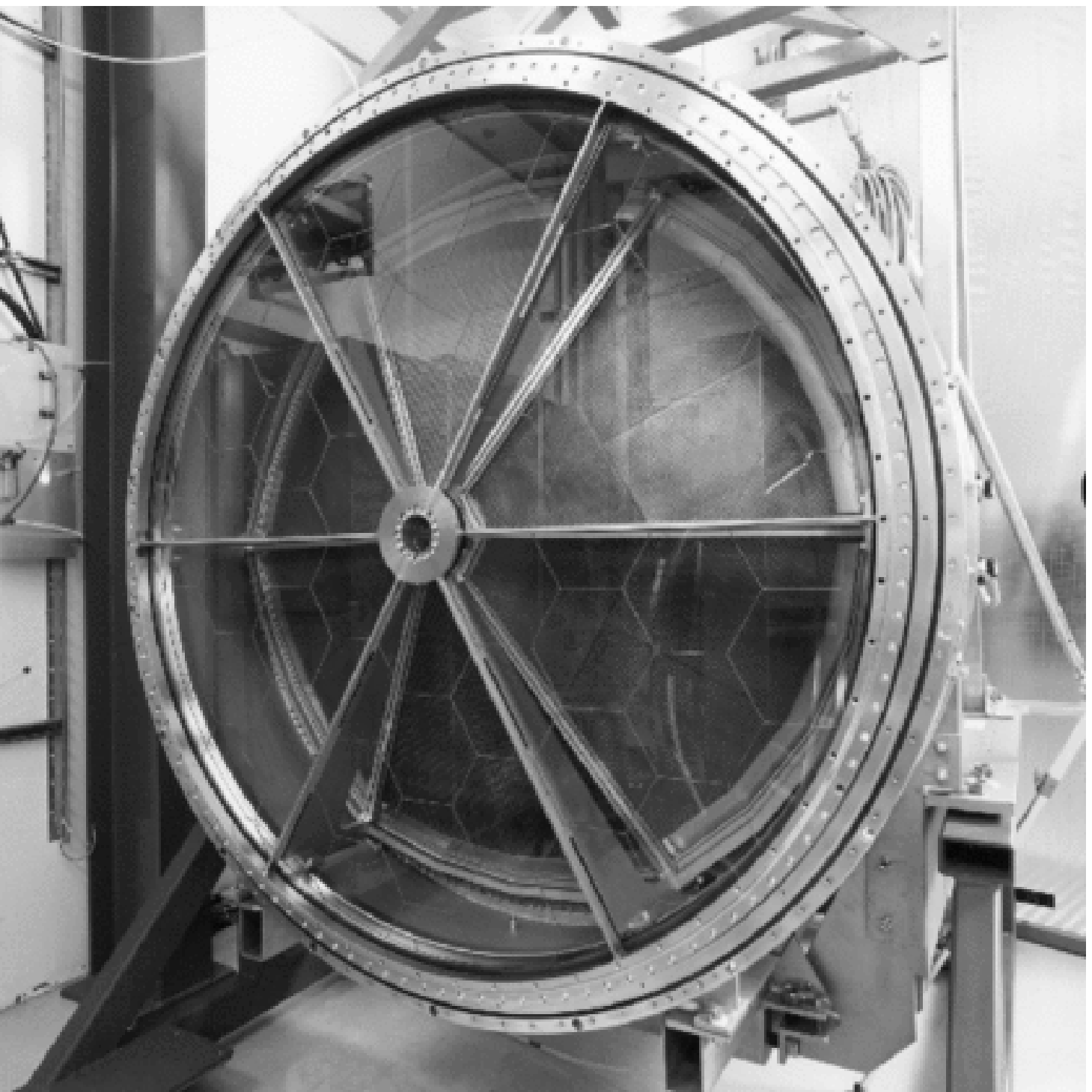}
\end{center}
\end{minipage}
\hfill
\hspace*{0.5cm}
\begin{minipage}[t]{62mm}
\begin{center}
\includegraphics[width=62mm]{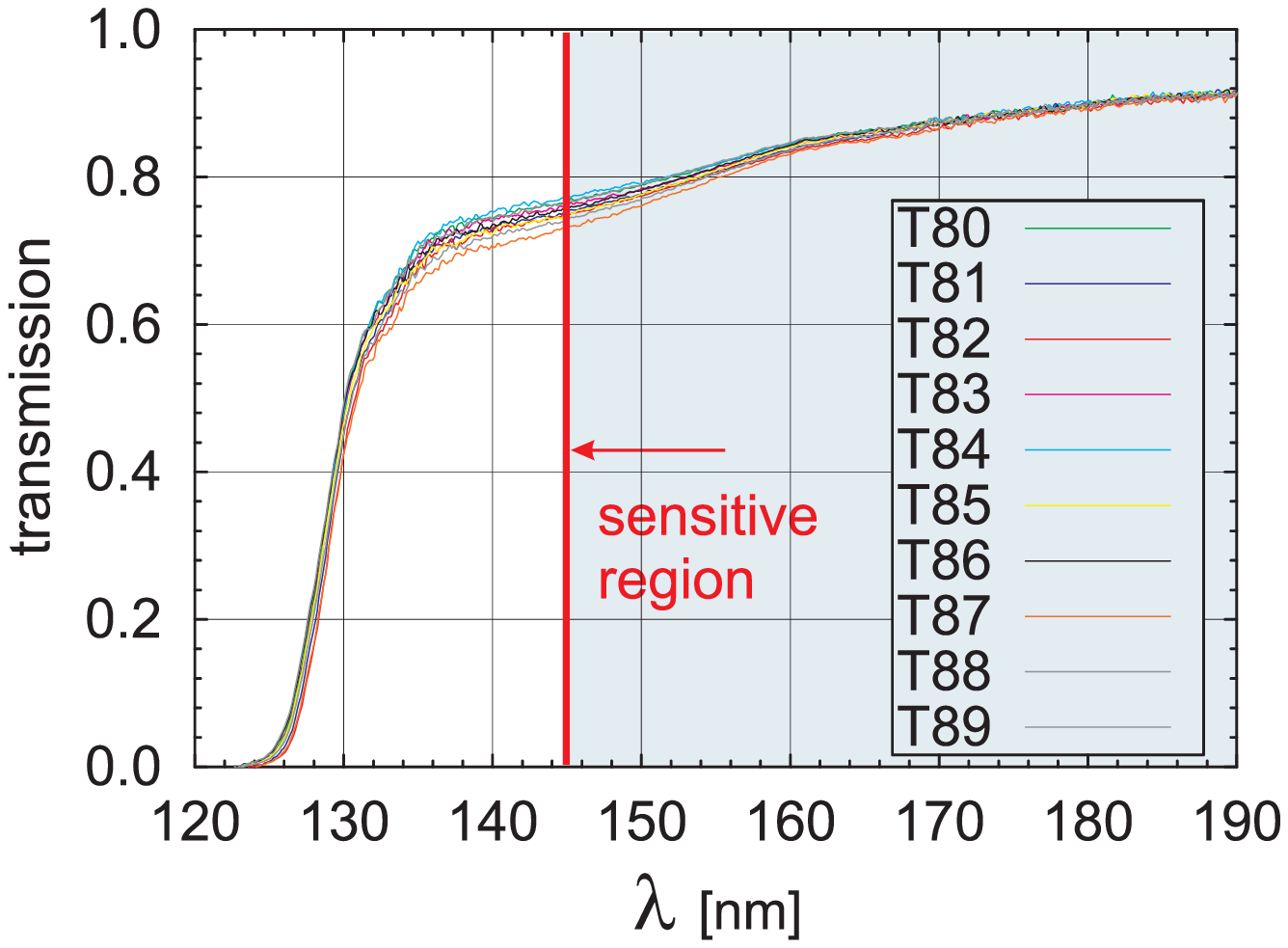}
\end{center}
\end{minipage}
\caption[]{Left: View of the photon detector with the assembled CaF$_2$ entrance window.
Right Transmissions of individual CaF$_2$ crystals T80 to T89.}
\label{window}
\end{figure}

The CaF$_2$ window separates the gas volumes of radiator and photon
detector. It was assembled from altogether 64 hexagonally shaped
single crystals\footnote{Korth Kristalle, D-24161~Altenholz.}
(200\,mm in diameter, 5\,mm thick each). A high VUV
transmission was achieved through proper selection of the CaF$_2$
raw material followed by an optimized polishing technique and was
verified for each crystal individually (fig.~\ref{window}).
The polished crystals were glued together ($\simeq 100~\mu$m
3M Scotch DP190) to one single disk of 1500 mm diameter (see
fig.~\ref{window}) with the central hexagon containing a hole for
the beam tube. Installed in vertical position, the disk is connected
to stainless steel mounting frames through a 2~mm thick and 10~mm
wide Viton$^{\textrm\textregistered}$
interface to damp the influence of the different thermal
expansion coefficients and to minimize mechanical stress effects due
to gravitational forces. Six thin spokes support the window on both
sides against pressure differences.

\begin{figure}[h]
\begin{center}
\includegraphics[width=90mm]{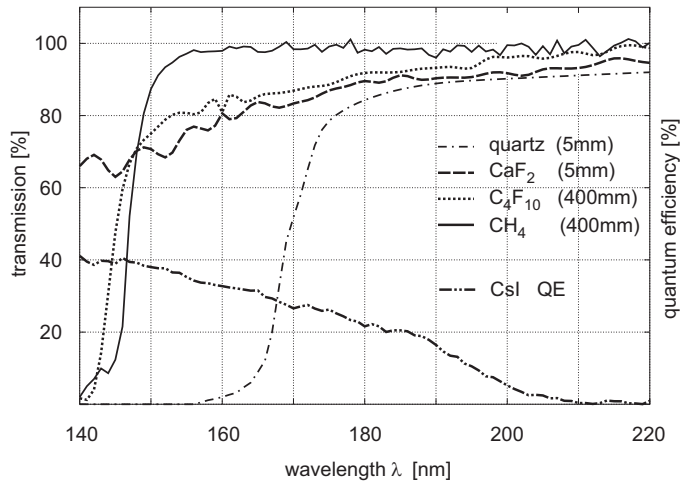}
\end{center}
\caption[]{Measured transmissions of radiator, detector gases, CaF$_2$ window and quartz (for comparison) are plotted (left abscissa) together with the the CsI quantum
efficiency (right abscissa).
}
\label{transall}
\hfill
\end{figure}

The gas volumes of radiator and photon detector amount to about 700
liters each and are filled to $\simeq 10\,-\,40$\,hPa above ambient
pressure. The photon detector  is supplied from commercially
available bottles with Methane (CH$_{4}$, purity 4.5) through an
open system at a typical gas flow of 300 - 350 l/h. For the
radiator, a batch of prepurified radiator gas C$_4$F$_{10}$ (3M
CEA410) is stored in a liquid reservoir and recirculated in a closed
system via an evaporation and compression-liquefaction chain. The
observed gas losses amount to about  1.5 - 2 l/h at a standard flow
of 300 l/h. Absolute pressure and gas flows of both systems are
steered and monitored by a PLC-controlled gas supply
system~\cite{pegasus} which also keeps the pressure difference
$\Delta P$
between radiator and MWPC below 3\,hPa to protect the
fragile CaF$_2$ window. The purity of the gases is monitored on the
inlet and outlet side through frequent VUV transmission measurements
utilizing a D2 light source combined with a standard monochromator
grid-photodetector setup. The achieved gas transmissions are
compiled in fig.~\ref{transall} together with the ones for CaF$_2$,
fused quartz, and the CsI quantum efficiency.

\subsubsection{Performance and results from in-beam measurements}\label{chapter_richperf}
The performance of the RICH is governed by the average number of
photons detected for each e$^{\pm}$ induced ring. The number of
detected photons per ring is
$\, N^{\gamma}_{det} \approx N_{0} \, Z^{2} \, D \, \overline{ \sin^{2} \Theta_{C}(\lambda)}$.
For relativistic e$^{\pm}$  ($Z=\pm 1$, $\beta \simeq 1$, $\Theta_{C}
\simeq 3.15^\circ$ is the Cherenkov angle) it depends on both the available
radiator length $D$
and a figure of merit
$N_{0} = k \,
\int \limits_{E_{1}}^{E_{2}} \epsilon (E_{\gamma}) \cdot dE_{\gamma}$,
where $k$\,=\,379~eV$^{-1}$cm$^{-1}$ and
 $\epsilon (E_{\gamma})$ is the global detection efficiency including the optical and
 electrical properties of the system.  The optical transmissions
 (fig.~\ref{transall}) and the parameter values measured for small samples
translate to an optimum N$_{0}\approx$\,109
corresponding to 10 $ < N^{\gamma}_{det}<$18.

Simulations have shown that for values $N^{\gamma}_{det}<9$ the
efficiency for e$^{\pm}$ identification and hence the on-line trigger
efficiency are significantly affected. To check the achieved
performance of the fully assembled system we have measured $N_{0}$
for all six sectors with a dedicated ion beam setup. As a
"calibrated" VUV light source we have used the known amount of
Cherenkov photons radiated from single relativistic Carbon ions
($E$\,=\,0.6~AGeV, $\beta = 0.794$) when passing through SiO$_2$ and
MgF$_2$ crystals placed close to the normal target position. The
radiator thickness and hence the light output was chosen such that
individual photons could be spatially resolved on the cathode pad
plane. The method is described in detail in ref.~\cite{fabbietti03}.
The measured pulse height spectra and angular distributions of the
photons were used to quantitatively model the electronic detector
response needed for the general HADES detector simulation package.
From the analysis of the single photon induced pad clusters the
aforementioned gas amplifications of about $(3-9)\cdot10^4$ and a
single electron detection efficiency $\epsilon \simeq 95$ \% were
deduced. Counting the number of registered photons leads to
experimental figure of merit values $80<N_{0}<100$ varying
slightly from sector to sector. The measured average light yield is
about 10\,-\,20~\% lower than expected from the small sample
laboratory measurements. We cannot identify a well defined reason
for this discrepancy and rather attribute it to deficiencies and
inhomogeneities across the whole detector area for all involved
components and parameters.

\begin{figure}[\hbt]
\centering \mbox{
\subfigure{\includegraphics[width=0.48\linewidth]{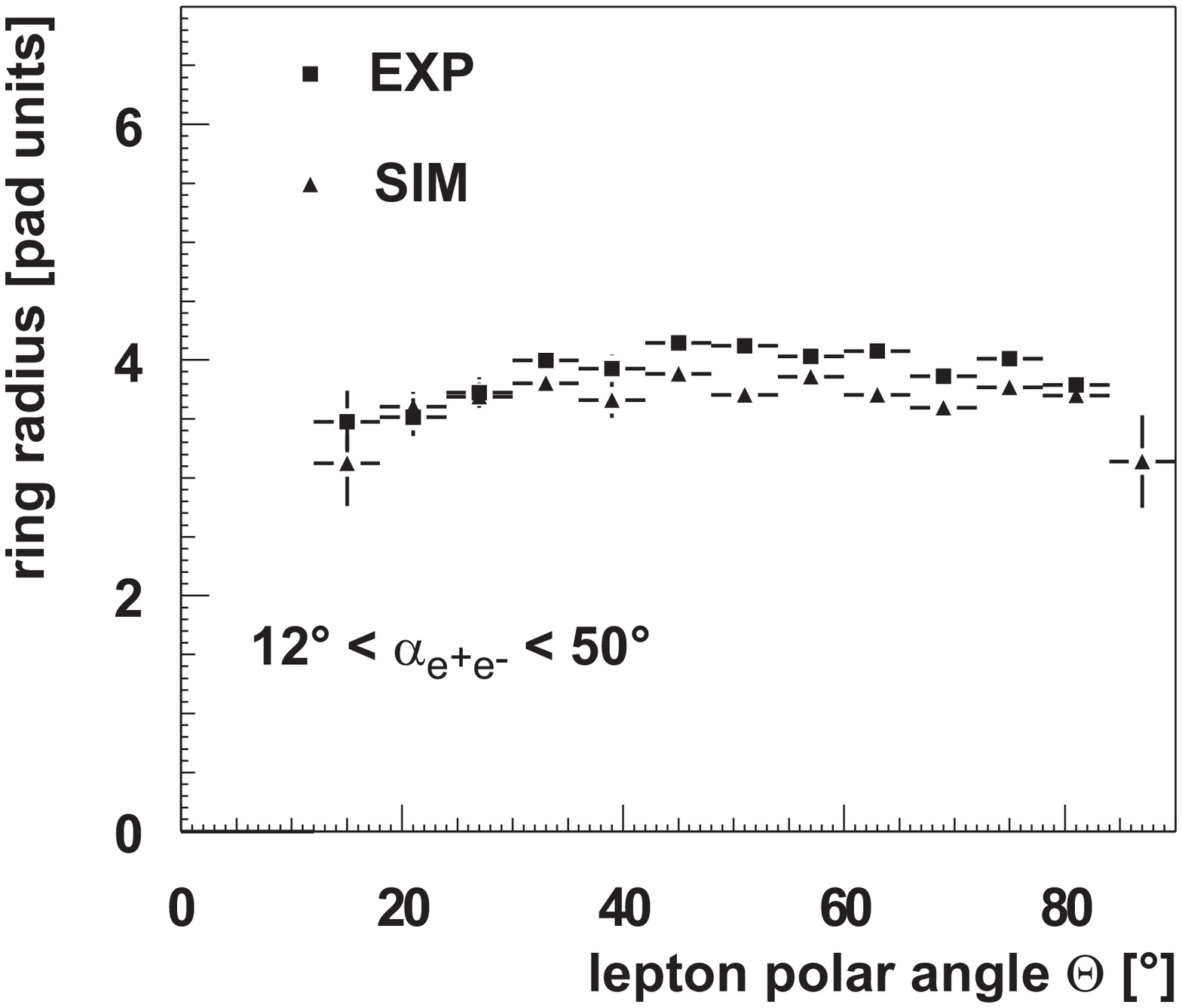}}
\subfigure{\includegraphics[width=0.52\linewidth]{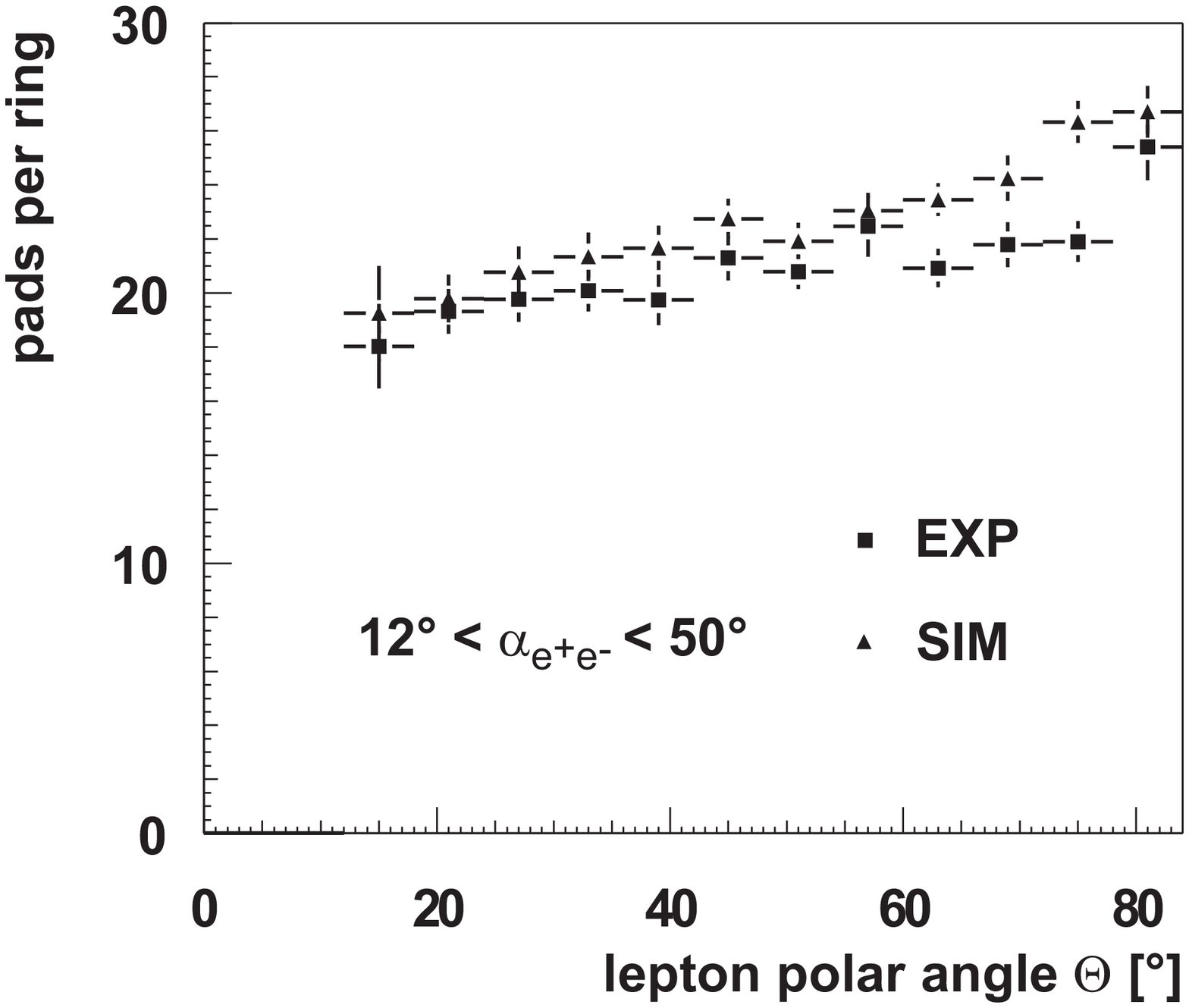}}}
\caption{Polar angle distributions of Cherenkov ring properties for
single electrons from open e$^{+}$e$^{-}$- pairs produced in the
reaction C\,+\,C at 1~AGeV (exp). Simulation results (sim) are shown for comparison.
The error bars reflect the widths of the nearly Gaussian
multiplicity distributions at each angle. Left: Ring radius vs lepton polar angle 
Right: Mean number of pads within a region of 13$\times$13 pads around ring centers.}
\label{rings}
\end{figure}

To verify the obtained results we have analyzed the detector
response also for relativistic electrons and positrons emitted as
open pairs (opening angle $\alpha_{e^+e^-}>12^{\circ}$) in fully
reconstructed $\pi^0$ Dalitz decays from C\,+\,C collisions at 1 AGeV
beam energy. These dielectrons are emitted from the target and
hence exhaust the full radiator length, in contrast to the dominant
background of e$^\pm$ candidates from external pair conversion. In
fig.~\ref{rings} we have plotted the measured ring radius and
the average number of fired pads within a
region of 13$\times$13 pads around the ring center for various polar
angles of the emitted electrons. The data have been obtained by
averaging over the full azimuthal detector acceptance, i.e. all six
MWPC modules, and are characterized by nearly Gaussian multiplicity
distributions at each polar angle. The small variation of ring radius
nicely reflects the adjusted pad design of the cathode plane (see
sect.~\ref{phdet}). The continuous rise of pad multiplicity
with polar angle is a consequence of the increasing radiator length,
from about 36 cm to 65 cm. Averaged over all 6 sectors,
one observes experimentally a mean of
$19.3 \pm 2.8$ pads around $\theta=25^\circ$ and $22.3 \pm 2.0$
pads around $\theta=75^\circ$. The large widths partly reflect
variations of optical and electrical performance of the different
detector modules. Comparing the experimental values with those from
simulations based on the experimental $N_{0}$ values yields a
reasonable agreement. The differences at larger angles point to
possible losses in radiator transmission at longer path lengths
and/or to local imperfections in photocathode quantum efficiency or
mirror reflectivity.

In summary, the overall performance of the RICH detector system,
although slightly lower than in the original design, has proven to
be sufficient for e$^{\pm}$ identification in nuclear collisions (see
sect.~\ref{pair_reco}).

\clearpage

\subsection{Tracking system}
\label{Chapter_MDC}

\subsubsection{Overview} \label{Section_Tracking_overview}

The high-resolution  spectroscopy of vector mesons
($\sigma_{M_{e^+e^-}}/M_{e^+e^-} \simeq 2.5\%$) in heavy-ion collisions via their dielectron
decay channel defines the decisive design and performance
constraints on the HADES tracking system. To reach this goal,
an intrinsic spatial cell
resolution of the order or better than 150~$\mu$m
 along with the
reduction of multiple scattering in detector materials and air,
high efficiency and a large acceptance are crucial requirements
for the success of the experimental program. Extended design studies
and prototyping~\cite{tracking-1,tracking-2,tracking-4} preceded the production
of the 24 Mini
Drift Chambers (MDCs) of four different sizes, conducted by GSI Darmstadt,
LHE/JINR Dubna, FZ Dresden-Rossendorf~\cite{tracking-3}, IPN Orsay
and University of Frankfurt.

The HADES tracking system consists of 24 trapezoidal planar
MDCs symmetrically arranged in six identical sectors.
It provides a polar angle coverage between $18^0$ and $85^0$
around the beam axis, forming four tracking planes (I-IV) of
increasing size. In each sector, two modules (planes I and II) are
located in front of and two (planes III and IV) behind the toroidal
magnetic field of the superconducting magnet, as shown in
fig.~\ref{tracking-setup}. The region between the six coils of the
magnet dictates the active area of the chambers.

During the construction special emphasis was put on the use of
low-mass materials for window foils (aluminized Mylar), wires
(bare aluminum) and counting gas (Helium-based) in order to
minimize multiple scattering. The total detector thickness per
chamber is about 5$\cdot$10$^{-4}$ in units of radiation length,
giving a total close to 0.2 $\%$, whereas the air in the tracking
system represents 0.3~$\%$. These values allow to achieve the
momentum resolution needed to accomplish the physics demands as will
be shown in sect.~\ref{Chapter_tracking_performance_momentum}.

The technical features are summarized
in sect.~\ref{Section_detector_characteristics}. Section \ref{Section_Tracking_alignment} briefly
sketches the aspects concerning the alignment of the different
modules with respect to each other.
In sect.~\ref{Section_tracking_performance}, the performance parameters of
the drift chambers regarding detection efficiency, track reconstruction efficiency and
spatial resolution are discussed. Section~\ref{Chapter_tracking_performance_momentum}
presents the corresponding achieved momentum resolution.

\subsubsection{Detector characteristics} \label{Section_detector_characteristics}

\begin{figure}[t]
\begin{minipage}[t]{65mm}
\begin{center}
\includegraphics[height=62mm,width=58mm]{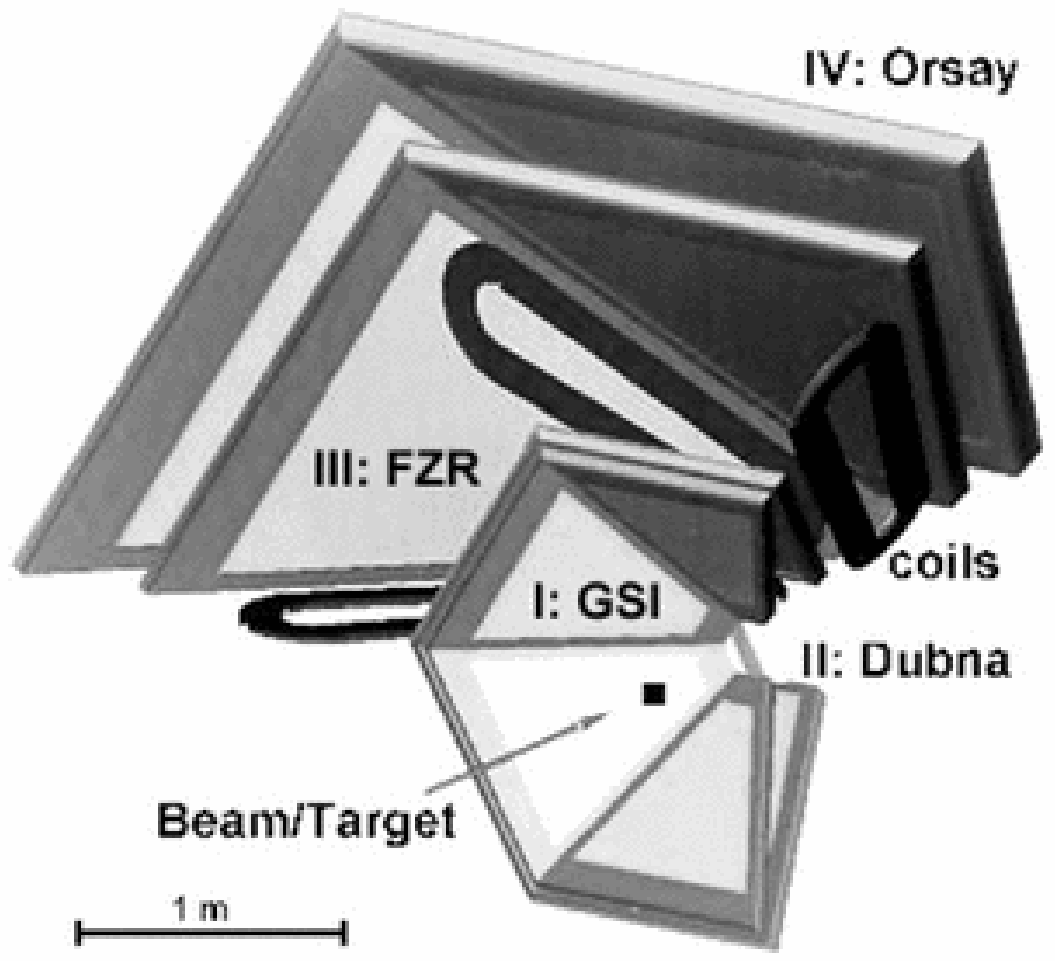}
\end{center}
\end{minipage}\hfill
\begin{minipage}[t]{65mm}
\begin{center}
\includegraphics*[height=65mm,width=58mm]{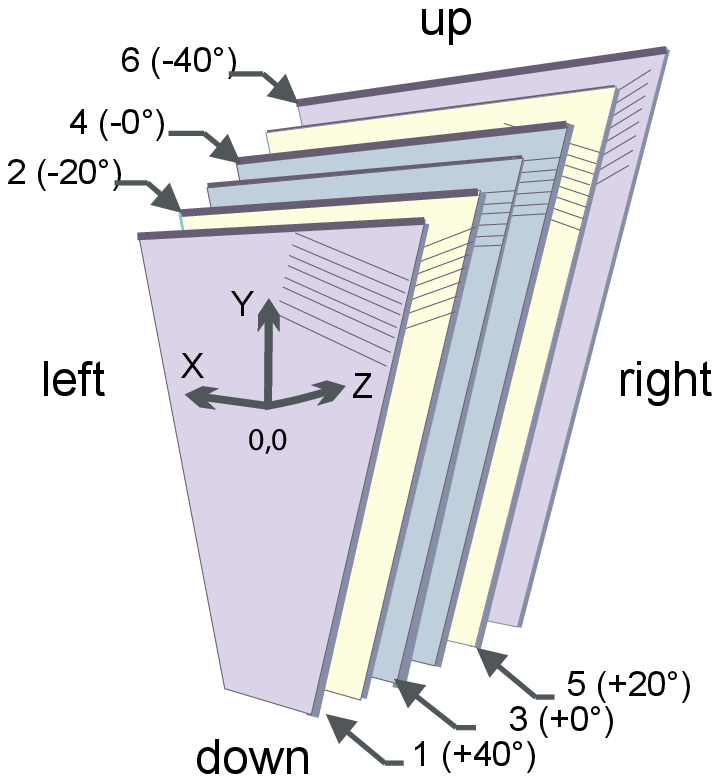}
\end{center}
\end{minipage}\hfill
\caption[]{Left: Schematic layout of the HADES tracking system. Two
sets of Mini Drift Chambers (MDCs) with 24 wire planes per sector
are placed in front and behind the magnetic field to measure
particle momenta. Right: Schematic view of the six anode wire frames
inside a HADES MDC, representing the six stereo angles.}
\label{tracking-setup}
\end{figure}

To cope with ambiguities in track reconstruction in the high
multiplicity environment of a heavy-ion reaction (for central
Au\,+\,Au collisions at 1~AGeV incident energy a maximum cell occupancy of
30~\% is estimated), all chambers are composed of six sense/field
wire layers oriented in six different stereo angles, {\emph i.e.}~$\pm
0^0$, $\pm 20^0$, $\pm 40^0$, see fig.~\ref{tracking-setup}. This
favors maximum spatial resolution in polar direction, which points
in the direction of the momentum kick. All four chamber types
contain about 1100 drift cells each, with increasing size from
5$\times$5~mm$^2$ (plane I) to 14$\times$10~mm$^2$ (plane IV). The
chambers provide active areas from 0.35~m$^2$ up to 3.2~m$^2$, thus
covering the same solid angle per sector. The main
feature of the design and the operation parameters of the chambers
is the consequent implementation of the low-mass concept, as already
pointed out above. Resulting from the extensive R\&D
phase~\cite{tracking-1} these requirements are met by three
preventive measures: (i) cathode and field wires made of annealed
aluminum (planes I-III: bare, IV: Gold-plated) with 80~$\mu$m and
100~$\mu$m diameter spanned with tensions between 80 and 120~cN,
depending on the chamber type, (ii) a Helium-based counting gas
(Helium:Isobutane = 60:40) and (iii) entrance windows made of
12~$\mu$m aluminized Mylar.

The 20~$\mu$m (planes I\,-\,III) and 30~$\mu$m (plane IV) thick
Gold-plated Tungsten sense wires are strained with an initial
tension of 40 and 110~cN, respectively. To compensate for the
total wire tension after being released from the assembly table,
chamber frames of planes I\,-\,III have been pre-stressed before wire
gluing. Together with the requirements concerning the acceptance
this resulted in a sophisticated layer frame design with only 3~cm
width, given by the coil case shadows, for the inner-most chambers. The
chambers of the outer-most plane~IV, which did not use the
pre-stressed technique employ two extra Carbon bars to keep the
wire tension loss due to deformation below 10~\%.

The experiments with HADES are expected to run at least ten years.
Creeping of the Aluminum wires and ageing are the main concerns with
respect to the long-term stability of the chambers. Creeping has
been systematically investigated in tension loss test series,
yielding a 10~\% loss in tension within five years. This has been
confirmed by remeasurements in one chamber of
plane~III~\cite{tracking-3}. Ageing is mainly caused by the
accumulated dose in combination with the materials used for
construction and operating the chambers. For example, Epoxy from
Araldite$^{\textrm\textregistered}$ is used for
gluing the wires on the frames
from Stesalit. The gas system is running in a re-flow
mode with typically 10\,-\,20~\% fresh gas and continuous purification
employing two large volume Copper catalyzer-filled cartridges, which
keep the Oxygen contamination level below 15~ppm. Two drift velocity
monitors~\cite{tracking-cl} provide a sensitive control of the
gas quality by measuring the drift velocity with a precision of
better than 0.2~\%. In addition, the simultaneous monitoring of the
relative gains allows conclusions on the gas contamination, {\it e.g.}
due to Oxygen. The expected maximum charge is of the order of 10~mC
per year and centimeter of sense wire. An accelerated ageing
test with $^{55}$Fe using two prototype chambers exhibited no
noticeable gain drop ($<$5~\%) for a time period of two years of continuous
running~\cite{tracking-1}.\\
These results on creeping and ageing, together with the careful
selection of materials and running conditions, suggest that the
requirements for the projected long-term operation of the HADES
tracking system can be met.

The drift chamber signals are read-out and digitized by means of
dedicated, customized boards mounted on the chamber frames, not
extending into the active area. Hence, special emphasis was put on
the integration of the modular front-end electronics, realized with
analog boards (16 channels) mounted on digitization boards (64 or
96~channels). Four sense wires are connected by flexible printed
circuits to the analog boards~\cite{tracking-db}, housing ASD8-B
chips~\cite{tracking-asd8} (8 channels, 1~fC intrinsic noise,
30~mW/ch, adjustable threshold) for differential amplification,
shaping and discrimination. These chips deliver logical signals with
variable width, being equivalent to the time the shaped signal exceeds the
given threshold. The logical signals are fed to TDC chips (CMOS, 8
channels/chip, 0.5~ns/ch, common-stop, 1~$\mu$s full range) on the
digitization boards. This semi-customized ASIC is multi-hit capable
and thus allows to detect also the \textit{\textbf{T}ime \textbf{o}ver
 \textbf{T}hreshold} (ToT) of each hit.
Besides spike and zero suppression this chip offers the possibility
of internal calibration, activated by a separate trigger type. The
design of the front-end electronics was decisively influenced by
minimizing the noise level on-line. In addition, the
ToT information is a valuable tool to discriminate
remaining noise hits off-line.

\subsubsection{Alignment} \label{Section_Tracking_alignment}

The aimed-at performance of the HADES detector can only be achieved
with a very precise knowledge of the positions of all detector
components, and in particular of the MDC tracking chambers.  As
discussed above, geometric surveys can only provide part of the
necessary information, and additional correction parameters must be
obtained from the event data themselves. Presently we have
developed alignment procedures based on (i) straight-track
reconstruction of data taken without magnetic field, (ii) straight
tracks  from cosmic radiation, and (iii) kinematically constrained
events from proton-proton elastic scattering. For all of
those, iterative fitting algorithms allow to produce a set of
translation and rotation parameters for each of the 24 tracking
chambers, as well as for the other position-sensitive detector parts
(TOF wall modules, Pre-Shower chambers, and RICH).
Using straight tracks selected from a well defined experimental situation
(beam energy, target, trigger),
a global alignment can be obtained and the corresponding parameters
entered in the database. Although the precision with which the parameters
were determined is of the order or better than 0.1 mm, they represent only
a global alignment, possibly biased by the events topology, which has
to be further refined by tuning individual layer or wire groups
offset parameters for individual chambers. This is still an on-going work.
These parameters enter the geometry used in the track reconstruction.

\subsubsection {Tracking performances} \label{Section_tracking_performance}

\paragraph {Detection efficiency}

The detection efficiency of the wire layers has been investigated by
measurements using $\beta$ rays from $^{90}Sr$ sources, cosmic rays
and reconstructed tracks from physics runs. Consistent detection
efficiencies for minimum ionizing particles detected in the two
inner drift chamber planes of about 90~\% and 97~\% were obtained,
whereas the layer efficiency of the outer drift chambers reaches
almost 100~\%. Due to the small drift cell size of the inner drift
chambers, the efficiency depends strongly on the applied high voltage
and the read-out threshold. The lower efficiency of inner drift
chambers results from operation at a smaller high voltage of
\mbox{-1750~V}, respectively \mbox{-1800~V} as compared to the
prototype in-beam test at typically -2000~V. The operation points of
the experimental runs have been chosen for stable performance during
data taking.

\paragraph {Track reconstruction efficiency}

\begin{figure}[\htb]
\begin{center}
\includegraphics[width=0.7\linewidth,clip=true]{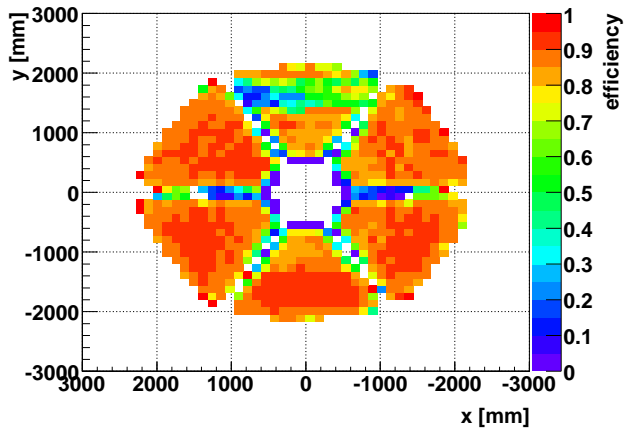}
\caption{Reconstruction efficiency in the HADES acceptance for
protons from p\,+\,p elastic collisions at 1.25~GeV kinetic energy.
Regions of reduced efficiency in the upper sector are due to read-out
failures and are fully taken into account through our GEANT simulation.}\label{fig_2d_eff_pp}
\end{center}
\end{figure}

The track reconstruction efficiency has been investigated
in proton-proton elastic scattering at 1.25~GeV kinetic energy
which allows to prepare a clean track sample using angular ($theta$ 
and $phi$) correlations of hits in the TOF
and Pre-Shower detectors only. Knowing that each detected
hit in the outer detector is caused by a p\,+\,p elastic scattering event, one
can obtain the reconstruction efficiency from the inner and
outer tracking system. Protons from elastic scattering are
reconstructed by the candidate search (see
sect.~\ref{Chapter_Clusterfinder}) with an efficiency
close to 100~\%. Requiring a fitted inner MDC segment (see
sect.~\ref{Chapter_track_fitting}) reduces the efficiency
to about 92~\%, a fit with both inner and outer MDC segments to 87~\%. A full
track reconstruction, including efficiency of the momentum determination via Runge-Kutta method,  
(see sect.~\ref{Chapter_runge_kutta}) results in an efficiency
of 86~\%. Figure~\ref{fig_2d_eff_pp} displays the reconstruction efficiency projected
on a plane perpendicular to the beam axis. The reconstruction inefficiency
visible in the upper sector is caused by readout electronics temporarily
failing during this run. The corresponding analysis and correction
factors were tuned accordingly and controlled
against the elastic scattering angular distribution.

\paragraph {Spatial resolution}

\begin{figure}[\ht]
\begin{center}\centering
\mbox{
\subfigure{\includegraphics[width=0.45\linewidth,clip=true]{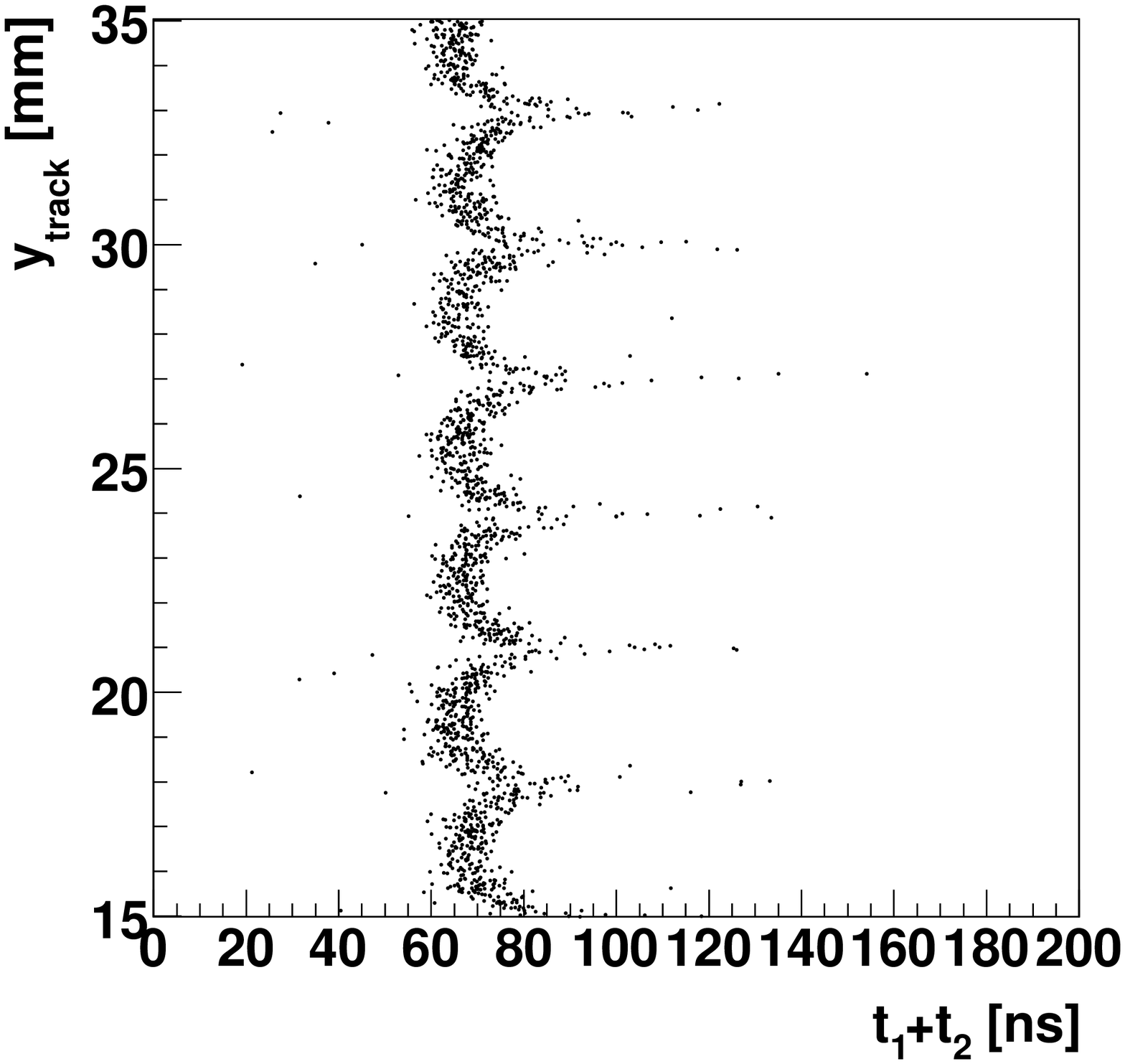}}
\subfigure{\includegraphics[width=0.45\linewidth,clip=true]{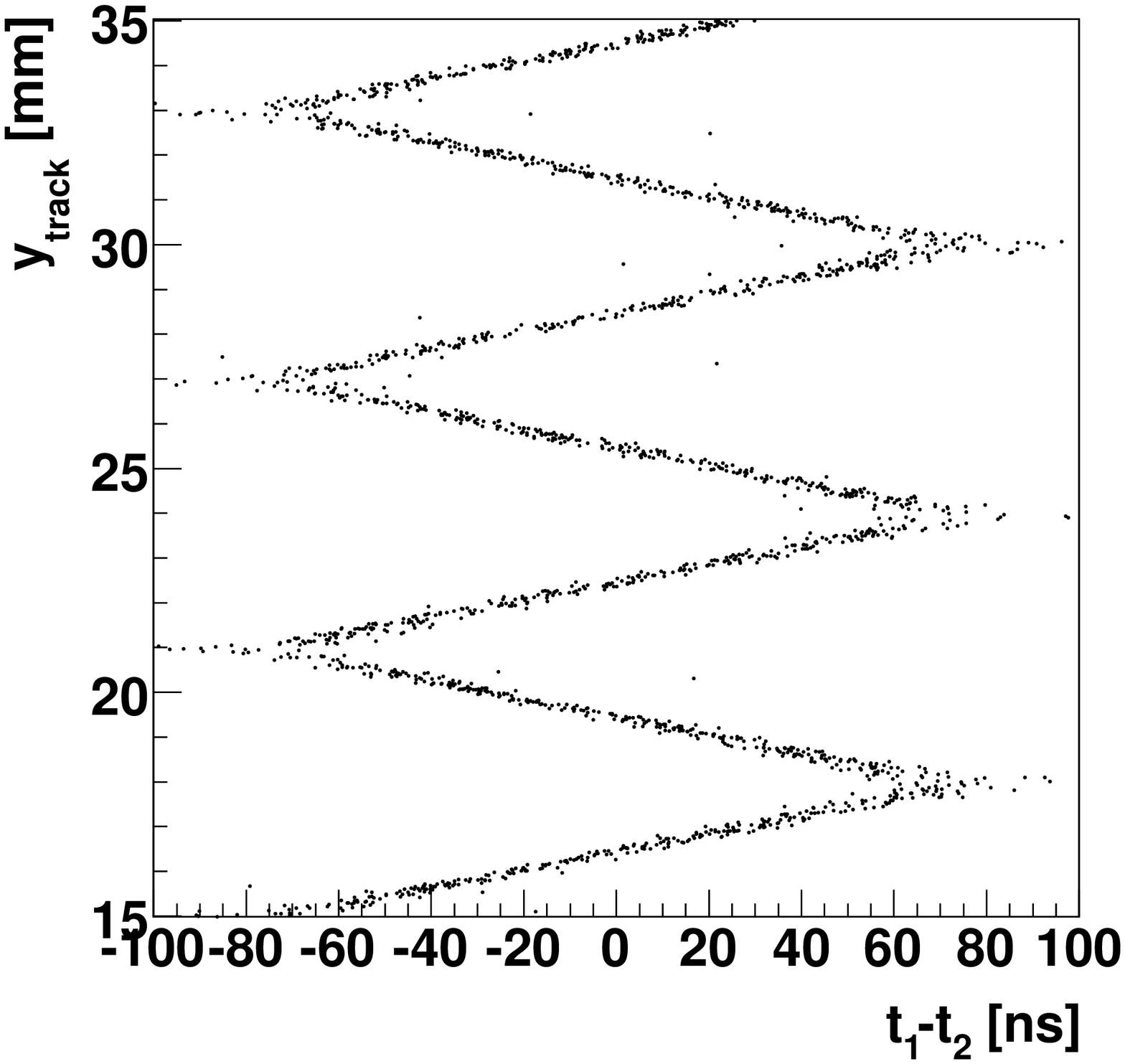}}}
\mbox{
\subfigure{\includegraphics[width=0.44\linewidth,clip=true]{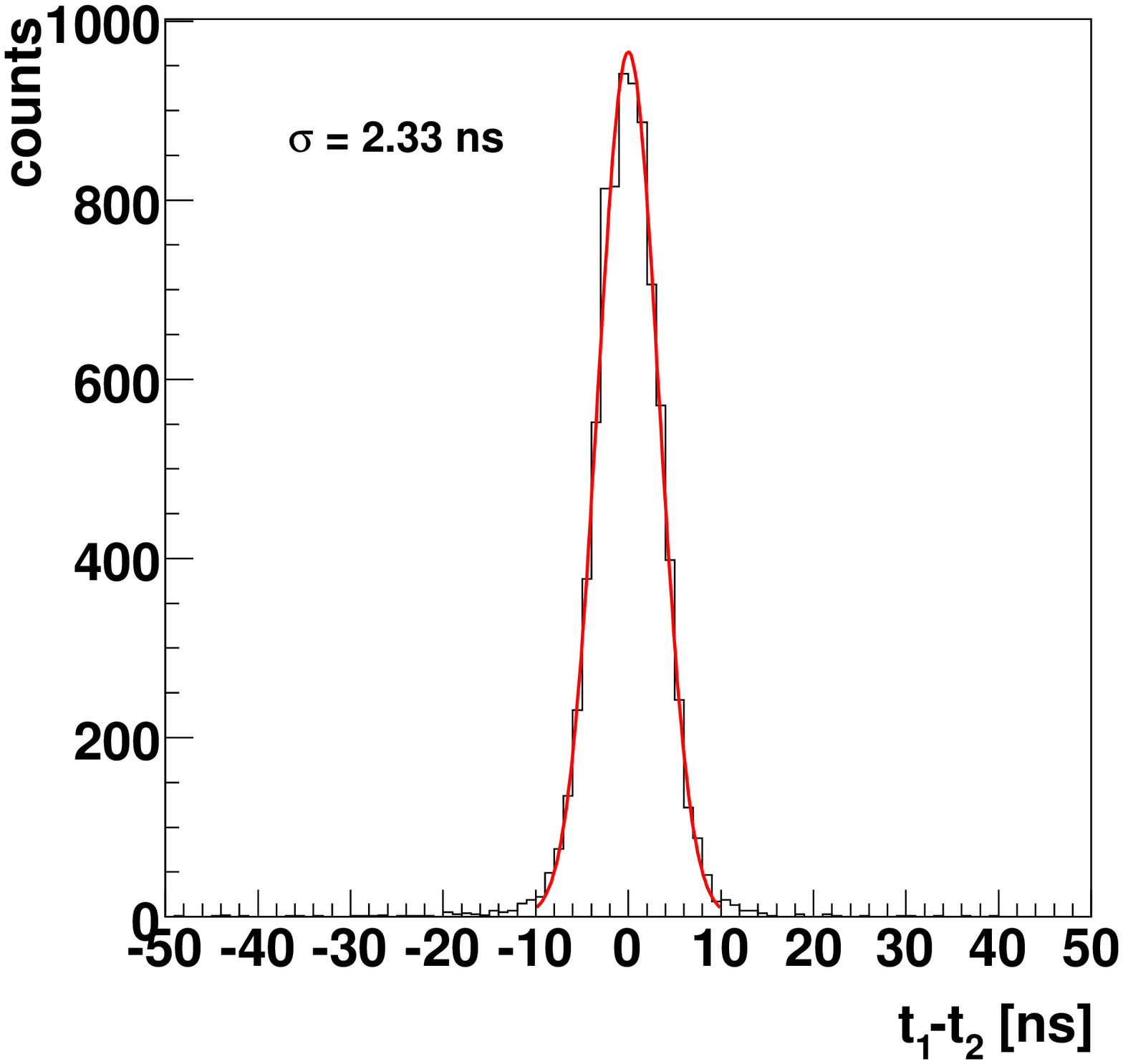}}
\subfigure{\includegraphics[width=0.48\linewidth,clip=true]{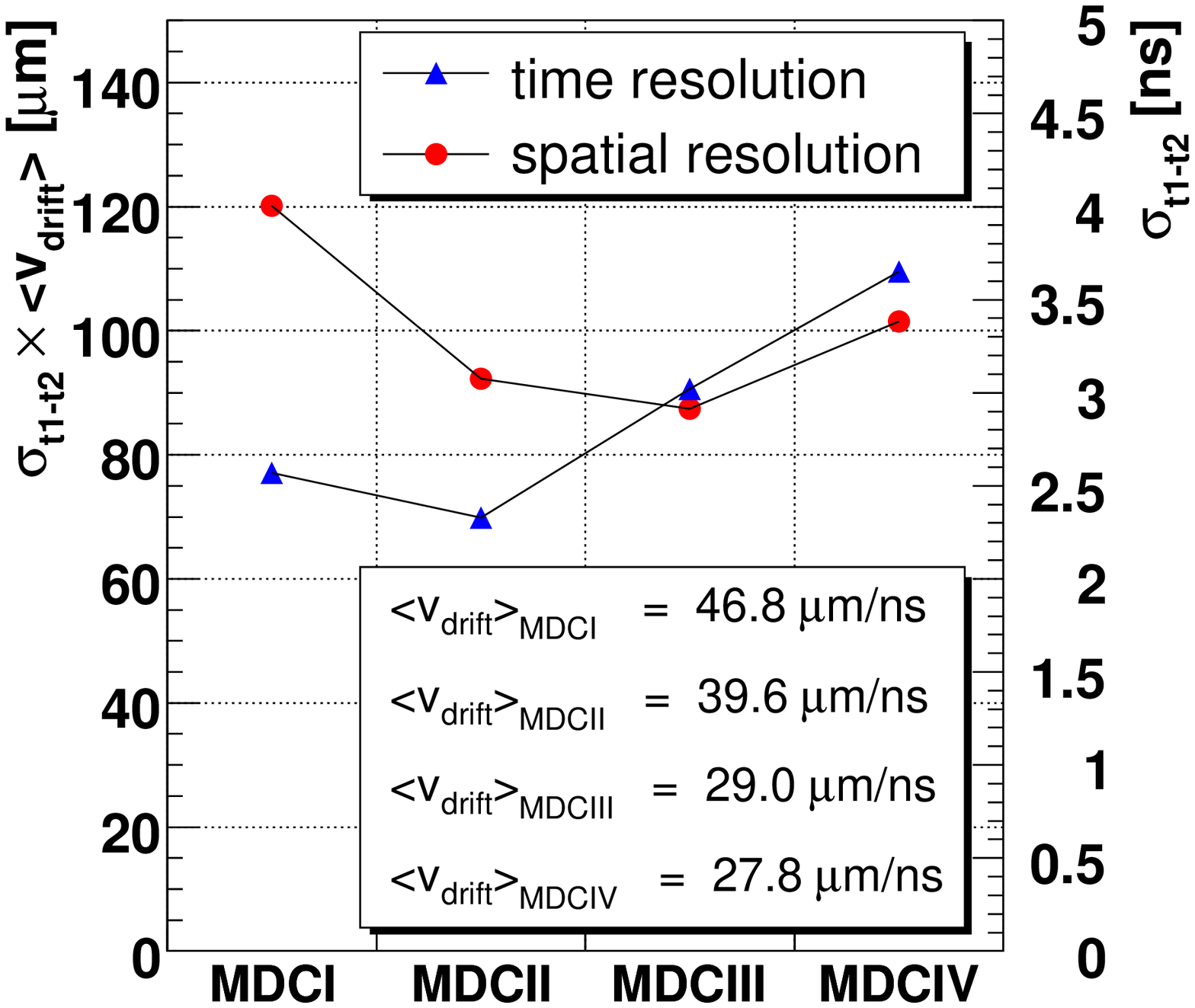}}}
\caption{Top left: Sum of the measured drift times of
two adjacent drift cells of the 0-degree layers versus the y-position
of the reconstructed trajectory in the MDC coordinate system.
Top right: Difference of the measured drift times of
two adjacent drift cells.
Bottom left: Projection, after transformation, of the measured drift time
differences, shown here for MDCII. Bottom right: Time resolution (blue triangle)
and corresponding spatial resolution (red dots) for the different
drift chamber types MDCI-IV. The spatial resolution has been obtained
by multiplying the drift time resolution by the average drift
velocity.}\label{fig_selftrack}
\end{center}
\end{figure}

First resolution measurements with a prototype of a
plane~II type chamber using a 2.1~GeV/c proton beam
and an external silicon tracker achieved a spatial
resolution $\sigma$\,=\,70~$\mu$m over 70\,-\,80~\% of
the drift cell~\cite{tracking-2}. These measurements
were performed at an optimum high voltage setting of
\mbox{-2000~V}. The spatial resolution of the drift
cells under in-beam conditions have been investigated
with the so-called \textit{self-tracking} method.

This method exploits time measurements of the same
track in two adjacent drift cells, one in each of the
two 0-degree layers (see fig.~\ref{tracking-setup}).
For a given impact angle of the particles on the chamber,
the sum of both drift time measurements is constant
over a large fraction of the cell. Once the distance
between the signal wires is known, the distance-time
correlation, the drift velocity and the time resolution
can be obtained. The drift velocity and distance-time
correlation have been found to be in good agreement
with {\sc Garfield} simulations of the drift cells \cite{Bormio02}.
Typical values of the difference between the simulated value
and the measured one stay below the
simulated time dispersion of 2~ns. The latter simulations
were then used to provide the distance-time correlation
for the track fitting.

Figure~\ref{fig_selftrack} shows
the results of the method for the MDCII case. Particle tracks (mostly from
protons and pions in the region of minimum ionisation) close to perpendicular
impact from a proton-proton experiment at 3.5~GeV kinetic beam
energy were selected. The sum
of the two measured drift times (fig.~\ref{fig_selftrack} top left)
shows nicely that the sum is almost constant (deviations occurs only at the wire positions), 
but it suffers from trigger time fluctuations. The time difference
(fig.~\ref{fig_selftrack} top right) shows a much cleaner correlation
pattern, which is not affected by this event-to-event
fluctuation of the trigger. By fitting the lines of this
z-shaped pattern and applying an appropriate transformation,
a straight pattern is obtained, which is further
projected. Assuming equal contribution of both drift cells,
time resolutions are obtained (fig.~\ref{fig_selftrack} bottom left),
ranging from 2.3~ns up to 3.6~ns, slightly depending on
chamber gas and cell geometry. Corresponding spatial
resolutions were deduced for all MDC using averaged drift velocities
(fig.~\ref{fig_selftrack} bottom right). They range from 70 up to 110
$m$m.

\paragraph {Energy loss measurement}\label{chapter_elos}

Due to the lack of ADCs there is no direct measurement of the
deposited charge in the MDC cells. From the width of the drift
time signal (\textit{\textbf{T}ime \textbf{o}ver \textbf{T}hreshold}
(ToT)) an unambiguous measure of the energy loss of the particle
can be extracted. The measured ToT depends on the gas mixture,
on the reduced electric field $E/P$, on the track geometry, on the
drift cell size and on the threshold setting of the ASD8 chip.
Special attention was paid to the unfolding of these dependencies
to allow averaging over all drift cell measurements contributing
to a particle track.

For the following discussion the particle tracks are parameterized
via minimum distance of the track to the sense wire and the impact
angle with respect to the wire plane.
\begin{figure}[\t]
\centering \mbox{
\subfigure{\includegraphics[width=0.48\linewidth]{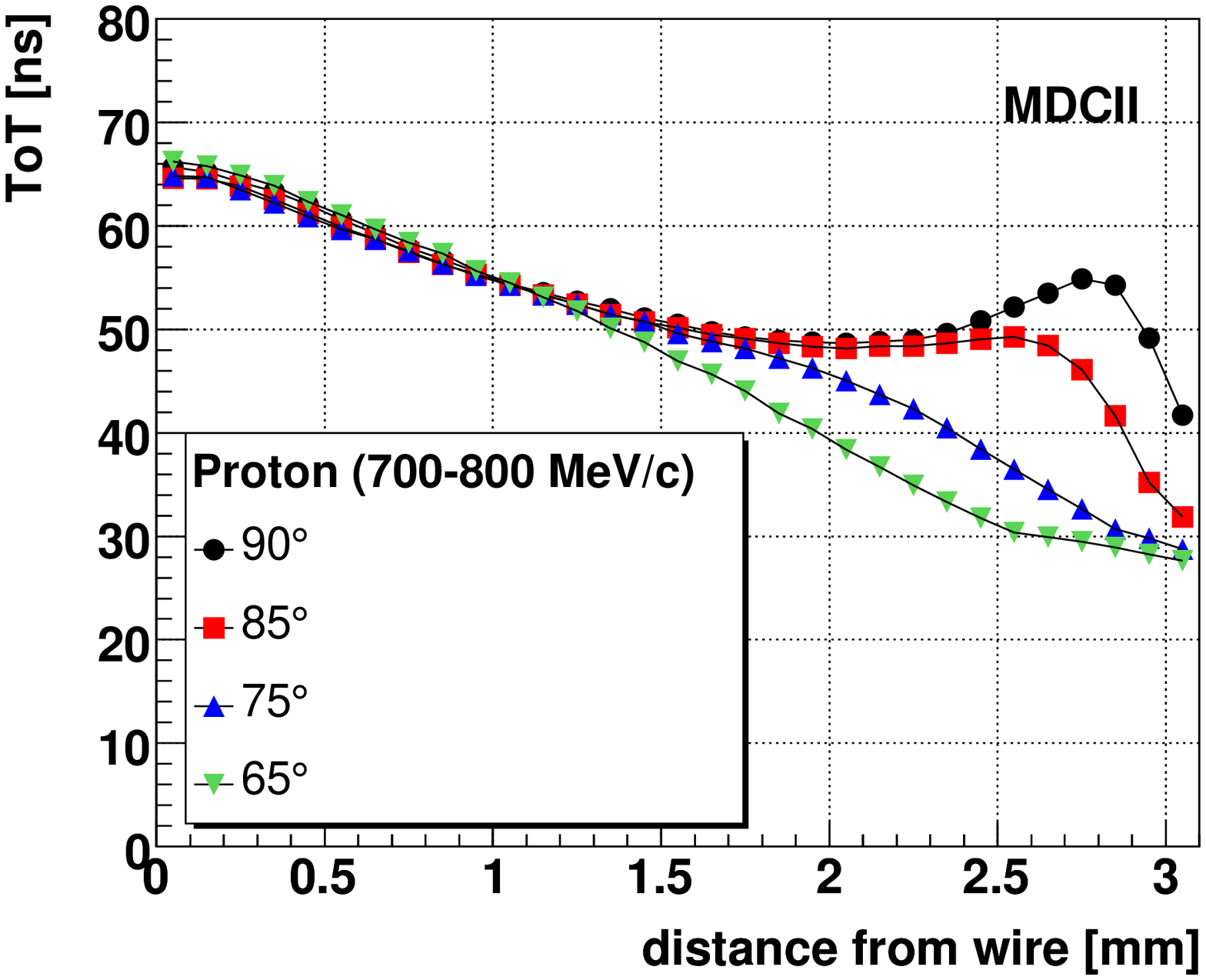}}
\subfigure{\includegraphics[width=0.48\linewidth]{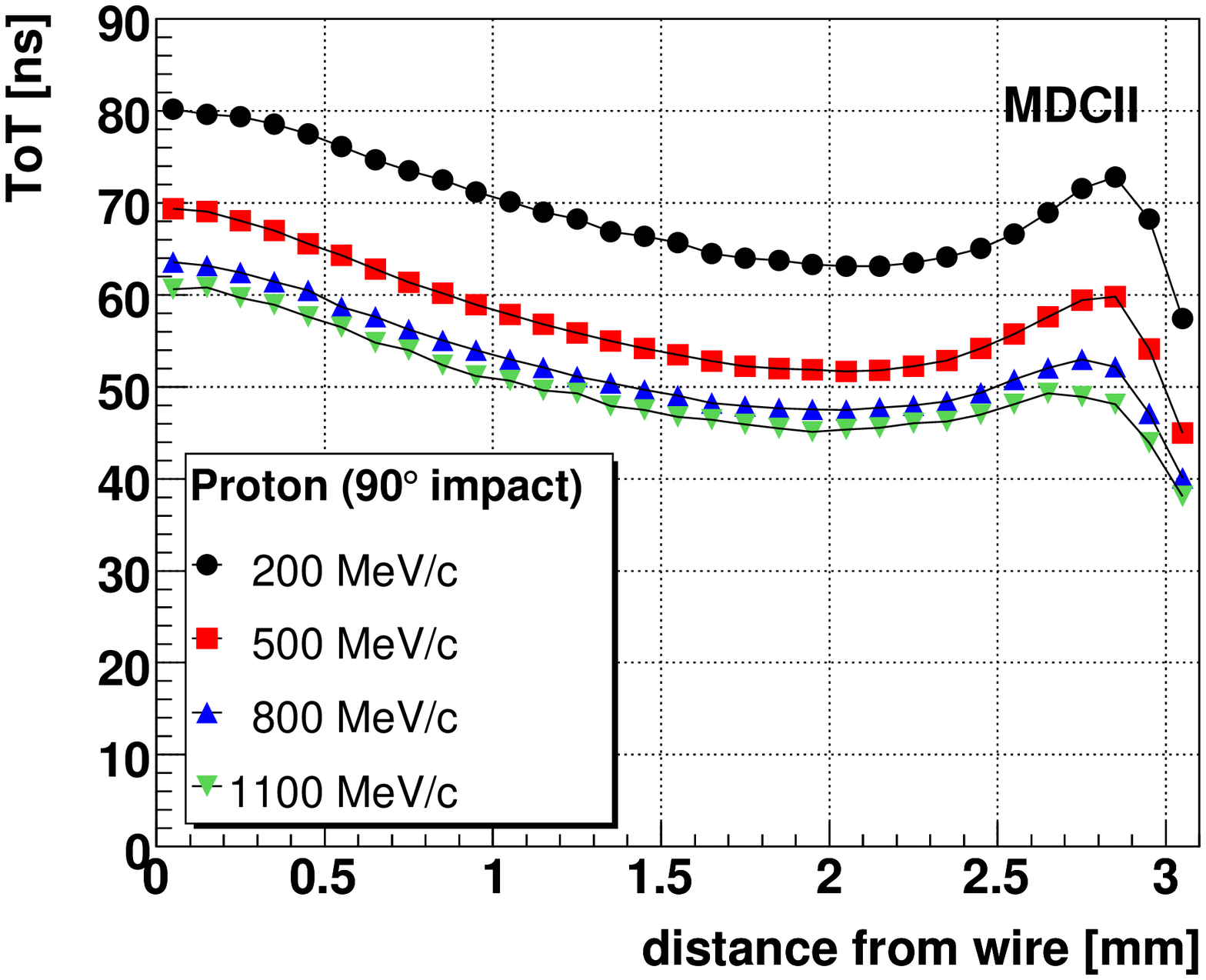}}}
\caption{Left: Dependence of ToT on the distance from wire to
particle track for 0.7\,-\,0.8 GeV/c protons at 4 impact angles. Right: Same for
protons with a average impact angle of $90^{\circ}$
and different momenta (shown for \mbox{MDCII}).}
\label{t1t2_norm_mindist_protons_energy}
\end{figure}
The measured ToT depends on the distance from the sense
wire, the impact angle (fig.~\ref{t1t2_norm_mindist_protons_energy} left)
and the momentum of the particle (shown for selected protons in
fig.~\ref{t1t2_norm_mindist_protons_energy} right).\\
To calibrate the measured ToT of the single drift cells
the correlation of the ToT with the particle energy loss
is fitted by the function
\begin{equation}\label{fitfunction_tot_vs_dedx}
ToT=f(dE/dx) =c_0+c_1 [\log_{10}(dE/dx+c_3)]^{c_2},
\end{equation}
with the parameters $c_0, c_1,c_2$ and $c_3$. Knowing the
value of the momentum and the
particle type from the other detectors, the energy loss of the particle
can then be calculated using the Bethe-Bloch formula.
\begin{figure}[\htb]
  \centering
     \begin{minipage}[c]{0.35\linewidth}
 \centering
        \caption[]{ToT versus energy loss correlation for protons at different angles of incidence. The distributions
        have been fitted by the function~(\ref{fitfunction_tot_vs_dedx}).}\label{dedx_t2t1}
  \end{minipage}
     \hspace{0.02\linewidth}
     \begin{minipage}[c]{0.60\linewidth}
       \includegraphics[width=\linewidth,clip=true]{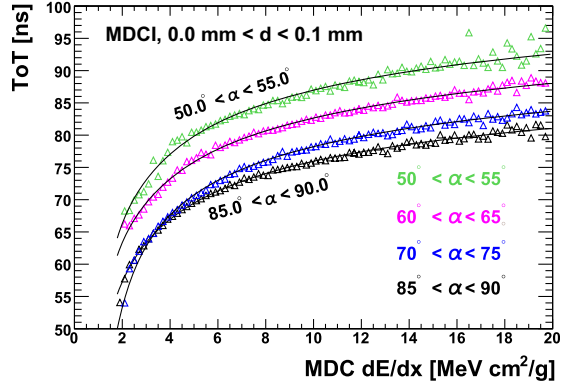}
    \end{minipage}
\end{figure}
The procedure is performed for
intervals of \mbox{$5^{\circ}$} impact angle and 100~$\mu$m
distance from the sense wire for all 4 MDC types
(fig.~\ref{dedx_t2t1}). \\
The correlation between the measured ToT and the energy loss
of the particles track is non-linear. This compression of the
correlation is most pronounced for strongly ionizing particles.
The shape of the correlation curve depends on the shape of
the signal and thus on the arrival time of the drifting electrons
on the sense wire and also on the signal shaping performed
in the readout electronics. \\
To allow for averaging over all drift cell measurements, all contributing
measurements are normalized to one reference measurement employing the
inverse correlation function (\ref{fitfunction_tot_vs_dedx}).
 Finally, a truncated
mean method has been applied to get a better resolution. After
truncation, on average 18 cells out of 24 remain, corresponding
to a 20~\% cut.

\begin{figure}[\htb]
  \centering
     \begin{minipage}[c]{0.35\linewidth}
 \centering
        \caption[]{Correlation of energy loss and polarity times
        momentum of particle tracks. The energy loss is averaged
        over all contributing drift cells of a track through the
        4 MDC planes and a truncated mean method has been applied.
        The solid lines represent the energy loss calculated from
        the Bethe-Bloch formula.}\label{fig_dedx_tot_final}
  \end{minipage}
     \hspace{0.02\linewidth}
     \begin{minipage}[c]{0.6\linewidth}
      \includegraphics[width=\linewidth,clip=true]{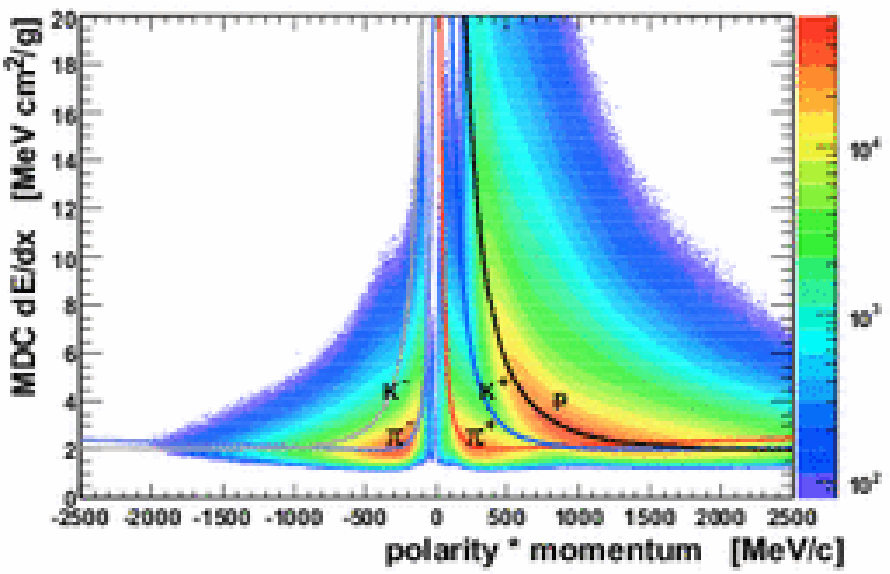}
    \end{minipage}
\end{figure}
Figure~\ref{fig_dedx_tot_final} shows the correlation of the
calibrated and truncated energy loss with particle momentum.
The calculated energy losses from the Bethe-Bloch formula are
represented as solid lines. Up to a momentum of about
0.7~GeV/c the MDC energy loss measurement can be exploited to
separate $\pi^{+}$ from protons or to improve the signal to
background ratio in a K$^+$ measurement.
\begin{figure}[\hb]
\centering
\mbox{
  \subfigure{\includegraphics[width=0.48\linewidth,clip=true]{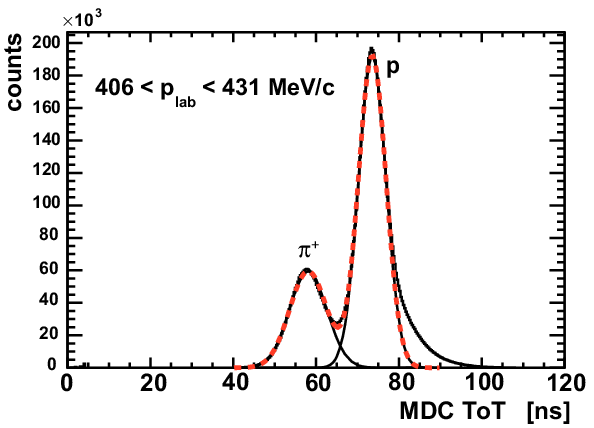}}
  \subfigure{\includegraphics[width=0.48\linewidth,clip=true]{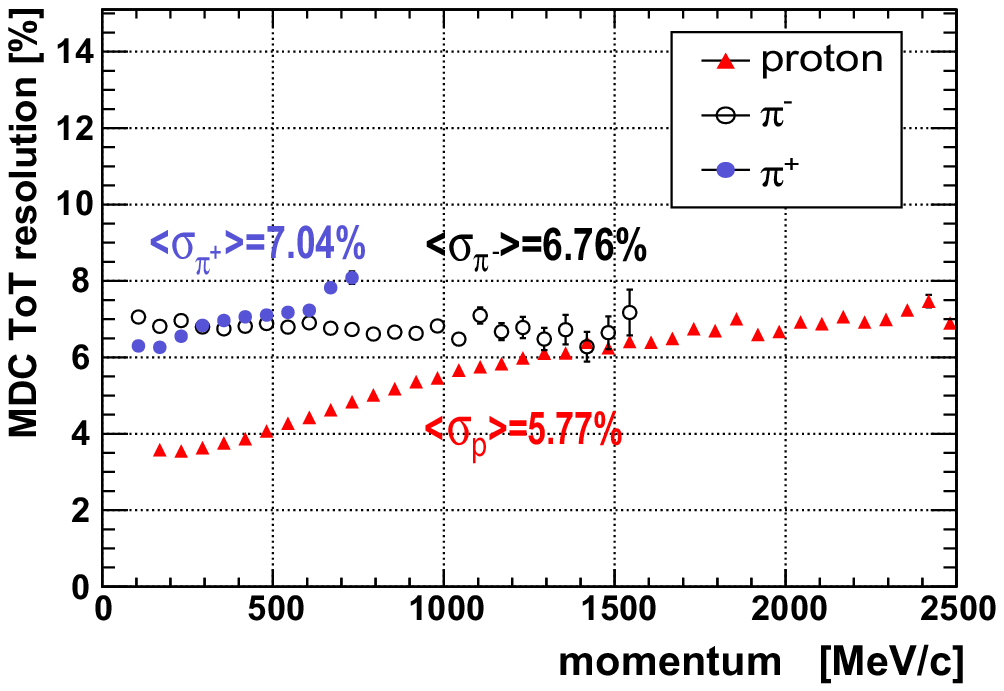}}}
\caption{Left: Measured ToT distribution from the MDCs for a
given momentum slice, averaged over the whole HADES acceptance.
Right: ToT resolution versus particle momentum. The merging of the
$\pi^+$ line into the proton line at high momentum is due to an
increasing number of wrongly identified protons.}\label{fig_ToT_resolution}
\label{fig_dedx_slice}
\end{figure}

The ToT resolution has been studied by making Gaussian fits
in each 25~MeV/c wide momemtum bin (fig.~\ref{fig_ToT_resolution} left).
The ToT resolution for different particle species is depicted in
fig.~\ref{fig_ToT_resolution} right. For minimum ionizing particles,
an energy loss resolution of around 7~\% has been achieved.
A better resolution of about 4~\% can be found for stronger
ionizing particles. The resolution strongly depends on the
number of drift cell measurements contributing to a particle
track.
The method described here yields similar
results regarding the $dE/dx$ resolution as compared to the
drift chambers of Belle \cite{belle} and BABAR \cite{barbar}.
It fits also nicely with the empirical formula given in
\cite{pdg_part_det}.


\clearpage
\subsection{Time-of-flight detectors}
\label{Chapter_tof}

\subsubsection{Overview}

The time-of-flight system of HADES consists of two scintillator
arrays, the TOF wall for larger polar angles ($44^\circ < \theta <
88^\circ$) and the TOFINO wall for the forward
region ($18^\circ < \theta < 44^\circ$).
Both detectors are placed in the region behind
the tracking system and they are used mainly for triggering
and particle identification purposes.

A fast determination of the charged particle multiplicity of the
event allows to select certain reaction classes. For heavy-ion
reactions it permits impact parameter selection and hence centrality
characterization. For elementary reactions it provides the
possibility to enhance inelastic channels with multi-particle final
states with respect to elastic scattering. Thus, the charged particle
multiplicity is a natural choice for a fast first-level trigger
decision. Moreover, the fast determination of the impact position of
each hitting particle, spatially correlated to the position in the
hadron-blind RICH detector, allows to perform a second-level trigger
decision in order to select events which contain lepton track
candidates. Both aspects will be discussed in more details in
sect.~\ref{Chapter_daq}.

The combination of the  time-of-flight measurement with
momentum determination provided by the tracking system and the
energy-loss measurement permits to perform efficient
particle identification (e,$\pi$,K,p), which is essential for the HADES
physics program.

\subsubsection{TOF}

\paragraph{Detector characteristics}

The TOF detector \cite{tof} follows the six-fold symmetry of the
whole spectrometer covering polar angles from
$44^\circ$ up to $88^\circ$. Each sector consists of
eight modules. Each module consists of a set of eight scintillator rods,
with a total of 384 rods, which are  enclosed in a
carbon fiber case.

In order to reduce the probability that two particles emitted in the
same collision hit the same rod (double hits) to less than 10~\%
the TOF granularity has been matched to the charged particle multiplicity
angular distribution. The rod cross section
is  20$\times$20 mm$^2$ for the innermost 192 rods and 30$\times$30
mm$^2$ for the outermost 192. The rod length varies from 1 m to 2 m,
respectively, from smaller to larger polar angles.

Each rod is made of BC408 plastic scintillator material from Bicron,
mainly for its  optical properties: good attenuation length (3.8
m in bulk) combined with high scintillation efficiency ($\approx
10^{4}\,\gamma$/MeV) and fast response (2.1 ns decay time). Each rod
end is glued to a light guide, bent by $65-67^{\circ}$  with respect to
the rod itself and coupled to a photomultiplier (PMT) by means of a
2.5 mm thick silicone disc. The light guide has an initial
square cross section in order to fit to the scintillator shape, reaching
progressively a circular cross section at the PMT photocathode side.
To optimize their optical properties, the rods
have been wrapped in aluminized Mylar sheets, along with
the light guides.

All sectors are equipped with 9133B PMTs from Electron Tubes
Limited.  Each PMT provides a signal from the last dynode for
the amplitude measurement and from the anode for time measurement.
The electronic chain includes a Constant Fraction Discriminator
(CFD), a Logic Active Delay (LAD) and a Time to Digital Converter
(TDC). The amplitude signal measurement is performed by an
electronic chain consisting of a shaper and an Amplitude to Digital
Converter (ADC). The signal from the CFD is split into two parts and
is connected to the first-level trigger electronics chain. By
performing an analog sum of all the CFD signals it is possible to
trigger on the charged particle multiplicity of the collision for
event selection purposes.

The time-of-flight ($tof$) of the particle, its hit position ($x$)
along the rod and the deposited energy ($\Delta E$) are measured as
explained below. Let us denote with $t_{left}$, $t_{right}$
the calibrated time
intervals between the instant when the reaction occurred (given
by the START detector) and the arrival of the two light pulses at
each rod end, and $a_{left}$, $a_{right}$ are the corresponding
light signal amplitudes. Then these equations can be used to
calculate the following variables:

\begin{equation}
tof = \frac{1}{2}\left(t_{right}+t_{left}-\frac{L}{V_{g}}\right),
\label{eq:tof}
\end{equation}
\begin{equation}
x = \frac{1}{2}\left(t_{right}-t_{left}\right)V_{g}, \label{eq:x}
\end{equation}
\begin{equation}
\Delta E = k\sqrt{a_{right}a_{left}e^{L/\lambda_{at}}},
\label{eq:de}
\end{equation}
where $V_{g}$ is the group velocity of the light inside the scintillator rod, $\lambda_{at}$
its attenuation length, $L$ the rod length, $k$ a constant and $x$ the hit
position along the rod.

\paragraph{Performance}

\begin{figure}[htb]
\begin{center}\centering
\mbox{
\subfigure{\includegraphics[width=0.5\textwidth, height=4.8cm]{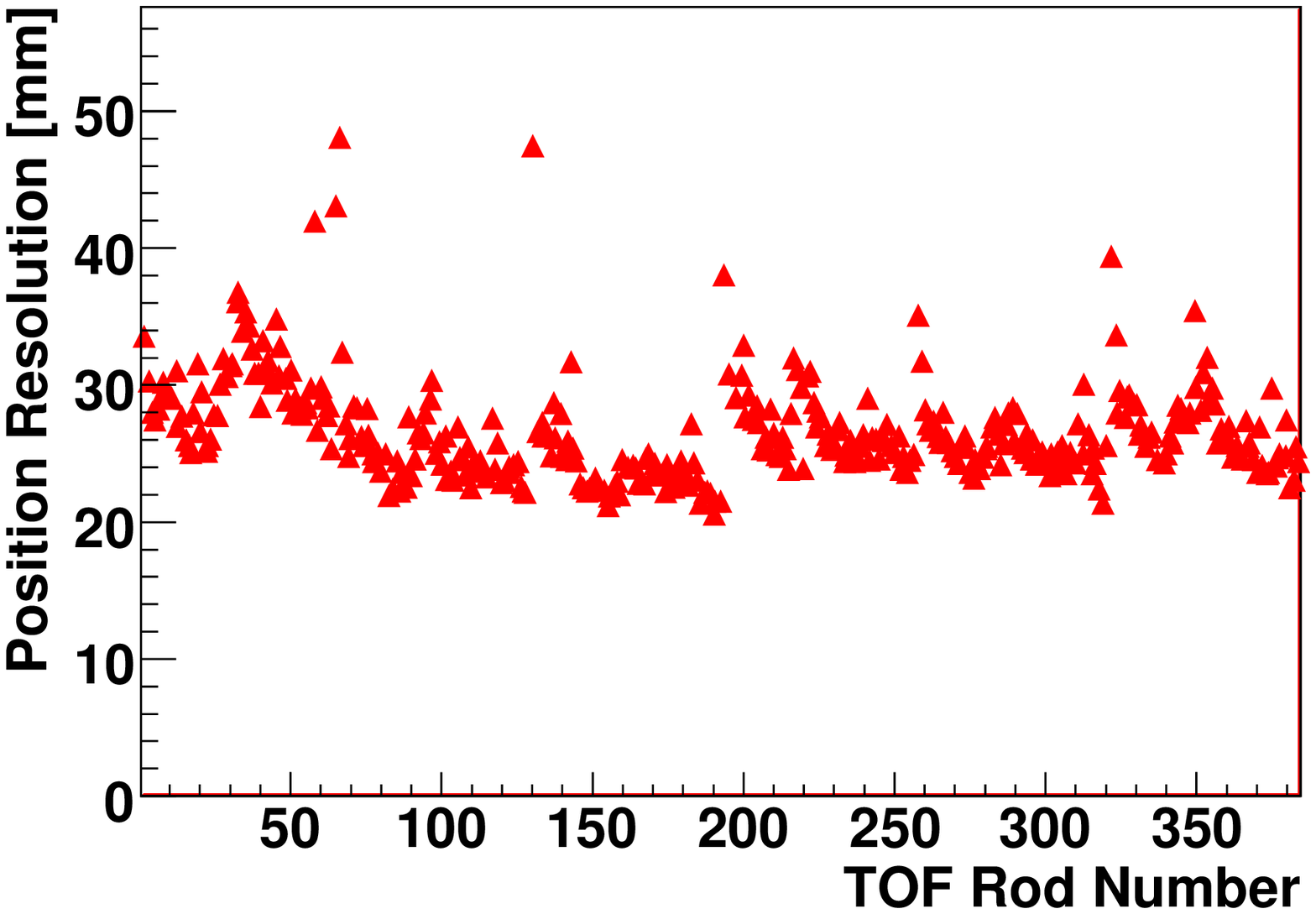}}
\subfigure{\includegraphics[width=0.5\textwidth, height=4.8cm]{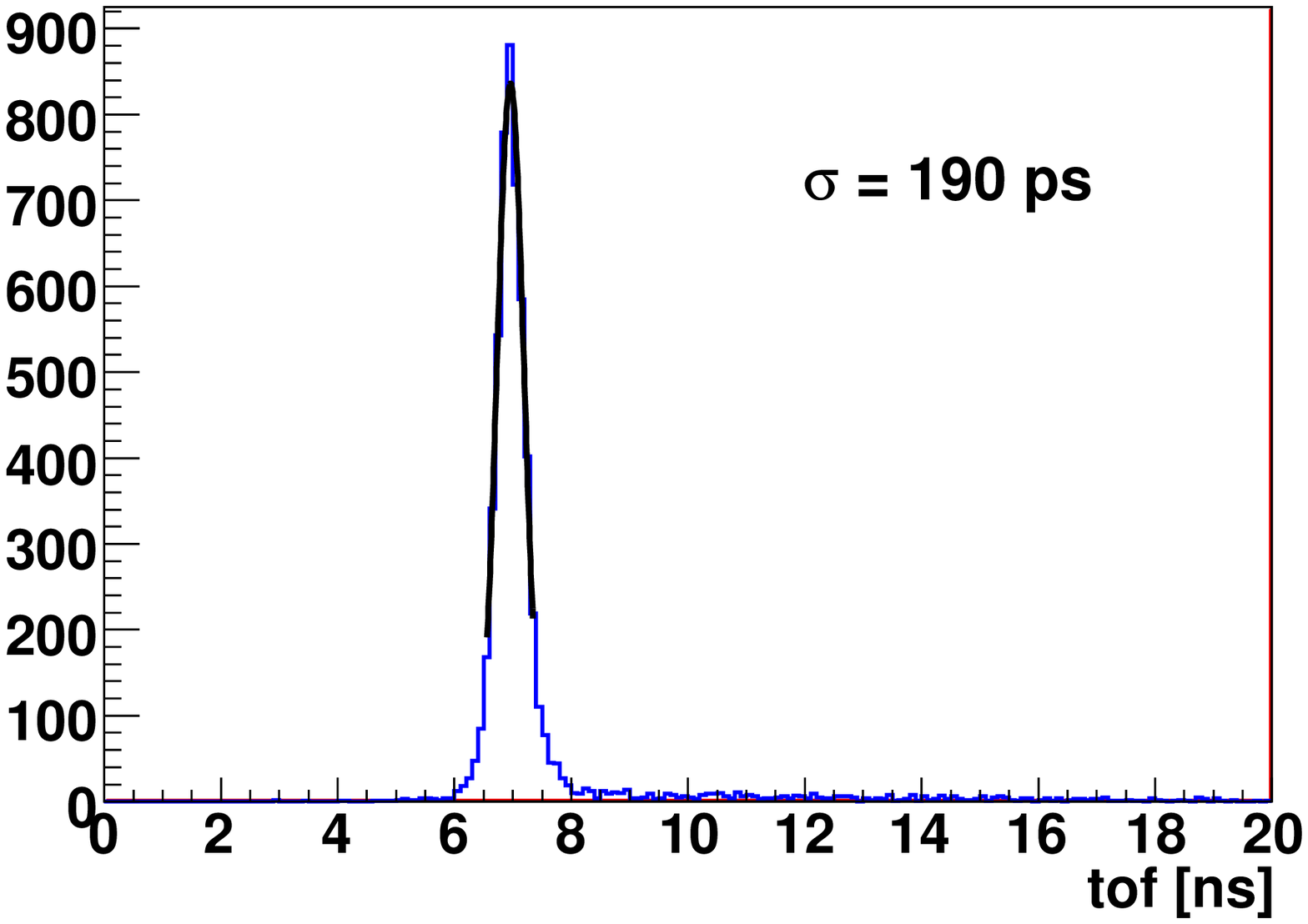}}}
\caption{Left: Position resolution $\sigma_x$ along the rod as a function of the TOF
rod number. Each group of 64 rods corresponds to one sector.
Right: Time-of-flight for lepton tracks in a C\,+\,C experiment.}\label{TOFres}
\hspace*{.5cm}
\end{center}
\end{figure}

The TOF wall has been running stably throughout several runs, and
its performance was evaluated in C\,+\,C and p\,+\,p experiments. The latter
case needs a separate treatment since it was performed without
start detector (sect.~\ref{Chapter_nostart}).

The average position resolution can be determined in correlation
with the MDC tracking system by taking data without magnetic field.
Let us assume that a particle emitted from the target crosses MDC
and TOF wall, creating useful signals in both detectors. By
projecting the segment reconstructed by the MDC on the TOF system,
one can calculate the position of the projected point, and its
distance from the hit as measured by the scintillator rod. The
distribution of the difference between the two positions
$x_{TOF}-x_{MDC}$ can be fitted rod by rod by a Gaussian function,
yielding the rod position offsets as well as the position resolution
which is shown in fig.~\ref{TOFres}. The average resolutions along
the rod $\sigma_x$ are 25 mm
and 27 mm, respectively for  20$\times$20 mm$^{2}$ and 30$\times$30
mm$^{2}$ rods.

Electrons and positrons can be used to evaluate the time resolution
 of the TOF detector. In the energy range studied by
HADES, the emitted leptons travel at velocities close to the speed
of light. Their time-of-flight is then essentially independent of
momentum and depends only on the traveled distance between the
emission point and the TOF wall which is known from tracking.
Figure~\ref{TOFres} shows the time-of-flight distribution for lepton
tracks, selected by using the spatial correlation with the
hadron-blind RICH detector, normalized to the same path length of
2.1 m. The distribution can be fitted by a Gaussian function with
a width of $\sigma_{tof}\approx190$ ps. This value depends on the
time resolution of the START detector. We used an additional method
in order to evaluate the TOF time resolution without this
contribution, by using dileptons. If we select events which contain
a lepton pair, their difference in time-of-flight should be centered
at zero after correction for the path length difference and does not
depend on the start time. The fit to Gaussian distributions obtained
from C\,+\,C data yields a characteristic resolution
 $\sigma_{TOF}$ of the TOF array of the order of 150~ps.

The energy loss information can be used for particle identification,
but it appears to provide smaller discrimination power as compared
to the time-based algorithm. For further information on energy
calibration and performances see~\cite{tof,KrizekPHD}.

\subsubsection{TOFINO}

\paragraph{Detector characteristics}

For time-of-flight measurements at polar angles
smaller than 45$^\circ$, HADES is currently equipped by a
low-granularity system called TOFINO, shown in fig.~\ref{shower_tofino}.
It is divided into six sectors each consisting of four BC408
scintillator paddles, arranged radially with respect to the beam axis.
Each scintillator has a trapezoidal form of 1.3 m height and
is 10 mm thick. The light is collected from the wide side of each
paddle (0.32 m wide) on a Hamamatsu 1949 PMT via a
light guide which is bent with respect to the paddle by
65$^\circ$.

The optical coupling between the light guide and the PMT is achieved
by means of a 2.5 mm thick silicone disk. The paddles along
with light guides have been wrapped with
aluminized Mylar. Both signals from each PMT are used for timing and
amplitude measurements using an electronic chain similar to the TOF
one. The TOFINO subdetector is mounted directly in front of the
Pre-Shower detector (sect.~\ref{Chapter_shower}), which provides the
particle hit coordinates on the paddle. In
order to calculate the time-of-flight ($tof$) of the charged
particle, the following equation is used:
\begin{equation}
tof = t-\frac{x}{V_{g}}, \label{eq:tofino}
\end{equation}
where $t$ represents the calibrated time interval between the reaction and the
arrival of the light pulse at the PMT, $V_{g}$ the light group
velocity in the paddle and $x$ the distance from the hit
position on the paddle to the PMT.

\begin{figure}[htb]
\includegraphics[width=0.5\textwidth]{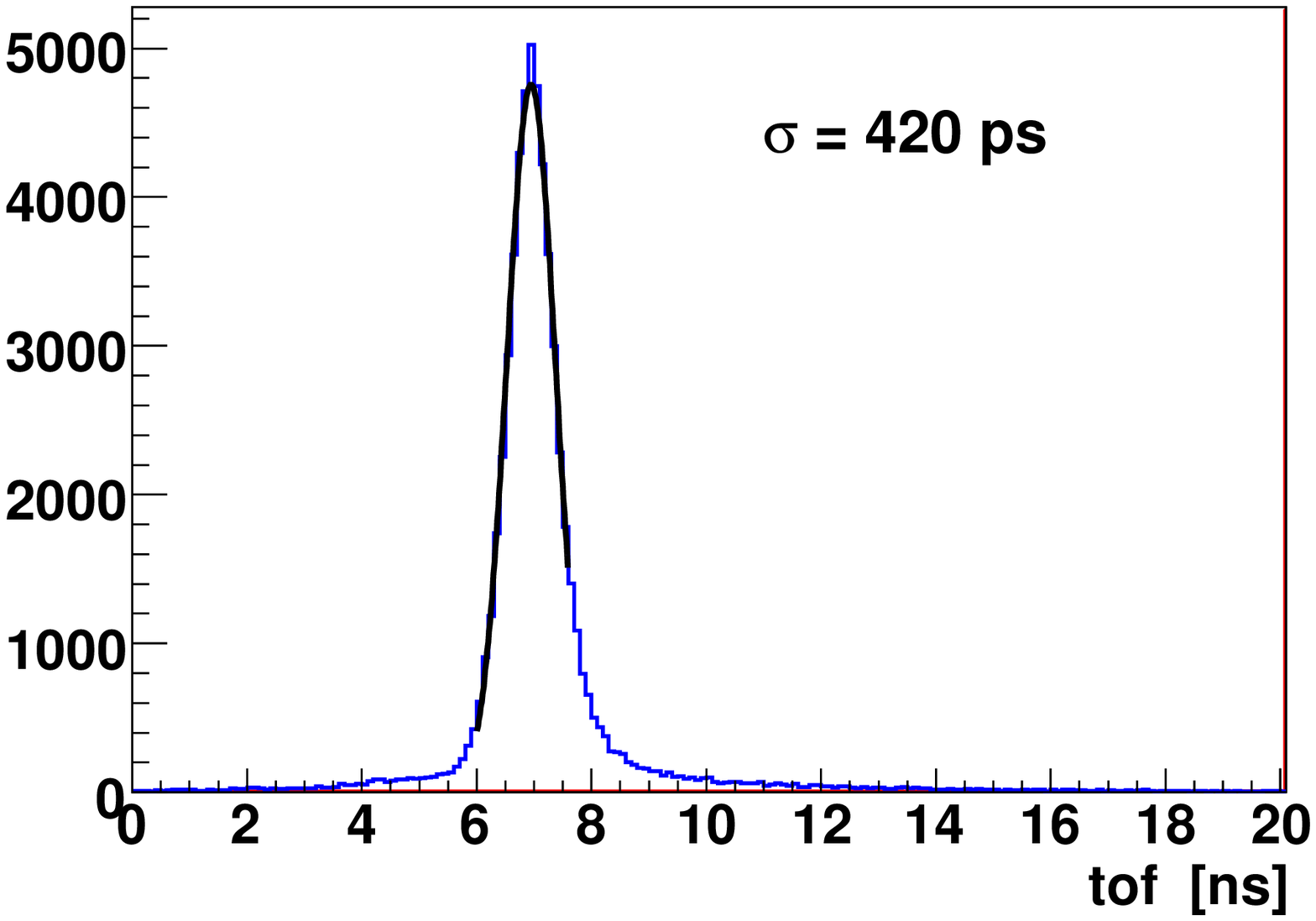}
\includegraphics[width=0.5\textwidth]{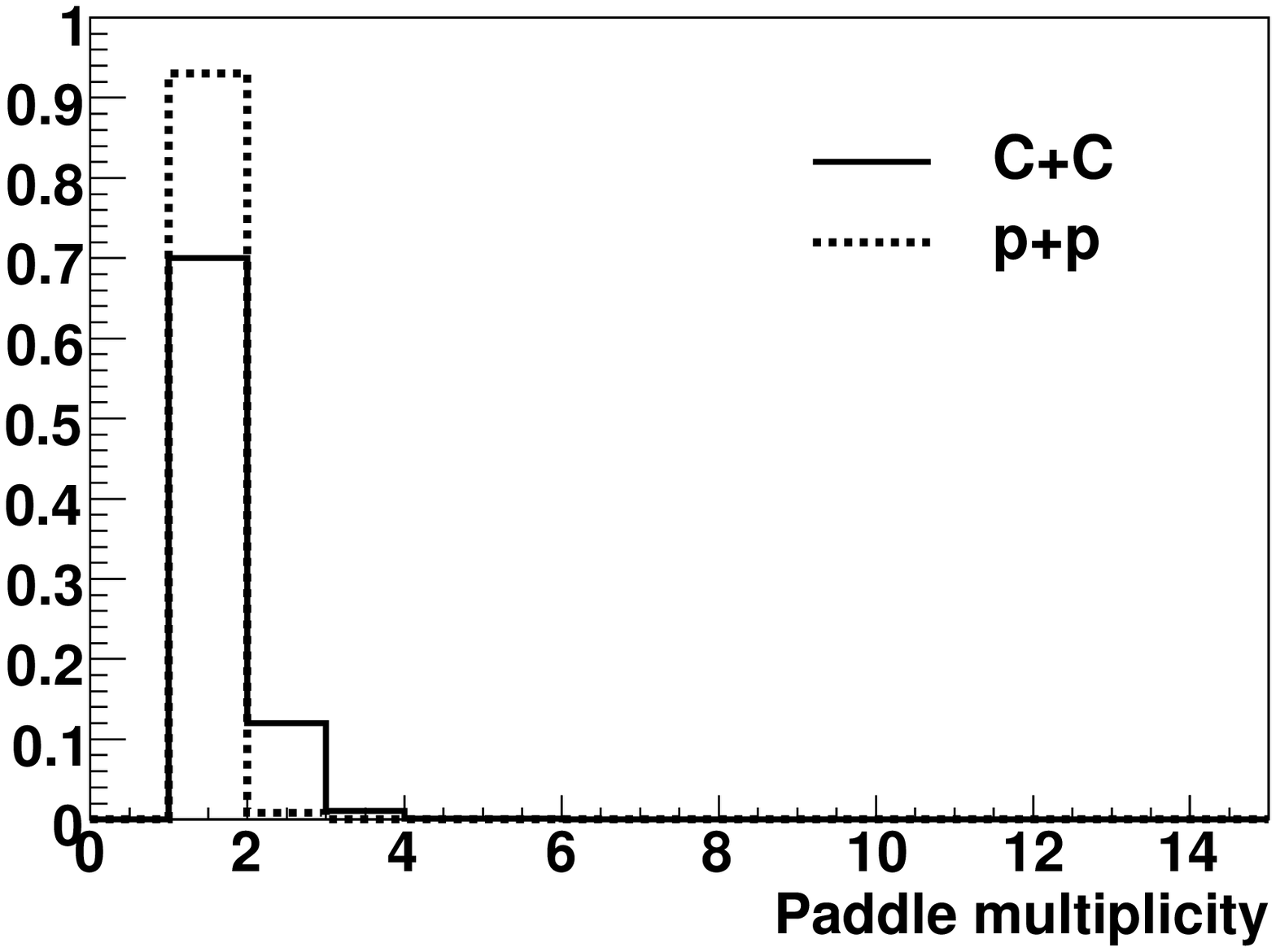}
\caption[]{Left: TOFINO time-of-flight for lepton tracks from
C\,+\,C collisions. The time resolution is about 420 ps. Right: TOFINO
hit multiplicities for C\,+\,C and p\,+\,p collisions. } \label{TOFINO_plot}
\hspace*{.5cm}
\end{figure}

\paragraph{Performance}
The time resolution and double hit capability of TOFINO are worse
than those of the TOF wall. Figure~\ref{TOFINO_plot} shows the
electron time-of-flight distribution for C+C collisions. The time
resolution of TOFINO is about 420 ps, which is determined mainly by
the geometry of light
collection system.\\
The reduced granularity increases the probability of multiple hits
on the same TOFINO paddle, which hampers time-of-flight extraction for
such events. The hit multiplicity depends on the reaction type as
shown in fig.~\ref{TOFINO_plot}. For C\,+\,C reactions about 15~\% of all
events have 2 hits in one paddle, while for p\,+\,p reactions
this number drops below 0.4~\%. The replacement of the TOFINO
detector by a Resistive Plate Chamber (RPC) wall \cite{rpc} is
foreseen in the near future to facilitate measurements with heavy systems
({\it e.g.} Au\,+\,Au).

\subsubsection{Measurements without start detector}
\label{Chapter_nostart}
For high-intensity proton beams ($\geq 10^7 \, s^{-1}$) it is not
possible to use a start detector since the induced background
hampers the stable
RICH operation. As a consequence, there is no common start time
reference for tracks in the same event. However, in this case one
can measure the difference in time-of-flight with respect to the
fastest particle, instead of the real time-of-flight.

A new algorithm for time calibration of TOF and TOFINO systems
as well as a procedure reconstructing the start time of the
reaction have been developed \cite{tofrec}. For the time calibration,
lepton pairs emitted in the same event were used. Assuming that both leptons
travelled over the same path length, their time-of-flight
difference $\Delta_{tof}$ must be equal to zero and does not depend on the start
signal timing. Thus by setting the $\Delta_{tof}$ distributions to
zero on a paddle-by-paddle basis, all the time offsets can be determined.

The start time reconstruction algorithm relies on the
assumption that the particle identity is known from other detectors.
From the assumed mass and the measured momentum, one obtains a
time-of-flight estimate, and thus the offset to the real start time
of the reaction. In HADES, there are two cases where such
identification is unique. In the first case we use the RICH
 detector which is offering powerful electron/positron
identification. For events without electrons or positrons we search for
a negatively charged particle, which can be deduced from the
track bending direction, and assume that it is a pion
(obviously the most copiously produced negatively charged
particle). In this way we
obtain an average time resolution $\sigma_{tof}$ of 340 ps and an efficiency of
about 92~\% for events with a lepton, whereas a resolution of 440 ps and
an efficiency of about 93~\% is obtained for events with a negative pion.

\begin{figure}[hbt]
\begin{center}
\centering \mbox{
\subfigure{\includegraphics[width=0.45\textwidth,height=130pt]{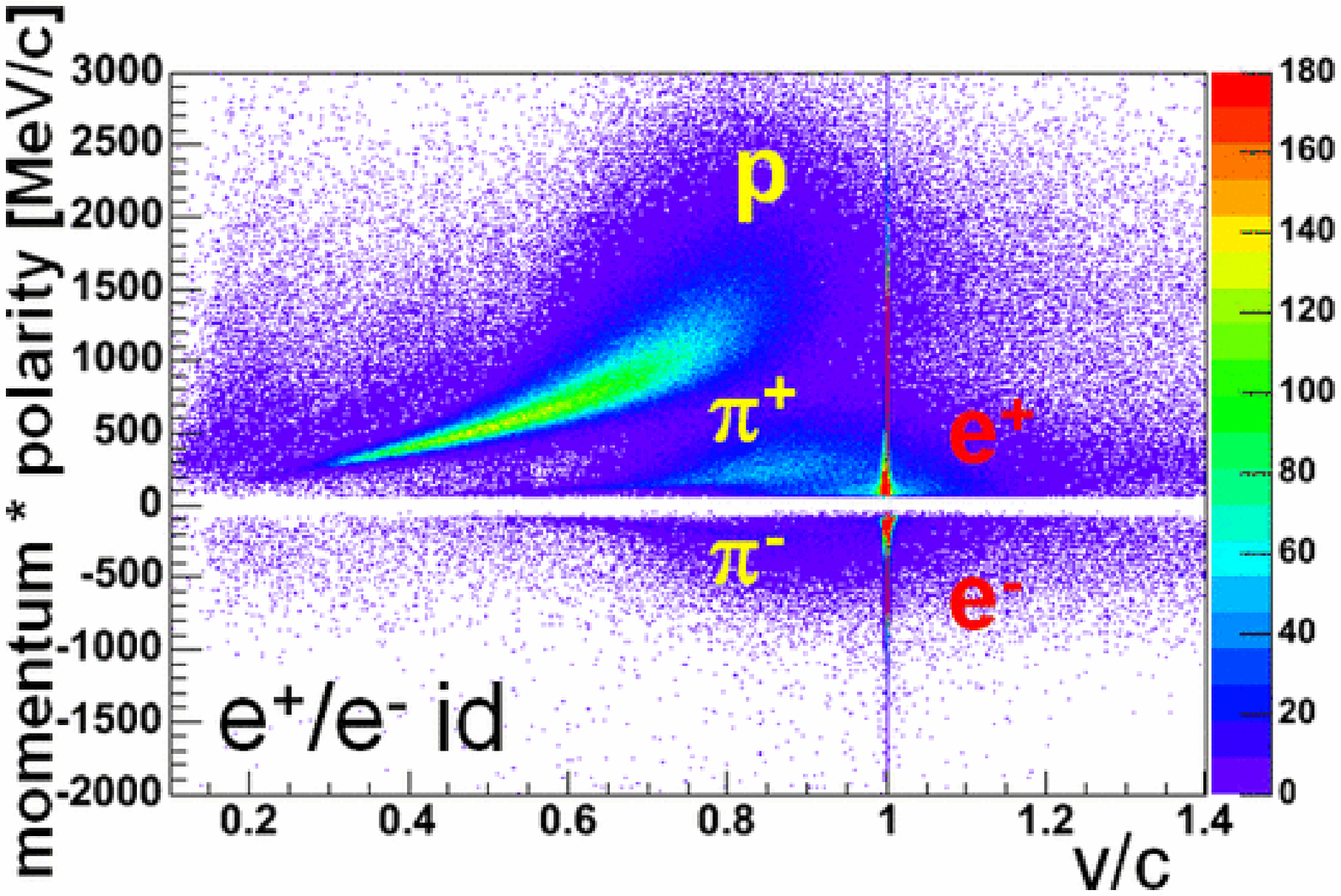}}
\subfigure{\includegraphics[width=0.45\textwidth,height=130pt]{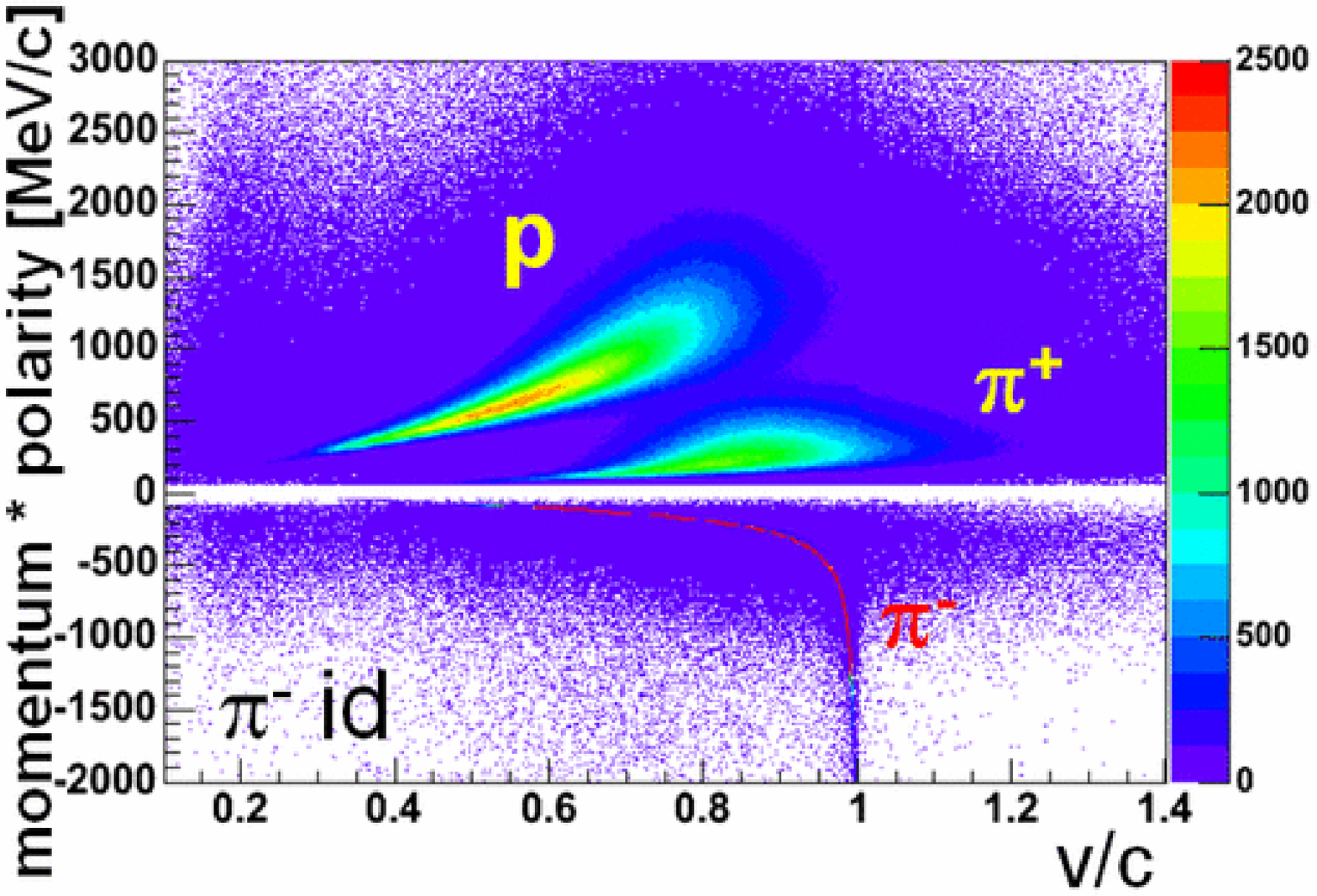}}}
\caption[]{Left: $Zp$ vs. velocity
for events from pp collisions at 2.2~GeV with an identified $e^{+}/e^{-}$. Right: Same but with an identified $\pi^{-}$ .}
\label{TOFREC_plot}
\end{center}
\end{figure}

After the start time reconstruction it is possible to use the recalculated
time-of-flight in order to identify particles for exclusive
analysis of decay channels, which includes either electrons
or negative pions. Figure~\ref{TOFREC_plot} shows momentum times polarity
versus velocity plots after start time reconstruction for the two
different cases. As one can see, protons and pions are well resolved.

\clearpage
\subsection{Pre-Shower}
\label{Chapter_shower}

\newcounter{counter1} \setcounter{counter1}{0}
\newcounter{counter2} \setcounter{counter2}{-5}

\begin{figure}[hbt]
\begin{center}
      \begin{picture}(370,220)(0,-10)
      \put(0,0){\includegraphics[width=0.5\textwidth]{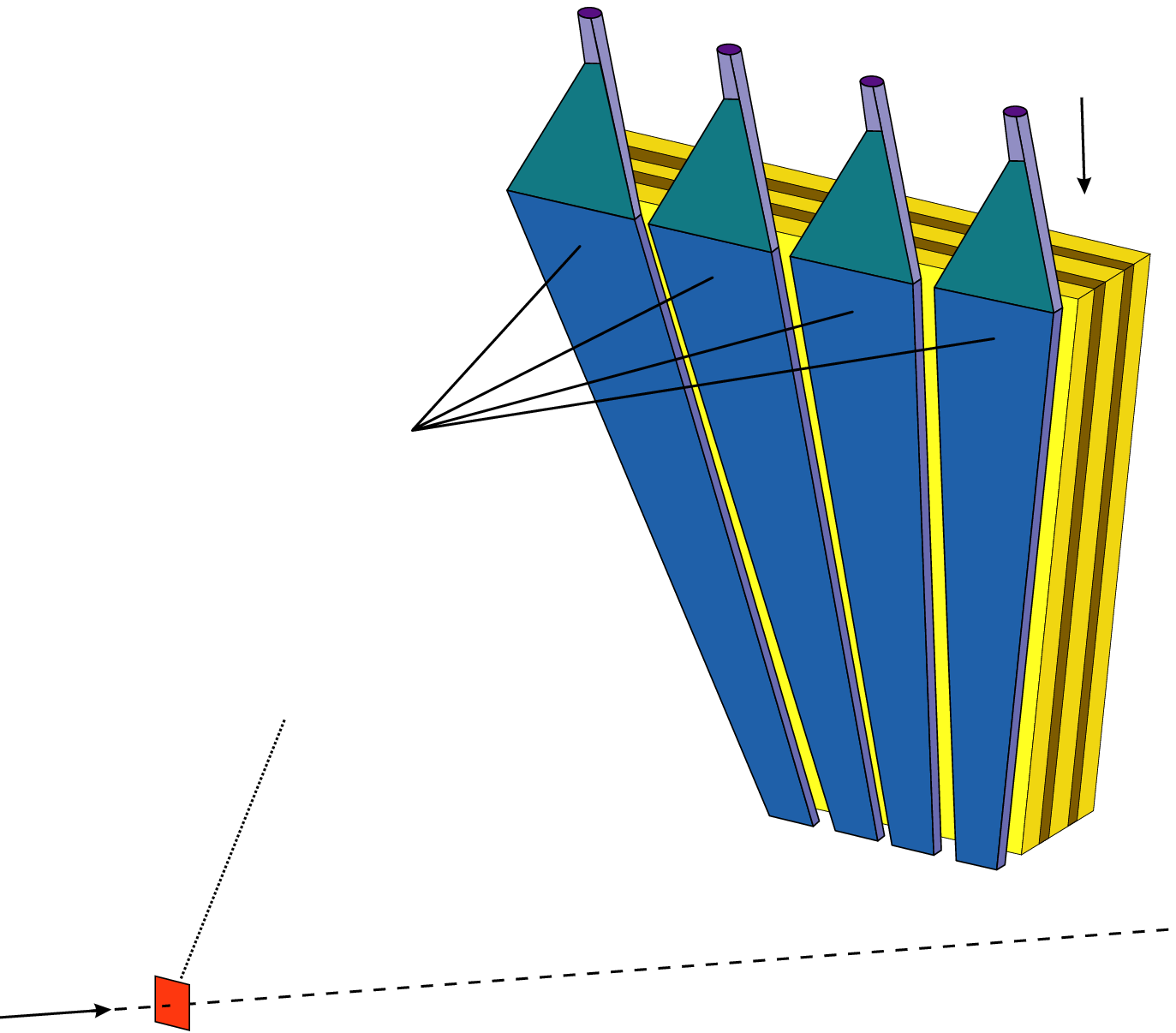}}
      \put(180,10){\includegraphics[width=0.35\textwidth]{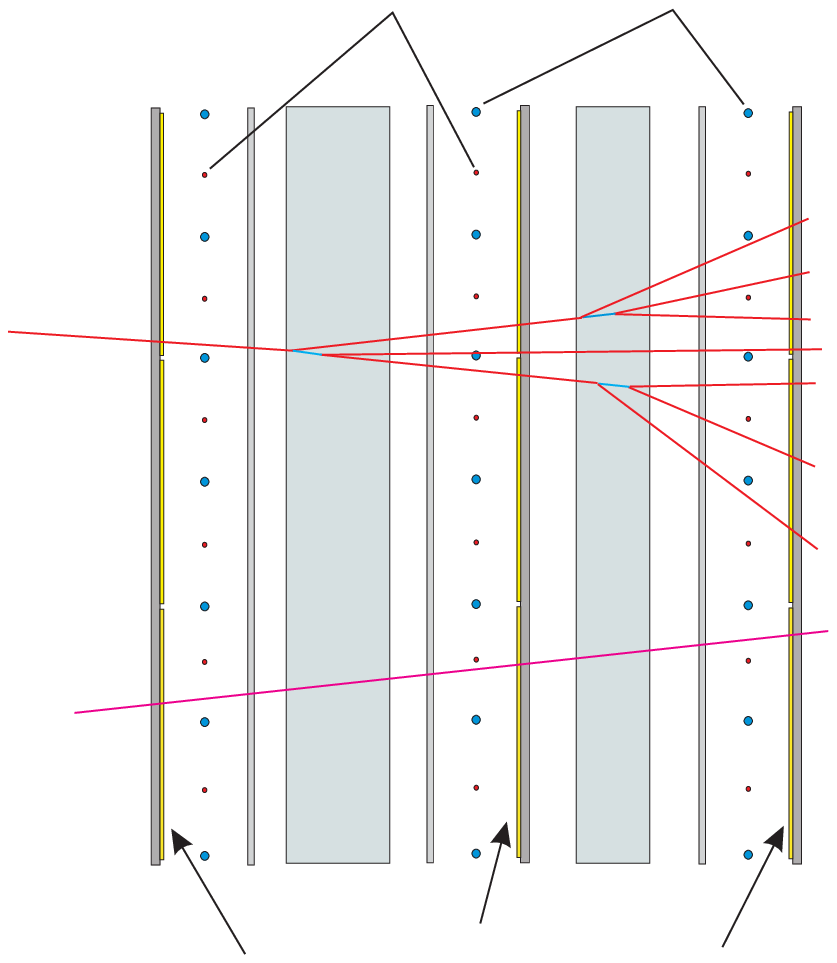}}
      \put(26,52){\scriptsize target}
      \put(0,72){\scriptsize 4 TOFINO paddles}
      \put(122,146){\scriptsize Pre-Shower}
      \put(200,166){\scriptsize potential wires}
      \put(260,166){\scriptsize ground wires}
      \put(230,7){\scriptsize read-out pad planes}
      \put(230,0){\scriptsize (cathode planes)}
      \put(48,0){\scriptsize beam direction}
      \put(227,37){\begin{turn}{90} \scriptsize Pb converter \end{turn}}
      \put(267,37){\begin{turn}{90} \scriptsize Pb converter \end{turn}}
      \end{picture}
\caption{Left: Schematic layout of the Pre-Shower detector and forward
TOFINO (one sector). Right: Side view of the Pre-Shower detector.
The three gas chambers and two Lead converters are shown. Electron-hadron
discrimination is performed by comparing the number of particles
measured in the chambers in front of and behind the Lead converters.
        \label{shower_tofino} }
\end{center}
\end{figure}

The Pre-Shower is located just behind the TOFINO at forward angles for
electromagnetic shower detection. Each sector module comprises three
trapezoidal wire chambers (pre-chamber, post1-chamber,
post2-chamber), separated by Pb converter plates as shown in
fig.~\ref{shower_tofino}. The basic idea of electron or positron
identification is schematically presented in
fig.~\ref{shower_tofino} (right): a charged particle passing through the
gas chambers is registered by measuring the induced charge on the
cathode pads with individual read-out. In order to obtain the
complete charge of the electromagnetic shower an integration over
several pads around the pad with the highest local charge value
(local maximum) is performed. The integration is done in parallel on
the corresponding pads of the three wire chambers. The comparison of the
integrated charges from the different layers is the basis of the
electromagnetic shower recognition algorithm, described in more
detail below.

The wire chambers are filled with an isobutane-based gas mixture
and are operated in the limited self-quenching streamer (SQS) mode
\cite{sqs1,sqs2}. The SQS mode guarantees that an avalanche charge
depends weakly on the particle specific energy loss because of the
charge saturation effect limiting the chamber gain. In this mode the
integrated charge is rather proportional to the number of particles
traversing the chamber than to their specific energy loss. This
mode of operation was selected to reduce fake contributions from
non-minimum ionizing protons. Such protons produce larger energy
losses in the post-converter chambers and would mimic
electromagnetic showers.

\subsubsection{Detector construction}

\begin{figure}[hbt]
\begin{center}
      \begin{picture}(370,200)(0,-10)
      \put(0,0){
          \includegraphics[width=0.41\textwidth,height=.45\textwidth]
          {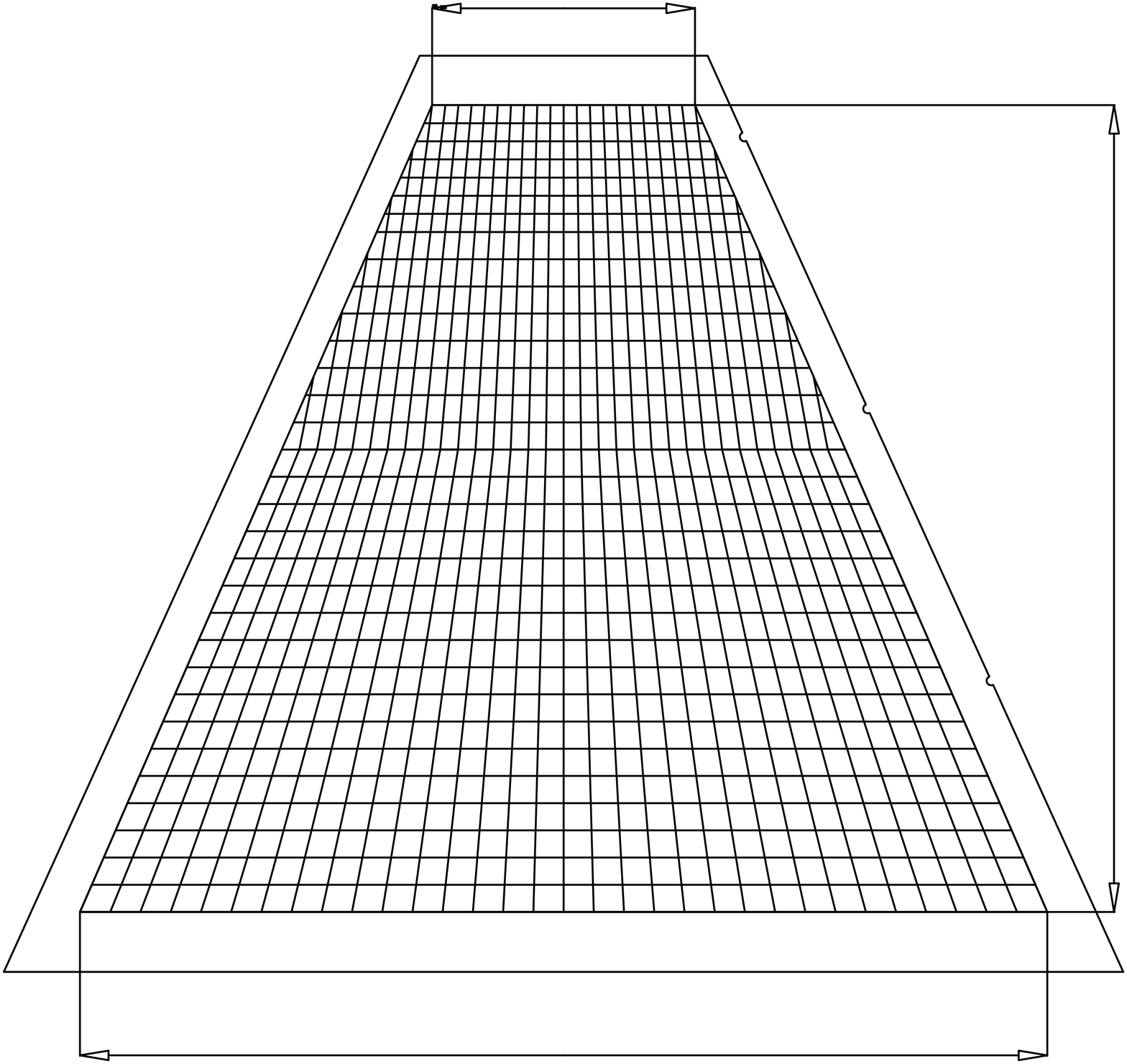}}
      \put(185,0){
          \includegraphics[width=0.41\textwidth,height=.45\textwidth]
          {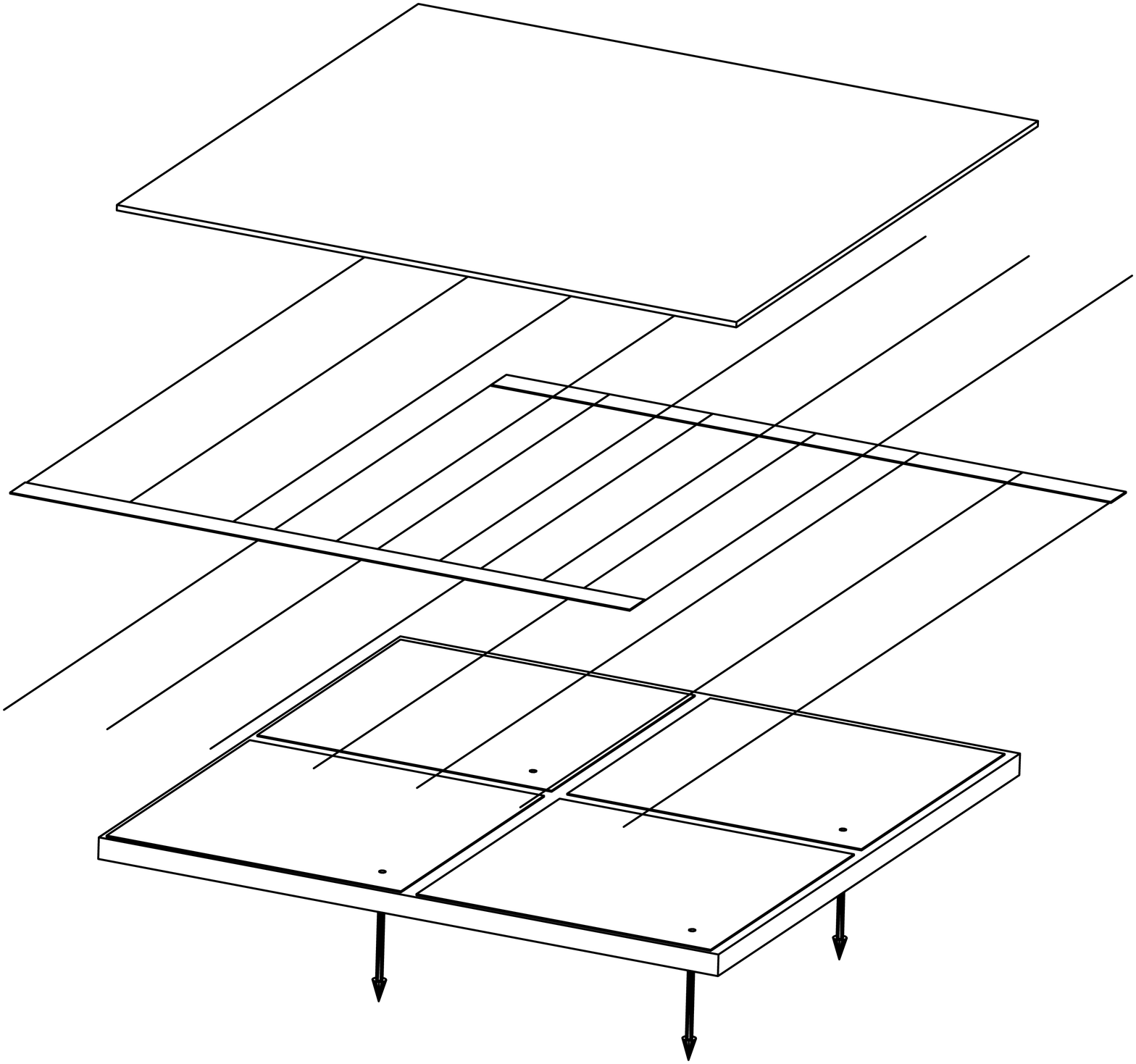}}
      \put(153,74){\begin{turn}{90} \scriptsize{126 cm, 32 rows} \end{turn}}
      \put(67,163){\scriptsize{42 cm, 20 columns}}
      \put(47,3){\scriptsize{160 cm, 32 columns}}
      \put(274,-10){read-out}
      \put(310,28){\scriptsize{pad plane}}
      \put(290,68){\scriptsize{anode wires}}
      \put(290,115){\scriptsize{cathode wires}}
      \put(240,129){\scriptsize{\shortstack[c]{cathode\\stainless steel}}}

      \end{picture}
 \caption{Left: Schematic drawing of a chamber of the Pre-Shower
 detector. The position-sensitive pad plane with 942 pads is shown. Right: The anode
wires operating at +2.7 kV and cathode wires at ground potential are mounted on
separate frames symmetrically positioned between two cathode planes.
        \label{shower_budowa} }
\end{center}
\end{figure}

A single sector of the Pre-Shower detector is composed of three wire
chambers and two Lead converters (fig.~\ref{shower_tofino}) of
$d_1\,=\,2\,X_0$ and $d_2\,=\,1.5\,\mathrm{X_0}$ thickness,
respectively, where ${X_0=0.56}$ cm is the Lead radiation length.
The wire chambers are constructed
identically, as shown in fig.~\ref{shower_budowa}, and consist of
one wire plane of equally spaced cathode and anode wires positioned
at \mbox{$\mathrm{4~mm}$} distance from two flat cathode planes. One
cathode plane consists of a 1 mm stainless steel sheet. The
second one is made of fiber glass with Copper
cladding. The latter one is divided into 942 pads to allow
individual read-out. The geometrical size of the three
wire chambers, internal wire separations, pad sizes and the
dimensions of the converters are scaled proportionally to achieve
the same solid angle coverage for all wire chambers of a given sector.
The construction parameters of the chambers were obtained from
simulations aimed at assessing the optimal geometry for minimizing
the double hit probability in a single pad (below 5 \% for Au\,+\,Au
collisions) and minimizing the number of pads involved in the charge
integration procedure in order to simplify the on-line electron
search. The resulting pad structure of the pre-chamber is presented
in fig.~\ref{shower_budowa}. A detailed description of all
dimensions and aspects of mechanical construction can be found in
\cite{ShwNim}.

The pads are arranged into rows (32 per chamber, see fig.~\ref{shower_budowa})
which are connected to front-end electronics
boards, based on a dedicated ASIC chip which represents a 32-channel
charge amplifier/shaper with output multiplexer, and digitization
with an 8-bit ADC \cite{noel}. In order to simplify the shower
pattern recognition algorithm (sect.~\ref{Chapter_daq}) full
digital information from 32$\times$32 pads per chamber is sent to a
dedicated Image Processor Unit (IPU).

\subsubsection{Performance}
\label{Chapter_showerid}

The main aim of the Pre-Shower detector is to identify electrons (positrons) by
means of the electromagnetic shower detection. The  electron candidate search consists of the following steps:
\begin{figure}[htb]
\begin{center}
\begin{picture}(370,160)(0,-10)
\put(5,-5){\includegraphics[width=0.8\textwidth]{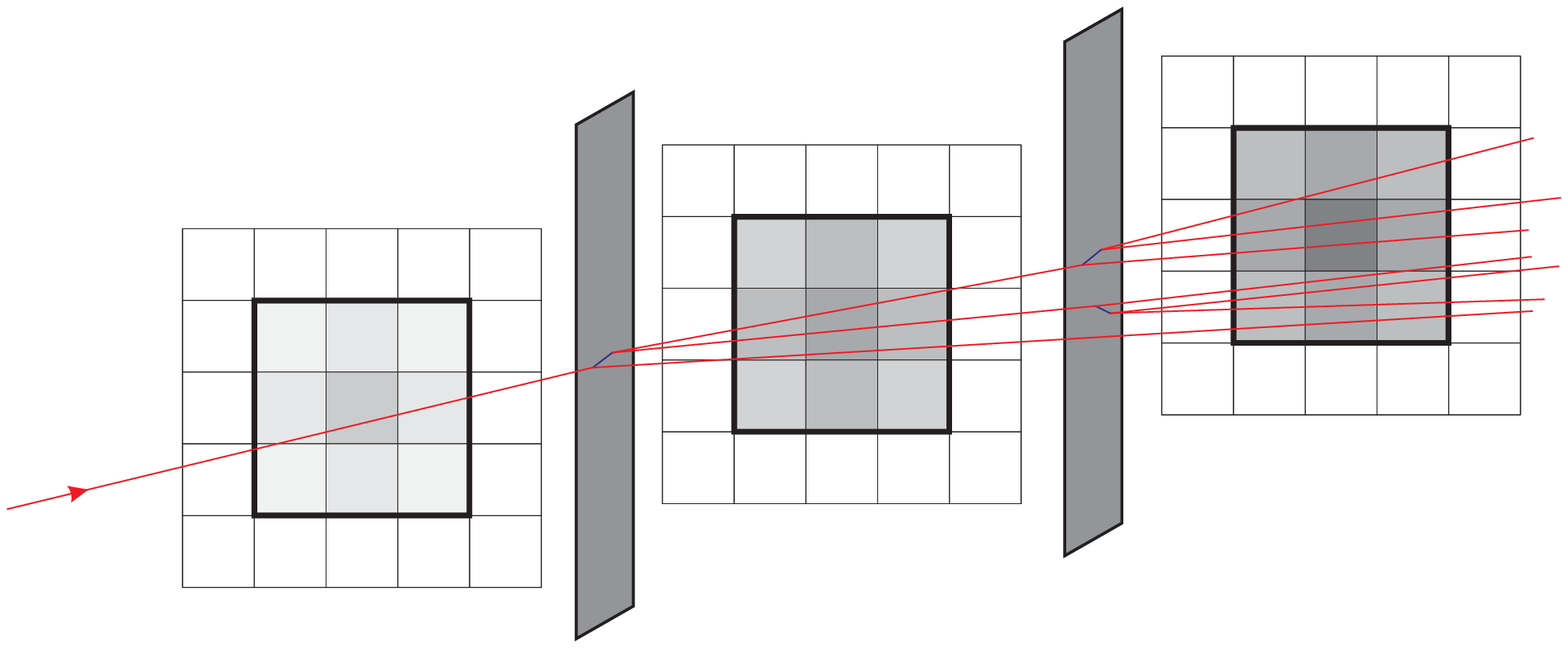}}
\put(135,-5){\footnotesize{Pb converters}}
\put(145,0){\vector(-2,1){25}} \put(160,0){\vector(2,1){25}}
\put(5,40){\footnotesize{electron}}
\put(45,100){\footnotesize{pre-chamber}}
\put(130,110){\footnotesize{post1-chamber}}
\put(225,130){\footnotesize{post2-chamber}}
\end{picture}
\caption{Schematic view of the shower algorithm. In each layer the
sum of the charge over $3\times3$ pads is calculated. The larger charge deposition
in the post-chambers is the signature of an electromagnetic shower.}
\label{shower_scheme}
\end{center}
\end{figure}

$(i)$ finding a local maximum of the charge distribution in the
pre-chamber (hit reference position) or if not present, a local
maximum of the charge in the post1-chamber, $(ii)$ integrating the
charge over 3$\times$3 pads (a local maximum pad and eight neighboring
pads) in the pre-chamber ($\Sigma_{pre}$), post1-chamber
($\Sigma_{post1}$) and post2-chamber ($\Sigma_{post2}$) around the local
maximum and $(iii)$ finally, applying one of the following electron
identification algorithms:

\begin{itemize}
\item[a.]Maximum Sum $T_{S1}(p), T_{S2}(p)$:
\setlength\arraycolsep{2pt}
\begin{eqnarray}
             \frac{\Sigma _{post1}}{\Sigma _{pre}}\geq
            T_{S1}(p) \quad or \quad
             \frac{\Sigma _{post2}}{\Sigma _{pre}}\geq
            T_{S2}(p),
\label{kaskada_algorytm2}
\end{eqnarray} where $T_{S1}(p)$ and $T_{S2}(p)$ are
momentum dependent threshold values.
\item[b.] Sum Difference:
\begin{eqnarray}
            \Sigma _{post1}+\Sigma _{post2}-
            \Sigma _{pre}(p) \geq T_{SD}(p).
\label{kaskada_algorytm3}
\end{eqnarray}
\end{itemize}

The thresholds for both algorithms have been optimized using a dedicated
Monte Carlo GEANT \cite{geant} simulation to get the best
ratio of recognized electrons to fake
events (protons, pions) and to maintain high electron efficiency
($\ge$ 80~\%) over a broad momentum range (0.1 GeV/c $\le$ p $\le$
1.5 GeV$/c$).

\begin{figure}[hbt]
\begin{center}\centering
\mbox{
\subfigure{\includegraphics[width=0.5\textwidth]{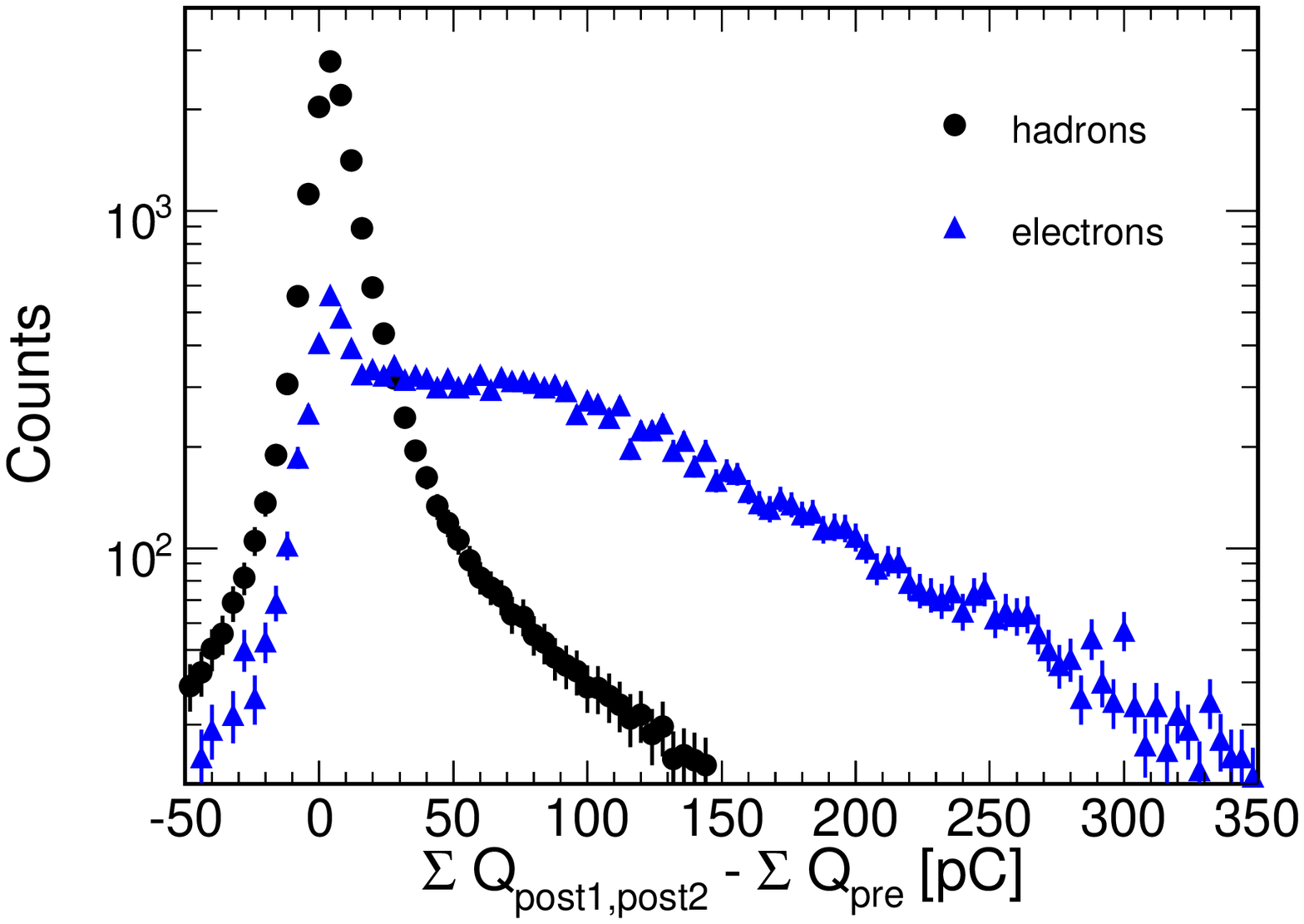}}
\subfigure{\includegraphics[width=0.5\textwidth]{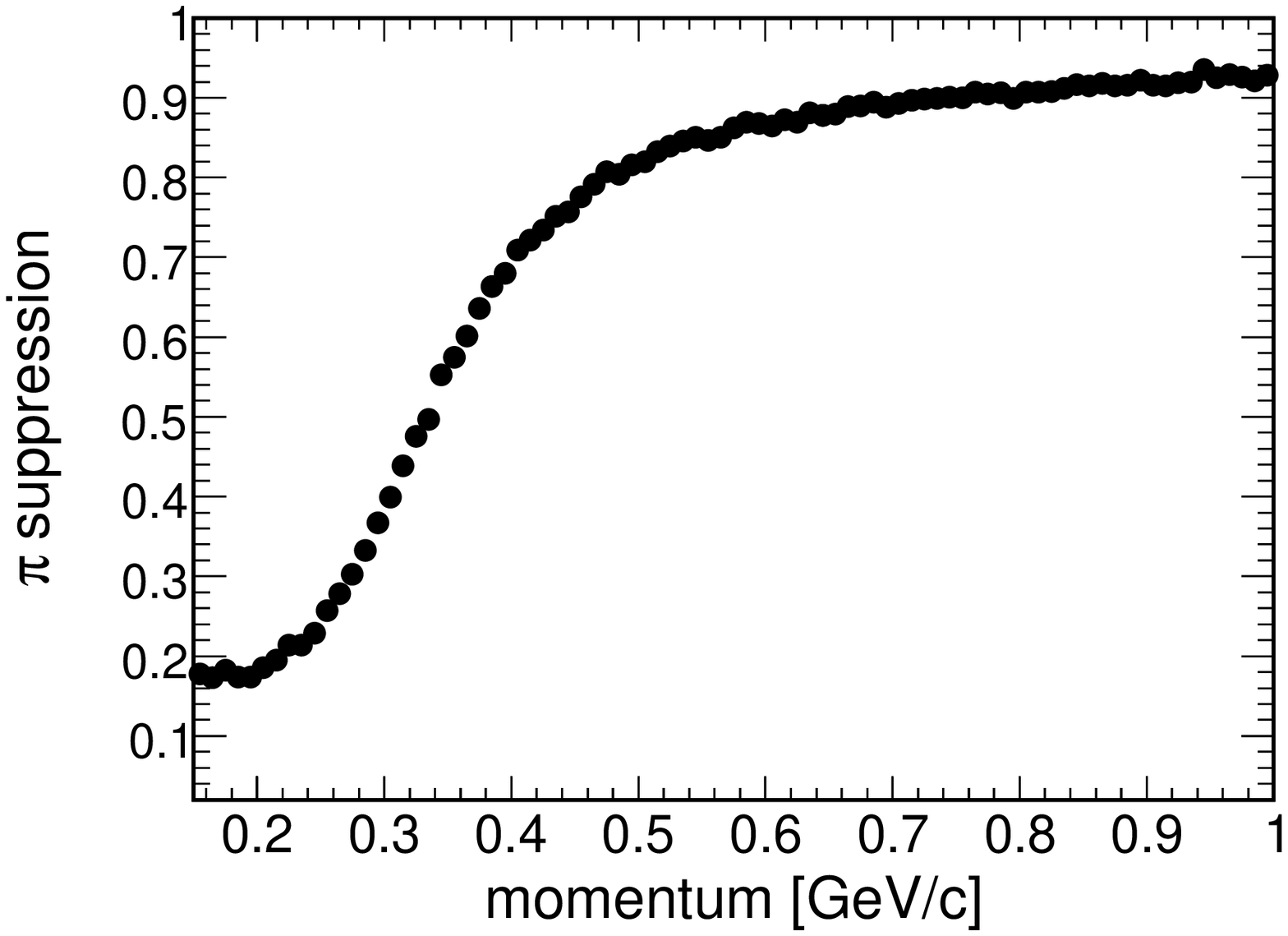}}}
\caption{Left: Sum of charge $\Sigma Q$, over 3$\times$3 pads, measured on
post1- and post2-chamber with subtracted charge measured in the
pre-chamber. Electrons with momenta $p \sim 0.5$ GeV$/c$ (triangles)
produce electromagnetic cascades in the Pb converters which
results in a larger charge deposit compared to the one from
hadrons, here $\pi^-$ (dots). Right: Pion suppression as a
function of momentum: the fraction of pions rejected after the
algorithm for lepton recognition (electromagnetic cascade) has been
applied. Data are taken from 3.5 GeV p\,+\,p collisions.}\label{shower_charge}
\end{center}
\end{figure}

Figure~\ref{shower_charge} (left) shows the Sum Difference
(eq.~\ref{kaskada_algorytm3}) measured for electrons with momenta
$p\approx 0.5$ GeV$/c$ (triangles) and minimum-ionizing pions. Using
these distributions the electron-hadron rejection can be evaluated
for a given charge threshold as a function of momentum. The right
panel of fig.~\ref{shower_charge} shows the pion rejection obtained
for a threshold yielding a constant electron efficiency of 80~\%.

The fraction of fake events is about 10~\%
for minimum ionizing protons or pions and increases for low momentum
protons. However, this can be reduced without affecting the electron
identification efficiency by applying a time-of-flight window of
8.7~ns on TOFINO hits. Using the
Pre-Shower condition (eq.~\ref{kaskada_algorytm3}) and the
time-of-flight window, more than 90~\% of the hadrons are rejected.
Finally spatial correlations of electron candidate hits in the Pre-Shower with reconstructed tracks in the MDCs and RICH rings provide a clean electron identification (see sec. ~\ref{chapter_pid})

\clearpage
\subsection{Target and beam detectors}
\label{Chapter_start}

\subsubsection{Overview}
The HADES physics program requires experiments with various
beam-target combinations. The investigation of elementary processes
leading to dielectron production is carried out with pion,
proton and deuteron beams incident on a liquid hydrogen target.
While the study of in-medium modifications of hadron properties at normal nuclear density
requires light projectiles on heavy targets, the effect of higher temperature
and compression can be investigated only with heavy projectiles.
For the elementary process studies, beam detectors based on scintillating
fibers have been developed.
For the beam monitoring, beam-profile,
time structure analysis and finally for the time-of-flight measurement for heavy-ion
experiments, polycrystalline diamond detectors are employed.

\subsubsection{The liquid hydrogen target}

\begin{figure}[htb]
 \center \includegraphics{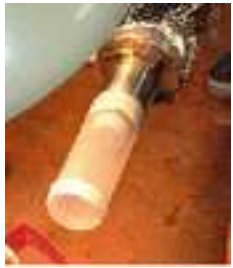}
\caption[]{LH$_2$ target. The
entrance window is glued on a stainless steel tube with a
diameter of 15~mm (visible through the transparent wall of
the cell), whereas the target vessel diameter is 25~mm. }
\label{lh2}
\end{figure}

The liquid hydrogen target (LH$_2$) has been developed at IPN Orsay to
fulfill the requirements for the study of elementary processes
through collisions of light projectiles (pion, proton or deuteron)
with protons. The target consists of a cell (50~mm long, 25~mm in
diameter) filled with liquid hydrogen at atmospheric pressure and a
temperature of 20 K. The liquid is contained in a vessel built out
of Mylar foils of different shapes glued together with
ECO-BOND 286 glue (see fig.~\ref{lh2}).

The target entrance and exit windows as well as the cylindrical part
are $100~\mu$m thick. An external carbon fiber housing, 4~cm in
diameter and 0.5~mm thick, is placed around the vessel which
is thermally insulated by 10 layers of superinsulation material
($6~\mu$m thick aluminized Mylar). The forward end cap of this
cylindrical carbon fiber cylinder is also made out of a $100~\mu$m thick Mylar
foil.

For proton-proton experiments, the interaction probability
of the in-beam housing material along the beam axis amounts in total
to less than 0.05~\%, whereas the 5~cm long liquid part of
the target yields a 0.7~\% interaction probability. The low energy
threshold for protons to escape the target is 15~MeV at $90^{\circ}$ and
reaches 30~MeV at forward angles. The cooling down from room
temperature takes 12~hours. The cryogenerator operation is
controlled by a Labview based interface.

\subsubsection{Light projectile beam detectors}

A production target for pion beams has been installed at the
heavy-ion synchrotron SIS \cite{SimonPions,DiazPions}. This opens up
the possibility for a broader research program at GSI, in particular
at HADES, by including the study of elementary reactions which is a
prerequisite for understanding the complex dynamics of heavy-ion
collisions.

\begin{figure}
\centering{
\includegraphics[width=1.0\textwidth]{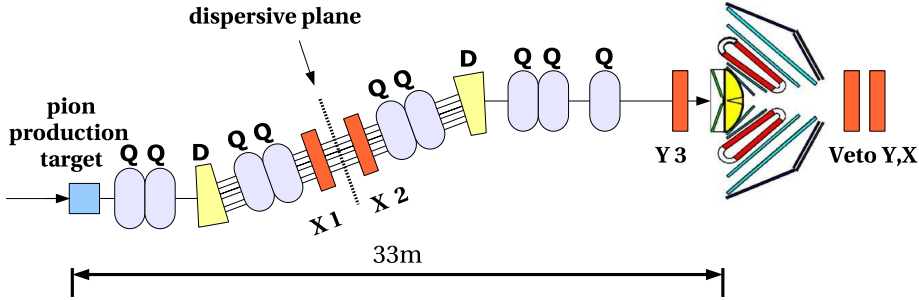}
\caption{The beam line to HADES in a configuration for
pion
 beams. Indicated are the positions of the pion tracking hodoscopes X1 and X2
 in the dispersive plane and
 Y3 in front of the target and of the veto Y,X behind the HADES setup.}
\label{fig:beamline} }
\end{figure}

For minimum-ionizing pion and proton beams, the diamond START
detector cannot be used since the deposited energy is too small.
Therefore, alternative detector concepts are required. In addition,
for secondary pion beams, a momentum measurement is mandatory for
each individual pion. This is achieved by measuring the position in
the dispersive plane of the beam line, shown in fig.~\ref{fig:beamline}. A system of
five detectors is available serving as START (Y3), VETO (Y,X) and
tracking devices (X1 and X2) for pion beams. All detectors
consist of scintillating plastic bars, strips or fibers, which are
read out individually with photomultipliers. Their properties are
summarized in table~\ref{pion_det}.

\begin{figure}
\begin{flushleft}
\includegraphics[width=0.65\columnwidth]{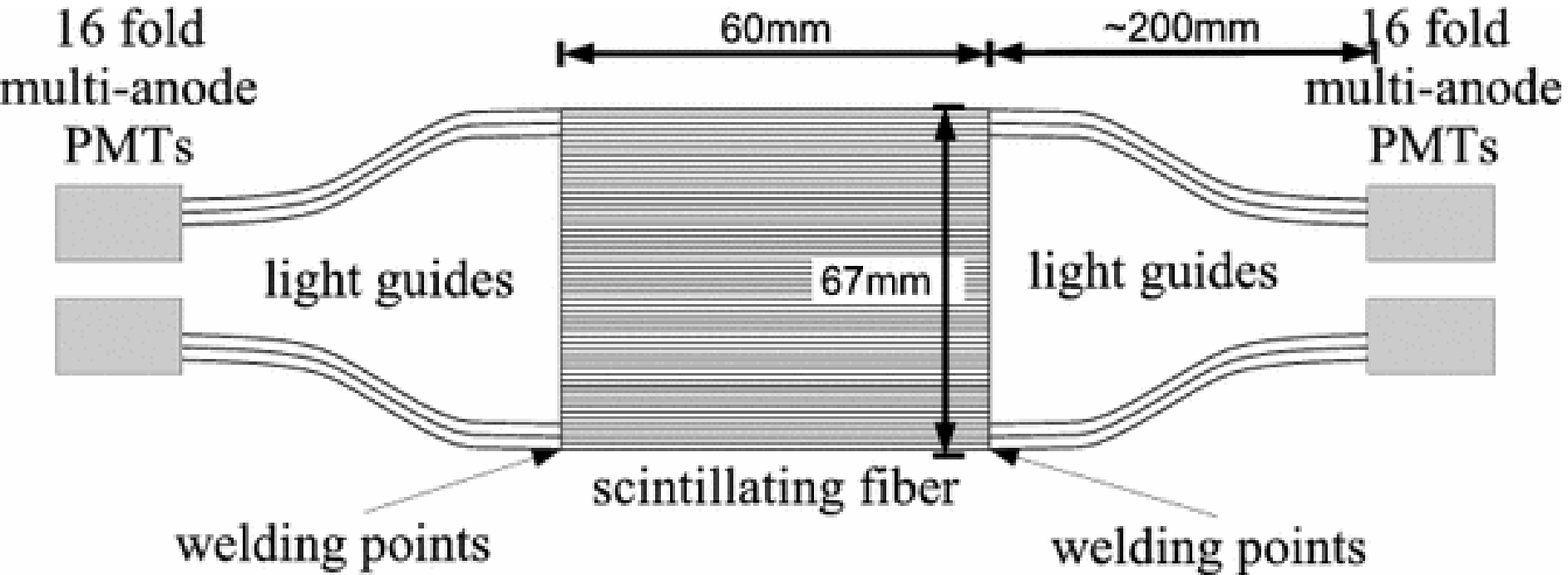}
\hfill{}
\includegraphics[bb=0bp 0bp 65mm 60mm,width=0.30\columnwidth]{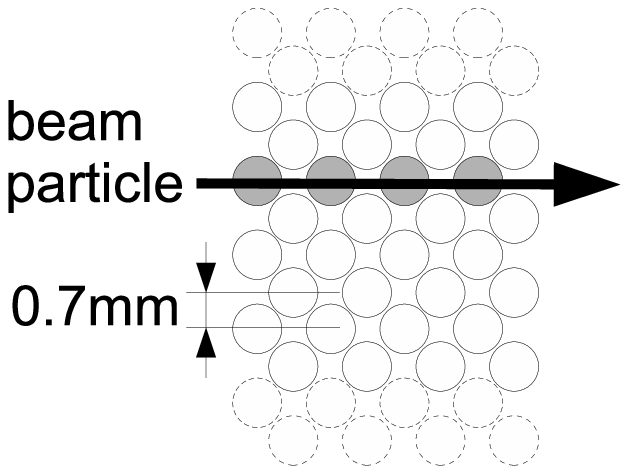}
\end{flushleft}
\caption{Left: Schematic view of the fiberplane Y3. Right: Section perpendicular to fibers.} \label{start_pion}
\end{figure}

The two scintillating bar hodoscopes used behind the pion production target
are described in detail in \cite{DiazPions}. Figure~\ref{start_pion}
shows a schematic cut of the Y3 fiber hodoscope consisting of 96
scintillating fibers of 1~mm diameter. Due to the finer granularity,
fiber detectors allow to run at a higher rate of $\approx$ 1~MHz/mm.
Neighboring fibers overlap by 30~\% to avoid geometrical efficiency
losses (close-packing, see fig.~\ref{start_pion}). To increase the
light output, in particular for minimal ionizing particles, four
layers of fibers are stacked behind each other. The fibers are read
out via light guides with 16-fold multi-anode photomultipliers
(\textsc{Hamamatsu} H6568).

\begin{table}
\begin{center}
\begin{tabular}{|p{1.1cm}|p{0.80cm}|p{1.65cm}|p{1.1cm}|p{1.55cm}|p{1.25cm}|p{1.65cm}|}
\hline
label & shape & material & channels & granularity (mm) & thickness (mm)& active area (mm$^2$)
\tabularnewline
\hline
\hline
X1, X2 & strips & EJ-212 & $\;\;$ 64 & $\;\;\;$  2  & $\;\;\;\;$  5  & 120 $\times$120
\tabularnewline
\hline
Y3 & fibers & BCF99-77 & $\;\;$ 96 & $\;\;\;$  0.7  & $\;\;\;\;$  5  & 67 $\times$60
\tabularnewline
\hline
Y, X & bars & BC404 & $\;\;$ 16 & $\;\;\;$  10  & $\;\;\;\;$  5  & 160 $\times$100
\tabularnewline
\hline

\end{tabular}
\end{center}

\vspace*{.5cm} \caption{Properties of the pion tracking hodoscopes. The labels refer to fig.~\ref{fig:beamline}.}
\label{pion_det}
\end{table}

In the intermediate focal plane the position of pions is determined
with two hodoscopes (X1/X2) of 128 scintillator strips (0.8$\times$5.0~$\textrm{mm}^{2}$)
each. They are read out with 32-fold linear phototubes
(\textsc{Hamamatsu} H7260), whereby the signals from two adjacent
strips are fed into one electronic channel.

Dedicated front-end electronics is used for the readout of the
hodoscopes. Fast signals are provided for timing and triggering
purposes. An estimate of the signal amplitude can be deduced from a
time-over-threshold measurement.

Using $\pi^-$ beams with a momentum of 1.17~GeV/c the efficiency of
the hodoscopes has been determined to be greater than 99~\%. The time
resolution is $\sigma<300\,$~ps. The position measurement at
the dispersive plane allows a determination of the pion momentum to
$\sigma_p/p \approx 0.1\,\%$.

\subsubsection{Heavy-ion beam detectors}

Heavy-ion beam intensities up to $10^8$ particles/spill for ions up to Au
and thin segmented targets (total thickness $\leq 5$ \% of the
interaction length) are being used. For this reason a START-VETO
detector system made of two radiation hard CVD (Chemical Vapour
Deposition) diamond strip counters has been designed. This solution
has the following advantages: (i) high rate capability (signal
base-width of about 2~ns), (ii) radiation hardness (better than
Silicon), (iii) fast signal collection time (hole mobility
1200~cm$^2$V$^{-1}$s$^{-1}$), (iv) low noise (band gap 5.5~eV).
Furthermore, no additional cooling is needed due to a thermal
conductivity of 1000\,-\,2000  Wm$^{-1}$K$^{-1}$; hence the detector can
be operated at room temperature.

\begin{figure}[hbt]
\begin{center}
\subfigure{\includegraphics[height=0.35\textwidth]{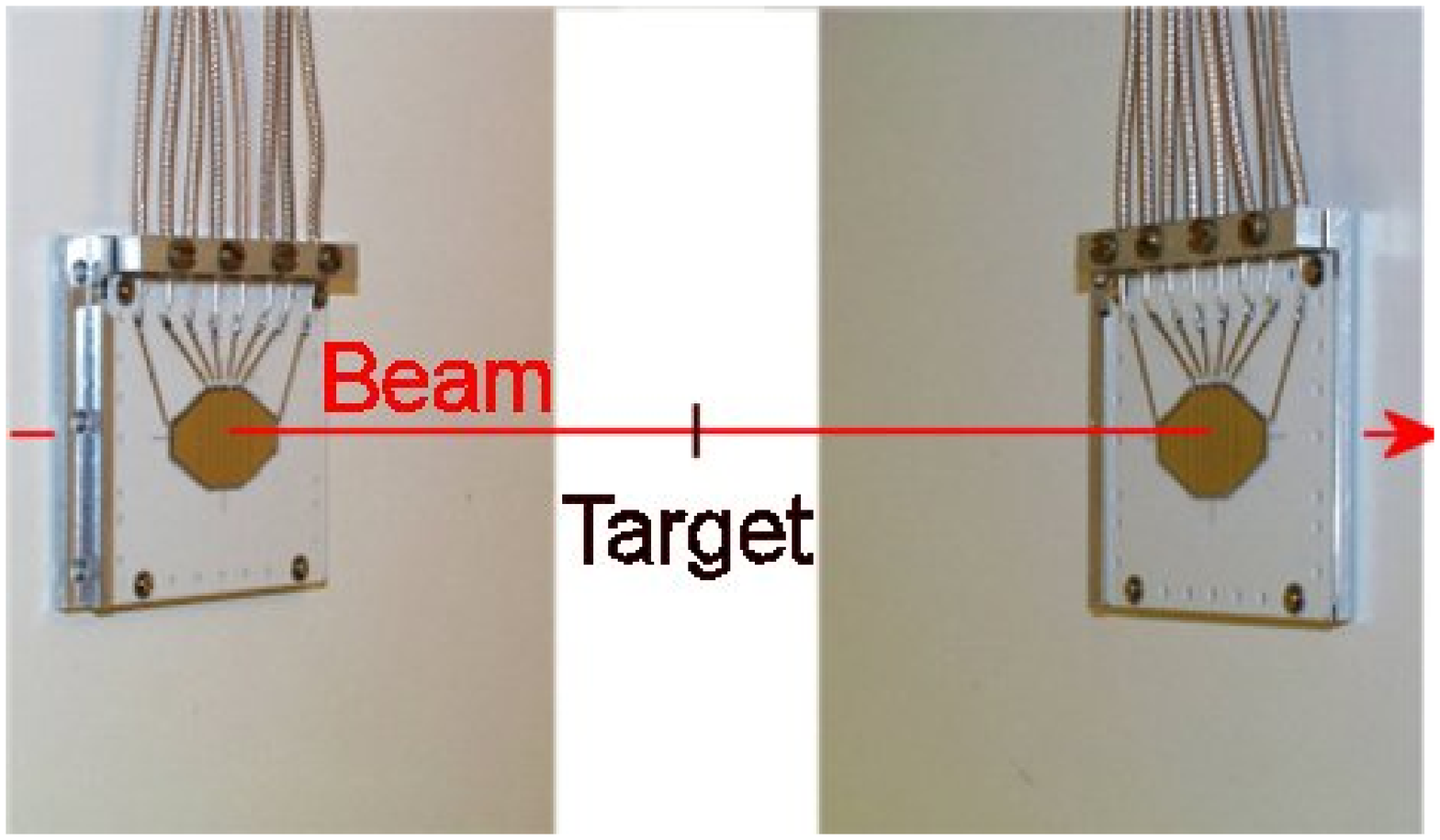}}
\subfigure{\includegraphics[viewport=15 12 541 520,height=0.35\textwidth]{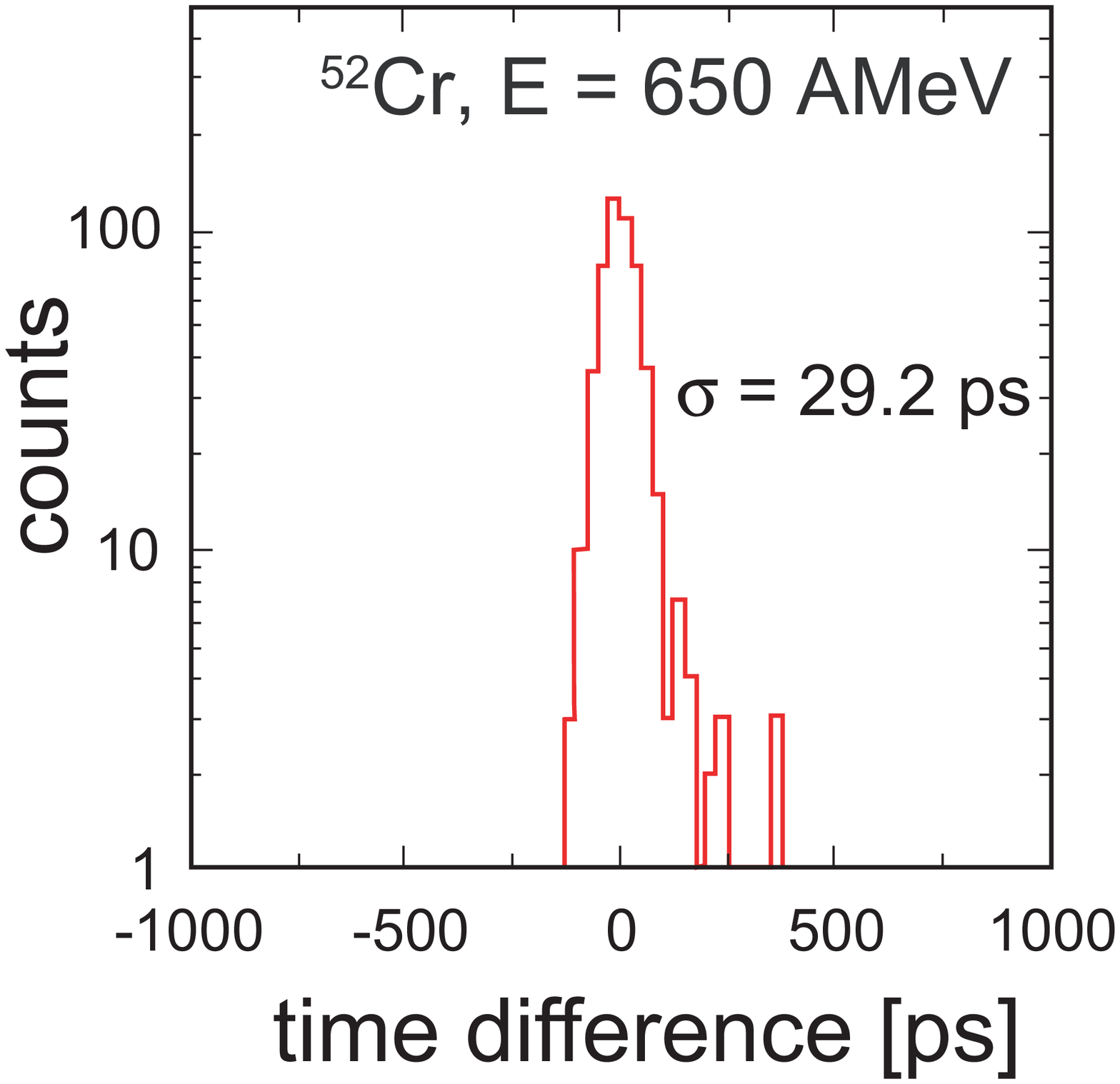}}
\caption{Left: Start-Veto system used in HADES heavy-ion
experiments. Each detector consists of 8 strips with widths varying
from 5.4~mm to 1.55~mm. {Right}: Time resolution measured with Cr
beam.}
\label{start_det}
\end{center}

\end{figure}

The CVD diamond with excellent timing properties (rise time below
500~ps) \cite{eleni} delivers a precise reference time signal for
time-of-flight measurement in HADES. The HADES START-VETO system
consists of two CVD diamond strip detectors located 75~cm upstream
(START detector) and 75~cm downstream (VETO detector) of the
target (see fig.~\ref{start_det}). Both detectors have identical
design, each consisting of 8 strips with variable width ranging
from 5.4~mm for the most outer strips to 1.55~mm for the inner
strips. The widths are optimized such that a coincidence of one
Start strip with 3 Veto strips is sufficient for a Veto efficiency
of 96~\%. Furthermore, with this design the count rate per strip
is nearly constant. In order to minimize secondary reactions in the
start detector, the thickness of the detector is 100~$\mu$m only.
The time resolution of the Start-Veto system is shown in
fig.~\ref{start_det}; it amounts to $\sigma=$ 29.2~ps.
The time resolution of the START-VETO system shown in fig.~\ref{start_det} 
was measured with $^{52}$Cr beam  at beam energy of 650 AMeV. It amounts to 
$\sigma=$ 29.2ps and is worse for lighter beam ions like $^{12}$C since the 
deposited energy is proportional to the Z$^2$. 
The time resolution measured with $^{12}$C at 2.0 AGeV beam is about 110ps.

Each strip of the detector is read-out by a low-noise
current-sensitive broad-band amplifier. The amplified signals are
further processed using leading-edge discriminators with two
outputs: (i) for the TDC (Time to Digital Converter) unit and (ii) for
the First-Level Trigger (LVL1) logic (sect.~\ref{Chapter_daq}).

\clearpage
\section{The trigger, data acquisition, and slow control systems}
\label{Chapter_daq}

As pointed out in sect.~\ref{sec:introduction}, $e^+e^-$ pairs arising from the
decay of vector mesons are rare probes. In order to avoid the data aquisition
system and the front-end electronics to be overloaded and to acquire the
statistics needed for the interpretation of the electron spectra, on-line data
reduction and event filtering have been used. The core of the event filtering
is a two-staged trigger system reducing the amount of purely hadronic
events, thus enhancing the electron yield. This strategy is outlined in
the following subsections. More detailed information on the trigger system
can be obtained from~\cite{lvl2-trigger-system-m-traxler}.

\subsection{Data acquisition and trigger distribution}

The trigger and data acquisition system of HADES is a
distributed system. Triggers are created and transmitted to the individual
subsystems from one place, the Central Trigger Unit (CTU) reacting on
external trigger input sources such as multiplicity triggers,
minimum bias or calibration triggers. As a result, a digital level-1 (LVL1)
trigger signal is generated by the CTU, consisting of a consecutive number
(trigger tag) and a trigger code containing the information about
the input source which has been activated. This information is forwarded via
Detector Trigger Units (DTUs) to readout modules where it is converted
into sub-detector specific signals and sequences, depending on the trigger
code.

After data has been read out from the front-end electronics, a level-2 (LVL2)
trigger algorithm selects events by searching for electron candidates. This is done
by Image Processing Units (IPUs) working on the data of the different detector
sub-systems using dedicated electron recognition algorithms.  The resulting hit
information is combined into a single LVL2 trigger decision in the
Matching Unit (MU) and then forwarded again via the CTU to all sub-systems.
As the LVL2 trigger decision arrives after a latency corresponding to several
events (5 to 10 events are typical values) the readout boards need to store the data
in buffers (LVL1 pipes) large enough to hold the data for this time.  After a
LVL2 trigger signal has been received the data is either copied into separate
memory (LVL2 pipe) or discarded, depending on the decision.

\begin{figure}[htb]
  \center \includegraphics*[width=115mm]{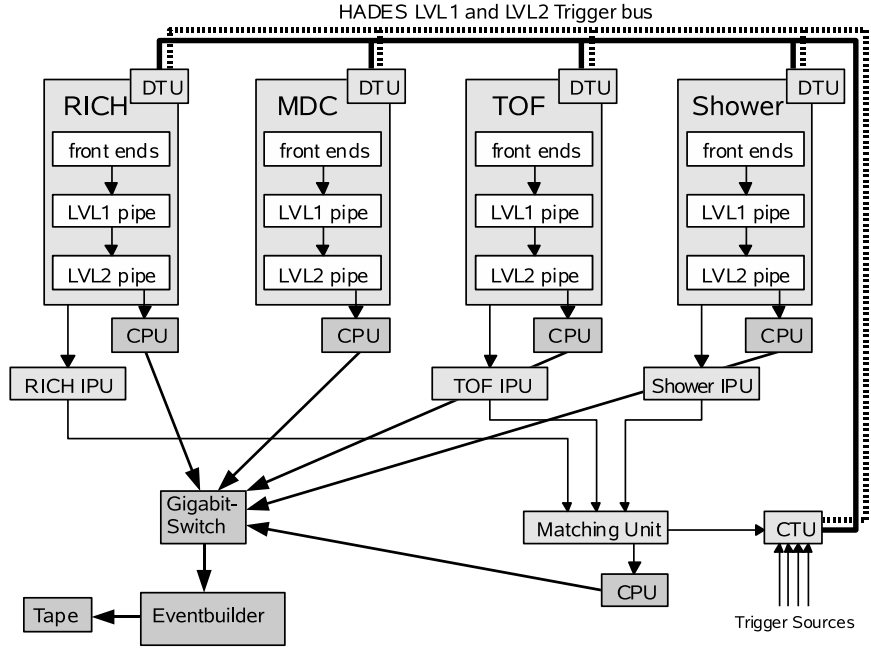}
  \caption[]{Block diagram of the trigger distribution and data
    acquisition system. Triggers are sampled by the central trigger unit (CTU)
    and transported via the trigger bus to the Detector Trigger Units (DTUs).
    The data is stored in the LVL1 pipe. At the same time pattern recognition
    algorithms are performed in Image Processing Units (IPUs), a Matching Unit
    (MU) finally combines this information.  After a positive decision, the
    data is read out via the VME CPUs and sent to the Event Builder.}
  \label{triggerdistribution}
\end{figure}

Readout programs finally transport the data (UDP network protocol is employed for this purpose)
to the Event Builder (EB), a PC which combines the data from different
asynchronous data sources into complete events and finally writes them to mass
storage.  As most of the systems are based on VME crates, these readout
programs are running on standardized VME-CPUs with Linux/LynxOS.
Figure~\ref{triggerdistribution} shows a complete overview of this concept.

\subsection{The LVL1 trigger}

The LVL1 trigger is created after one of the signal inputs of the
CTU has been activated.  At the same time a common start  signal is
provided for time measurements and serving as a source for gate
signals for charge measurements.  This signal is distributed to all
sub-detectors within typically 500\,-\,600~ns.  Meanwhile, the CTU
inputs are locked until all DTUs have released the trigger bus. As a
consequence, the HADES trigger system is not dead time free and the
rate of accepted triggers strongly depends on the readout
capabilities of the individual sub-systems. On the other hand, this
architecture makes sure to have only complete events written to disk
and no sub-detector (or part of its data) is missing.

The trigger inputs can be separated into two classes: calibration
and physics triggers. In the first case, the digitization is carried
out without zero suppression to monitor the time-dependent
performance of the detectors, obtained with 3~Hz to 4~Hz during data
taking using a pulser.

The physics trigger however is usually derived from the START detector
measuring all incident beam ions in coincidence with a reaction trigger from TOF/TOFINO.
In the case of a missing START detector
(like in the proton runs) the first particle observed in the TOF/TOFINO
detectors is used as a reference. Furthermore, as already mentioned in
sect.~\ref{Chapter_tof}, certain reaction classes have to be selected in order
to get enhanced statistics.  Besides the reaction triggers, downscaled minimum
bias triggers are recorded as well \emph{e.g.} to measure proton-proton elastic events for
absolute cross section determination.  Typical rate reductions obtained by
the reaction triggers are about one order of magnitude as compared to minimum
bias rates.  All first-level trigger decisions are implemented in fast ECL logic using
discrete components.

\subsection{The LVL2 trigger architecture}

\begin{figure}[htb]
  \center
  \includegraphics*[width=90mm]{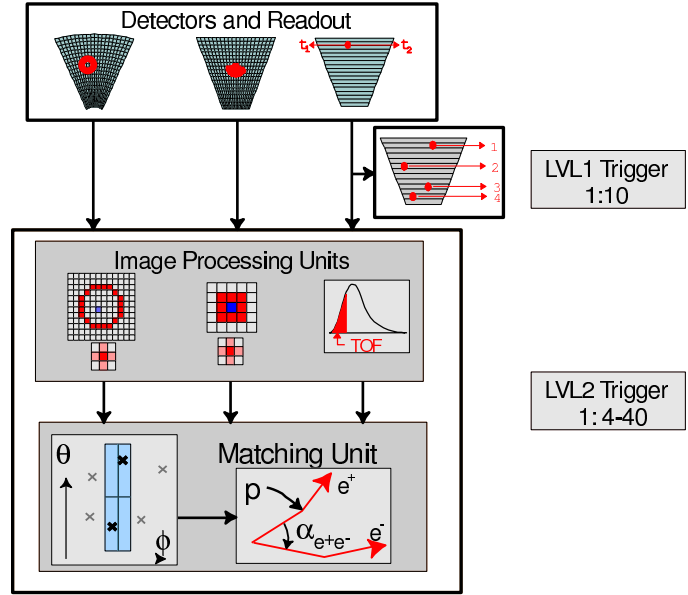}
  \caption{Overview of the reduction capabilities of the complete HADES trigger
    system.  The first-level trigger (LVL1) is generated via charge
    multiplicity measurement  with a certain admixture of minimum bias and calibration triggers.
    The second-level trigger is searching for electron hit candidates in the RICH,
    PreShower and TOF detectors which are correlated using the Matching Unit. The reduction factor
    of the LVL2 algorithm is 4 - 40, depending on the reaction under consideration.}
  \label{triggerconcept}
\end{figure}

The LVL2 trigger is based on two consecutive steps, which are sketched in
fig.~\ref{triggerconcept}.

In the first step each IPU is searching for either electron signatures in the data of the corresponding detector, respectively: Cherenkov rings in the
RICH~\cite{preformance-RICH-j-lehnert}, or fast particles in the TOF, development
of electromagnetic showers in the Pre-Shower or both. For each of these signatures
position and angle information is provided.

In the second step the Matching Unit
(MU)~\cite{dilepton-selection-thesis-m-traxler}, connected to all IPUs,
combines the angle information from the electron signatures of the
IPUs before (RICH) and after the magnetic field (TOF/Pre-Shower).  Electron
candidates are reconstructed by performing approximate tracking.  Here, we exploit the fact
that in first order particles are only deflected in polar direction. A narrow
window in the azimuthal angle allows to correlate RICH and Pre-Shower/TOF hits.
Second order corrections are taken into account by an angle-dependent matching
window. The same procedure applies for positrons.

Finally, electrons and positrons are combined into pairs
(dielectrons), which is done for like-sign as well as for
unlike-sign pairs.  The trigger conditions can be chosen in
each experimental run individually. Typical requirements are at
least one ring, one electron candidate or one pair per event. All
units are custom-built electronics, based on programmable logic
(FPGAs, CPLDs)  and digital signal processors as discussed in the
following sections.

\subsection{The RICH IPU}
\label{Chapter_RICHIPU}

The RICH IPU searches for possible ring centers on every pad of the
squared 96$\times$96 detector plane of a single sector, containing pads of
varying dimensions (see sec.~\ref{Chapter_RICH}). Thanks to the
special design of the photon detector pad plane the rings exhibit a
constant diameter of 8~pads. The ring search is challenging due
to the low photon statistics as well as by background from various
sources such as charged particle hits in the photon detector and
electronic noise. Therefore, several algorithms have been tested
\cite{RICH-IPU-diss-j-lehnert}, where the following was chosen.

For every possible ring center on one of the (96$\times$96\,=\,9216)
loci, a 13$\times$13
pad region is analyzed. All hits on a ring within a radius of 4 pads
are added. There are two veto sub-regions, inside and outside the
\emph{ring region}, where the pads are added. These hits account to
the \emph{veto region}. If a certain threshold value in the
\emph{ring region} is exceeded and the value of the \emph{veto
  region} is below its preset threshold, the center is assumed to be a
ring candidate. The found ring centres on the 96$\times$96 plane are
then cleaned using a local maximum search. All  six HADES sectors
are searched in parallel.  FPGA based pattern reconstruction modules
and a ring recognition module have been developed. They are
implemented as VME modules and obtain the hit pattern information
from the RICH readout electronics.

\subsection{The Pre-Shower IPU}

The purpose of the Pre-Shower IPU  is to find electron signatures in three layers of the
Shower detector, {\it i.e.} in all HADES sectors in parallel, as described in
sec.~\ref{Chapter_shower}. The Pre-Shower IPU searches for an increase in
charge of the post-chamber layers, which is done by summing up the charge over
3$\times$3 pad regions, as given by eq.~\ref{kaskada_algorytm2}.  Subsequently, a local maximum search is performed.
The full 8 bit pulse height information is used (3$\times$1024 pads). Additionally, the Shower IPU
performs detector readout.

The realization of the Pre-Shower IPU is similar to that of the RICH IPU. The
data is processed in parallel by FPGAs (Altera EPF10K100ARC), where the
algorithm is implemented in a pipelined fashion, row by row. Due to the
additional tasks of readout, pedestal correction, zero suppression and storage
of the data in pipes for readout after a positive LVL2 decision the
Pre-Shower IPU for the complete HADES detector consists of 12 VME boards
(motherboards with add-on card).

\subsection{The TOF IPU}

The Time-of-flight IPU (TIP) is a combined readout and image processing unit.
The TOF readout system is also handling the TOFINO, the START detector, pion
hodoscope and several other modules, like latches and scalers.

The complete TOF readout system is based on the VME standard. Analog data from
the detectors are sampled in VME based TDC/ADC modules (CAEN v878/v879). In
addition, the multi-hit TDC v1190 (from CAEN) and the latch/scaler 3820 (from
SIS) have been used. Here, the latch provides information about the fired LVL1
trigger source and the scaler monitores the numbers of triggers occured during
the dead time of the CTU.

Each of the VME crates houses up to 18 converter modules, the VME
CPU, one TIP module and one DTU.  The TOF IPUs are reading out the
TDC and ADC boards, which is done in 4 VME crates in parallel. After
readout on-line data analysis and image processing is performed by
the IPUs. In case of a positive LVL2 trigger, data is sent via one
more TOF IPU board, acting as a concentrator, to the EB.

\subsection{The trigger and readout board}

Unlike the other systems the Trigger and Readout Board (TRB)
\cite{ieee_report}) is a stand-alone ethernet based readout board replacing the
v1190 module with
on-board TDC functionality, based on the HPTDC~\cite{hptdc} with the advantage
to be closer to the detectors. It has been used in several experimental runs
to read out beam hodoscopes as well.

\subsection{Performance of the DAQ and the LVL2 trigger}

In addition to the events containing an electron candidate as
determined by the LVL2 trigger algorithm, some LVL2 trigger signals are
independently marked with a preset downscaling factor as  positively
triggered, irrespectively of their physics contents. These events
provide an unbiased event sample for a study of the LVL2 trigger
algorithm properties (see below).  The downscaling factor ranges
from 1:3 to 1:9 depending on the event size.  It was tuned to keep
the reduction of the accepted LVL1 trigger rate below 10 \%.

The performance in the LVL1 and LVL2 trigger rates are strongly coupled.  For
reactions with large particle multiplicities, where the DAQ system has to
transport more data, the LVL2 trigger rate is limited due to the bandwidth
 of the readout electronics.  On the other hand, the data
reduction factor, related to LVL2 trigger performance, depends strongly on the
collision system (background and data load) and the additional
downscaling factor for the unbiased LVL2 events. Therefore, the collision
system has a direct impact on the accepted LVL1 trigger rate.

Typical reduction factors are in the order of 1/10. They were
achieved for small collision systems including C\,+\,C for
the LVL2 trigger condition requiring at least one electron candidate
found by the MU. This LVL2 setting leads typically to 17~kHz
accepted LVL1 triggers, 2~MBytes/s transported LVL2 data (at
4~kHz) and trigger/DAQ deadtime of nearly 100 \%. For the
experimental runs, the beam intensity was chosen to result in a DAQ
deadtime of about 60 \%. Table~\ref{daq_perf} summarizes achieved
DAQ performances for the C\,+\,C and
Ar\,+\,KCl reactions.

For reactions with large particle multiplicities, where the DAQ
system has to transport more data, the LVL2 trigger rate is limited
due to the design of the data paths of the readout electronics, which is one
of the reasons for the ongoing upgrade project~\cite{ieee_report}.

\begin{table}
\begin{tabular}{|c|c|c|c|}
\hline
beam energy & accepted   &  reduction & accepted \\
/ system & LVL1 rate  & factor     & LVL2 rates \\
\hline \hline
2 AGeV/ C\,+\,C & 17~kHz & 12-20 & 4~kHz $\simeq$ 2~MBytes/s \\
1.76 AGeV/ Ar\,+\,KCl & 7~kHz & 3-4 & 1-2~kHz $\simeq$ 7-14~MBytes/s \\
\hline
\end{tabular}
\vspace{.5cm}\caption[]{DAQ and trigger performance for typical in-beam
conditions.}
\label{daq_perf}
\end{table}

However, one should note that the accepted trigger rates are
approximately  a factor of two smaller than the equivalent LVL1
rates when a random pulser is used. The reason is traced back to
short-time (10 - 150 $\mu$s) beam intensity fluctuations caused by
the extraction system resulting in a reduced effective duty cycle
and a larger dead-time.

For the above conditions we found electron pair efficiencies of
about 90 \%, in good agreement with the trigger emulation using the
off-line analysis ~\cite{Dilepton-Trigger-A-Toia}. No physics bias
is introduced by the second-level trigger, as will be discussed in
sect.~\ref{Chapter_simana}.

\subsection{Slow control}

The slow control system is based on the EPICS~\cite{epics} control
system. We use the EPICS toolkit to build the necessary client and
server programs which monitor and control the hardware of the
experiment.  As much as possible, applications available in the
EPICS community were reused and only the  parts for our custom
hardware were written from scratch. Our development included the
driver for CAMAC via VSB bus, high-voltage control with a LeCroy
1440 system, driven via CAMAC, readout of a custom temperature
monitoring system, and readout of gauges for gas bottles via serial
lines. In addition, many types of genSUB records were produced to
control VMEbus-based custom hardware. Some devices were designed
using the field bus CANbus for crate control, for control of
low-voltage regulation boards, for power supplies, and for the fast
RICH current monitor.

On the client side, we use the common tools for user, namely a
graphical user interface, an alarm handler and a backup and restore
tool. To store values of read-back parameters the ChannelArchiver tool is
used. The stored data can be used for reconstructing gains and detector efficiencies.
A more powerful means of storage than the internal file
format of the Archiver was developed at SLAC, the OracleArchiver, which has been
adapted to our needs and extended to provide summing and mean value generation
during runs.


\clearpage

\section{Data analysis and detector performance}
\label{Chapter_simana}

\subsection{Framework}

The HADES on-line/off-line analysis is realized within the HYDRA
framework \cite{sanchez}, {\it i.e.} the {\bf H}ades s{\bf Y}stem for
{\bf D}ata {\bf R}eduction and {\bf A}nalysis, based entirely on
the C$^{++}$ class package ROOT \cite{root}.
This approach allows full and consistent
use of all built-in features of the ROOT software developed and
maintained at CERN, and which has meanwhile become a {\it de facto}
standard in most high energy and nuclear physics experiments.  The
object-oriented design chosen for HYDRA is modular, which means that
detector-specific and/or task-specific classes are all derived from
a common set of base classes, and it is therefore very flexible and
extendable. Input can be taken from different data sources, namely
event servers, list-mode data files in various stages of analysis,
data summary files or simulation files. The execution flow of the
data processing is realized via freely configurable task lists,
controlled via ROOT user macros.  The initialization of
geometry, set up and calibration parameters is possible from an
Oracle data base \cite{oracle} and/or from ROOT files, with full
version management implemented.  The analysis can be run in
stand-alone batch mode or from interactive ROOT sessions.  Indeed,
the developed analysis code is routinely used, both for off-line
and on-line data processing during data-taking runs.  In the on-line
mode, convenient control of the basic functionality is available via
a Graphical User Interface implemented with the Qt widget
library \cite{qt}. Details of the HYDRA class design and
implementation are given in \cite{sanchez}.

For simulation studies, the analysis is interfaced to the detector
simulation package GEANT3 \cite{geant} from CERN via ROOT event
files that are read and digitized by HYDRA. Event overlay, {\it i.e.} the
embedding of simulated tracks into real events for efficiency and
performance investigations, is supported as well. Finally, a
comprehensive and modular ROOT-based event generator, called
PLUTO$^{++}$ \cite{pluto,pluto2}, has been developed for fast simulation
studies, but also as an input source for detailed GEANT simulations.

The HYDRA and PLUTO$^{++}$ frameworks have been implemented to run
on various flavours of the Linux operating system.  Apart from the
(freely available) Oracle data base client, only open-source
software has been used.

In the following subsections we present the aspects of the
high-level data analysis which are specific to HADES experiments,
namely the track reconstruction algorithms, the particle
identification (hadron and lepton) procedures, and the dielectron
reconstruction procedure. The description of the specific analysis
packages needed for elementary reactions, such as the event hypothesis
and the kinematic refit, will be part of a future publication as
they go beyond the scope of the present paper.

\subsection{Track reconstruction} \label{Chapter_tracking}

The reconstruction of the particle trajectories in the tracking
system of HADES is accomplished in several steps:
\begin{enumerate}

    \item The spatial correlation of fired drift cells in the
    drift chambers (MDCs) is performed by a \emph{track candidate}
    search (sect.~\ref{Chapter_Clusterfinder}) based on the
    identification of so-called \emph{wire clusters}. The wire
    clusters are defined using only the geometrical positions of the
    fired drift cells and define \emph{track segments}.
    \emph{Track candidates} are finally obtained through the matching
    of track segments in the inner and outer drift chambers within
    one sector.

    \item
    The corresponding space positions of the \emph{track candidates} are
    fitted by a model function taking into account the drift time
    information of the cells (see sect.~\ref{Chapter_track_fitting}).

    \item For electron identification, the inner \emph{track segments}
    are matched with rings in the
    RICH detector, while for all tracks the outer \emph{track segments} are matched with
    hit points in the TOF or TOFINO and Pre-Shower detectors.

    \item The particle momentum  is determined by various
    algorithms making use of the bending of its trajectory inside
    the magnetic field region (see sect.~\ref{Chapter_track_momentum}).

    \item Particle identification is supplemented by the information of the particle momentum, its time of flight and energy loss in the TOF/TOFINO and the MDCs. 
        Furthermore, the correlation with a reconstructed ring in the \mbox{RICH} detector and a detected electron shower in the Pre-Shower detector provide an efficient
    lepton identification.

\end{enumerate}

\subsubsection{Track candidate search}\label{Chapter_Clusterfinder}

\begin{figure}[\htb]
\begin{center}
\includegraphics[width=0.8\linewidth]{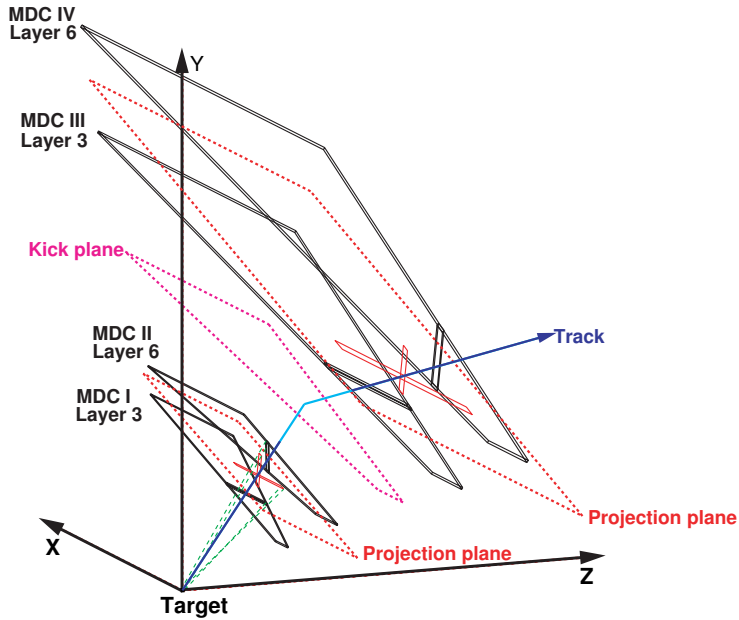}
\caption[Schematic view of the track candidate search]{Principle of
the \emph{track candidate} search in the track reconstruction
procedure. For an easy view, only one layer
 is shown in each MDC} \label{cluster_projection}
\end{center}
\end{figure}

The \emph{track candidate} search is based on the projection of the area of
fired drift cells along a certain direction onto a common projection
plane and the localization of maxima in this plane. The projection
plane chosen (see fig.~\ref{cluster_projection}) is the centre
plane of two coplanar outer chambers, while for the non-coplanar
inner geometry the plane is chosen such that the projections of the
drift cells are of similar size.

\begin{figure}[\htb]
\begin{center}
\includegraphics[width=0.9\linewidth]{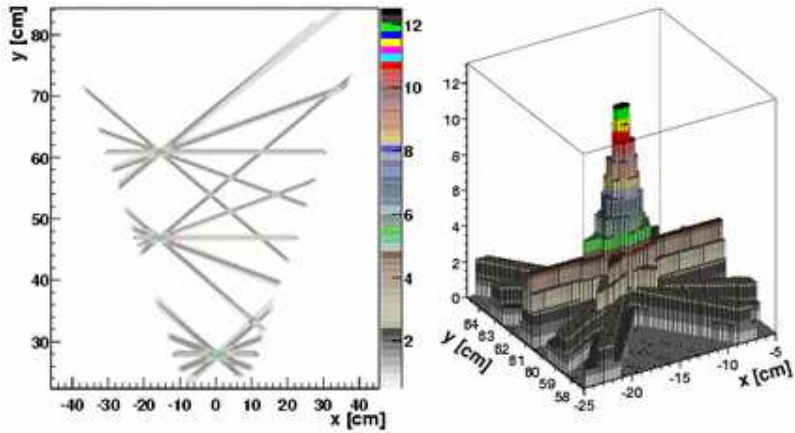}
\caption[Projection of the drift cells in the cluster finding]
{Left: $x-y$ detector coordinate space projection of
the drift cells in the cluster finding procedure. Right:
2-dimensional histogram with a peak at the location where the drift
cell projections have maximum overlap. In this example, the $z$ axis indicates the peak height, corresponding here to a track totalling 12 hit layers in the inner drift chambers \protect\cite{pech1}.}
\label{cluster_2d_projection}
\end{center}
\end{figure}

For the inner drift chambers the projection is performed with
respect to the centre of the target. Here, only the extension
of the target along the beam axis is taken into account.
The impact point of the track on the projection plane is given
by the local maximum of two-dimensional distributions built
from the slices spanned by fired drift cells, as shown in
fig.~\ref{cluster_2d_projection} \cite{pech1}. To reduce the number of fake
candidates created by accidentally crossing hit wires, all fired
drift cells belonging to either one or both inner chambers are
simultaneously projected onto one plane. The target position
and the location of the maximum in the projection plane then define
a straight track segment in space.

The deflection of a charged particle by the toroidal magnetic
field of the HADES magnet can be approximated by a
momentum kick on a nearly flat virtual \emph{kick plane}~\cite{sanchez}
in the field region (see sect.~\ref{Chapter_kickplane}).
Hence, when searching for wire clusters in the outer drift chambers,
the same strategy is followed as for the inner ones, except that the
target position is replaced by the intersection point of an inner
segment with the virtual momentum kick plane. This inherent matching
of inner and outer segments defines a track candidate. This is shown
in fig.~\ref{cluster_projection} which depicts a schematic representation
of the candidate search. Note that this procedure neglects the
additional deflection due to weak fringe fields reaching into the
MDC volumes.

The spatial resolution of the \emph{track candidate} search is defined by
the wire spacing.  The stereo angles of the wire planes have been
optimized for best resolution in the direction of particle
deflection.  Thus, the position resolution is worse along the
$x$ coordinate of the chamber than on the $y$ coordinate. For
the inner drift chambers the resolution along the $x$ coordinate is
\mbox{$1.12-1.5\:$ mm} and along the $y$ coordinate \mbox{$0.8-1\:$
mm}. Due to their larger cell size,  the
resolution  in the outer drift chambers is \mbox{$3.9-4.8\:$ mm}
and \mbox{$2.9-3.3\:$ mm} in $x$ and $y$ directions respectively.

\subsubsection{Track segment fitting}\label{Chapter_track_fitting}

The precision of the reconstructed hit points in space is improved from a few mm
down to typically 0.1\,-\,0.2 mm by making use of the drift time measurement and by
fitting the space coordinates of the track to a track model.  This
requires to convert a measured drift time into a distance to
 the sense wire. The distance to time correlation for each drift cell
 geometry (\emph{$x-t$} correlation) was obtained from {\sc
Garfield} \cite{Garf} simulations and checked against test measurements
\cite{MarkertPHD}.

The drift time measured by the \mbox{TDC} connected to a drift cell
is the sum of the real drift time, the propagation time
of the signal along the sense
wire up to the readout electronics and the time of flight of
the particle to the drift cell relative to the stop signal derived
from the trigger.  The propagation time of the signal
can be subtracted from the measured time since it can be computed
from the known position of the hit on the wire.  As the remaining
time value still contains a part due to the time of flight of the
particle, it is suitable to perform a track model fit with respect to a
time variable, simultaneously extracting the time of flight of
the particle.

Fitting two chambers simultaneously is performed employing a
straight-line-track model. Again, we neglect the residual fringe
fields in the chamber region. However, since only the coordinates of the hit points at each
chamber mid plane are used in the final momentum determination, the relative small
bending effect between the two chambers is a second order effect and
can be neglected. The following function $F$ for
the least square minimization is evaluated for all drift
cells of a \emph{track segment} \cite{ier1}:
\begin{align}\label{fitter_functional}
    F&=\sum_{i}\frac{(t_{drift}^{i}+t_{shift}-t_{TDC}^{i})^{2}}{(\Delta_{TDC}^{i})^{2}} w_{i}, \\
                                                                                  \notag \\
    t^i_{drift}:   \quad \quad  &  \textrm{drift time from the {\sc Garfield} model}, \notag \\
    t_{shift}:   \quad \quad  &  \textrm{time shift of all cells},                  \notag \\
    t^i_{TDC}:     \quad \quad  &  \textrm{measured drift time (after proper TDC calibration)},     \notag \\
    w_i:           \quad \quad  &  \textrm{weighting constant (Tukey weight)},        \notag \\
    \Delta^i_{TDC}:\quad        &  \textrm{error of drift time measurement},          \notag
\end{align}
where $i$ labels the individual drift cells and runs over all cells
in the \emph{track segment}.

In the first iteration, the time shift $t_{shift}$ is calculated as
the mean deviation of all drift time measurements from the known
$x-t$ correlation.  This constant, determined independently for each
particle track in an event, contains the time of flight of the
particle to the drift chamber, as well as deviations which are
common to all drift cells considered in the fit. Those can
originate from errors in the determination of the calibration
coefficients as well as from changes in the operating conditions of
the chambers which modify the $x-t$~correlation of the drift cells.
Individual deviations (time offsets, calibration parameters,
high-voltage operating conditions)
among the layers which are used for a given track fit
are not taken into account by this constant and
therefore, they contribute directly to the
deterioration of the resolution.
The error of the drift time measurement $\Delta_{TDC}$, as used in
the functional $F$, is taken from the {\sc Garfield} simulations.

The weighting factors $w_i$  are
calculated according to a Tukey weight distribution \cite{rusta06}.
The weight serves to minimize the influence of outliers (like
uncorrelated noise or drift time measurements belonging to another
track) on the fit result. It is evaluated dynamically at each step from
the difference between the calculated drift time and the measured one. 

\subsection{Momentum determination} \label{Chapter_track_momentum}

The HADES data analysis employs three different momentum
reconstruction algorithms. The \emph{kick plane} method provides a
fast and robust estimate of the particle momentum with limited
resolution using reconstructed inner track segments and hit points
on the META detector only.  With better precision, as well as
moderate computational requirements, the \emph{spline} method
obtains the particle momentum from matched reconstructed inner and
outer \emph{track segments}. Finally, the \emph{Runge-Kutta} method
provides the best precision in reconstructing the particle momentum,
but requires more computational resources.
Whereas {\it kick plane} can provide a momentum at order zero in case
of no outer MDCs (experiments done before 2004), {\it Spline} and {\it Runge-Kutta} are the standard
methods to reconstruct the momentum. {\it Spline} provides first guess
 momentum and particle polarity which are then used
for the iterative {\it Runge-Kutta}.
All three momentum
algorithms are applied from the lowest to the highest precision
level, each step providing a starting value for the next one with
the needed accuracy.  The results of all algorithms are stored in
parallel, allowing for a detailed monitoring of the procedure.

\subsubsection{Kick plane} \label{Chapter_kickplane}

Schematically, the progressive deflection of a charged particle
on its way through the toroidal field of the HADES magnet can
be substituted by a single kick occurring on a two-dimensional,
almost flat virtual surface, called kick plane~\cite{sanchez}.
This surface is determined in ray-tracing simulations using
GEANT3 and corresponds roughly to the center plane of the magnetic
field in any given sector of HADES. Within the \emph{kick plane} approach,
the momentum $p$ of a deflected particle is obtained in a
straightforward way from its deflection angle via pre-computed
look-up tables. This approach provides a very fast initial value
of $p$, to be used as starting point in more refined track fitting
algorithms (see below).

In general, the momentum change of a particle in the field, {\it i.e.}
the momentum kick, can be written as
\begin{equation}\label{kick_formula}
 |\overrightarrow{p_{in}}-\overrightarrow{p_{out}}| = |p_k| = 2 \; p \;
 \sin(\Delta\theta_k/2),
\end{equation}
where $\Delta\theta_k$ is the deflection angle in the bending plane,
and $\overrightarrow{p_{in}}$ and $\overrightarrow{p_{out}}$ are the
incoming and outgoing three-momentum vectors, both of magnitude $p$.
For large momenta (i.e. small deflections) one can set $2
\sin(\Delta\theta_k/2) \simeq \Delta\theta_k$ and hence the momentum
kick is approximated by
\begin{equation}
   |p_k|\simeq p\;\;\Delta\theta_k.
\end{equation}
On the other hand, the total deflection is given by the integral of
the Lorentz force acting on the particle of charge $q$ on its way
through the magnet
\begin{equation}\label{integral_kick}
    \Delta\theta_k=\frac{q}{p}\int_{l_{in}}^{l_{out}} B \; \sin\alpha \; dl=K (l_{out}-l_{in})=f(sin(\Delta\theta_k/2)),
\end{equation}
where $\alpha$ is the angle between the particle trajectory and the
magnetic field,  $K$ is a constant depending on particle polar and azimuthal angle.
 For the HADES field configuration,
$\sin\alpha \approx1$ and $|p_k| \simeq \int{B(l) dl}$
is a function of the path length, {\it i.e.} it depends at first order on the
particle polar and azimuthal emission angles but also at second order on the momentum.

Performing a Taylor expansion in $sin(\Delta\theta_k/2)$
of eq.~\ref{integral_kick}, keeping only terms up
to second order, one obtains
 \begin{equation}\label{expan}
 a+b\sin(\Delta\theta_k/2)+c\sin^2(\Delta\theta_k/2),
 \end{equation}
which finally, with eq.~\ref{kick_formula}, gives the result~\cite{sanchez}
\begin{equation}\label{momentum2}
 p={\frac{1}{2}} {\frac{p_{k0}}{{\sin(\Delta \theta_k/2)}}} + p_{k1} + 2 p_{k2} \sin(\Delta \theta_k/2),
\end{equation}
 where p$_{k0}$ can be interpreted as
the $0^{th}$-order momentum kick. The parameters p$_{k0}$, p$_{k1}$,
p$_{k2}$ depend only on the entry and exit
points of the particle track in the magnetic field; they are stored
in look-up tables as function of polar and azimuthal angles.

Besides its computational simplicity and speed, the \textit{kick-plane}
method has also the advantage that it can even be applied when only
position information from the META detector is available. In that case, the
limited META position resolution (\mbox{$\sigma_{META}$= 6\,-\,13~mm})
dominates the momentum resolution $\sigma _p /p$, which goes
from about 2~\% for 0.15~GeV/c electrons, up to values of about 15\,-\,18~\% for
1.4~GeV/c electrons, slightly depending on the polar angle.

\subsubsection{Spline method} \label{Chapter_spline}

\begin{figure}[\htb]
\begin{center}
\includegraphics[width=100mm, height=100mm]{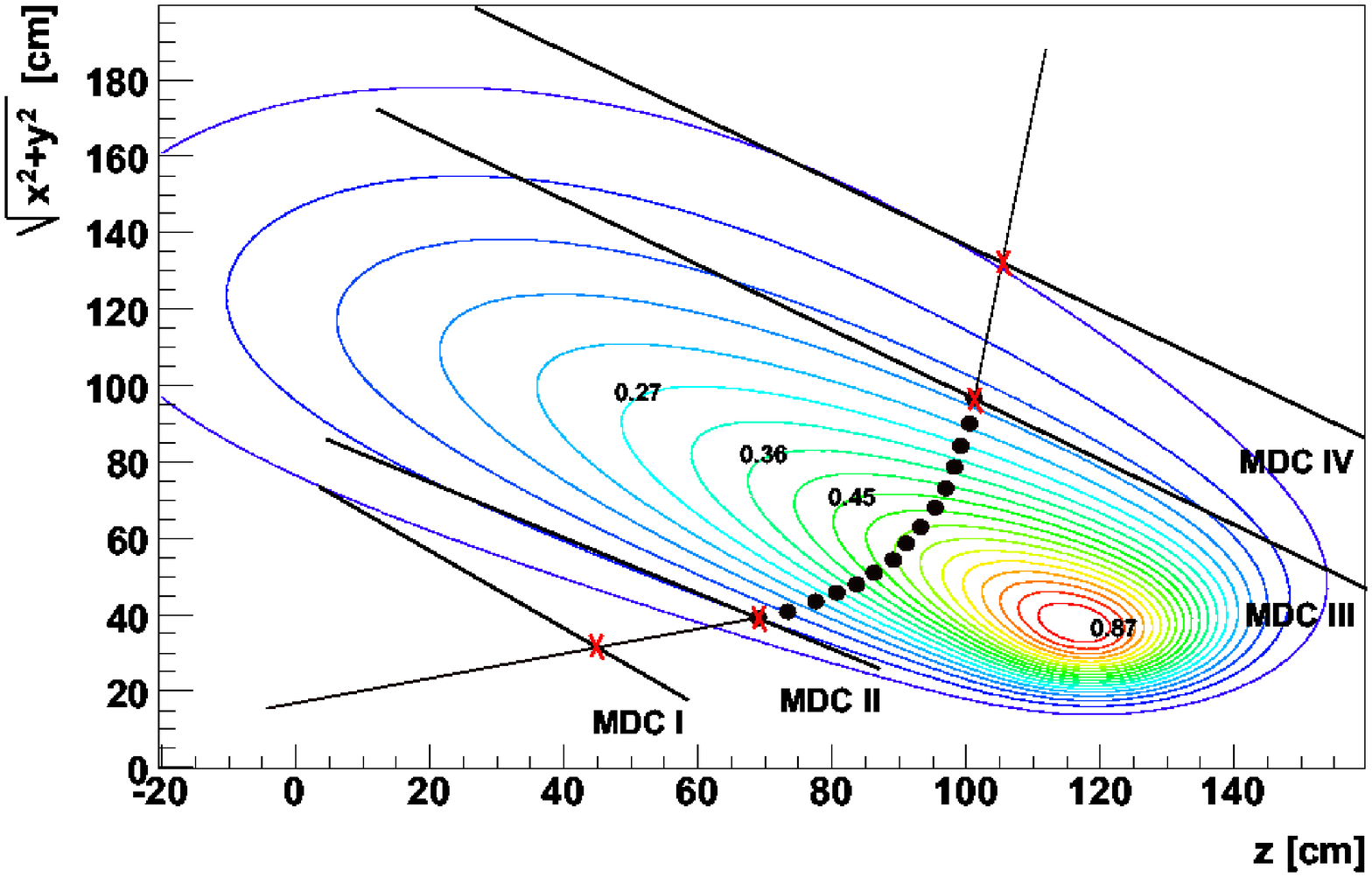}
\caption[splineExplain]{The track as modelized by a cubic spline in the plane
defined by the three-momentum at the target and the beam axis.
The 50 points (only 15 are shown here for clarity) run from MDCII up to MDCIII.}
\label{splineExplain}
\end{center}
\end{figure}

In the framework of spline momentum reconstruction method, a cubic
spline is taken as a model for the particle trajectory through the
magnetic field. The spline provides a smooth curve passing through
the detector hit points, as well as smooth first- and second-order
spatial derivatives at the given hit points.  As an input, this
algorithm employs track candidates containing inner and outer track
segments.  They are used to calculate intersection points of the
reconstructed segments with the chamber mid-planes. Using four such
intersection points in space, a cubic spline function in the
($z,r=\sqrt{x^2+y^2}$) plane
 is applied. Fifty equally-spaced points are selected and the
corresponding derivatives are calculated. The magnetic field
strength is computed at the selected points using the
three-dimensional HADES magnetic field map (see fig.~\ref{splineExplain}).

By minimizing the squared deviations
 $d^2=(r_{spline}-r_{field})^2$
between the spline abscissa $r_{spline}$
and the one obtained from GEANT simulations
$r_{field}$ - this
one being momentum dependent - one gets a momentum value at each of the 50 points.
By averaging over the full set of selected points, a momentum is
determined~\cite{rusta06,wind74}. However, since a cubic spline does not necessarily
exactly model the trajectory of a charged particle through the
magnetic field, systematic corrections, determined through GEANT
simulations and parameterized, have to be applied in order
to achieve sufficient resolution.  With these corrections, the
relative momentum resolution $\sigma _p /p$ of the method, as obtained from
simulations, is ranging from 1.5~\% to 4.5~\% for 0.15~GeV/c
electrons in the $\theta $ range $[20^{\circ}\,-\,80^{\circ}]$,
decreasing to values ranging from 1~\% to 2.8~\% for 1.4~GeV/c
electrons in the given angular range.

\subsubsection {Runge-Kutta method}\label{Chapter_runge_kutta}

The standard method to determine the momentum of a particle
traversing a known magnetic field consists in solving its
equations of motion in the field region. In general, the
system of second-order differential equations is handled
by the fourth-order Runge-Kutta method of Nystrom in a
recursive way \cite{NRC}. The numerical solution of a
differential equation requires, however, initial conditions
of the function and its first derivatives, provided
in our track reconstruction by the spline method (momentum
and polarity) and the track-segment fitter (vertex and
direction). The track parameters - $x, y$ at $z=0.$ and
two direction cosines in the MDCI chamber coordinate 
system and the momentum $p$ -
are iteratively optimized
to fit to the hit points measured in the MDCs.

\begin{figure}[\htb]
\centering
\mbox{
\subfigure{\includegraphics[width=0.49\textwidth]{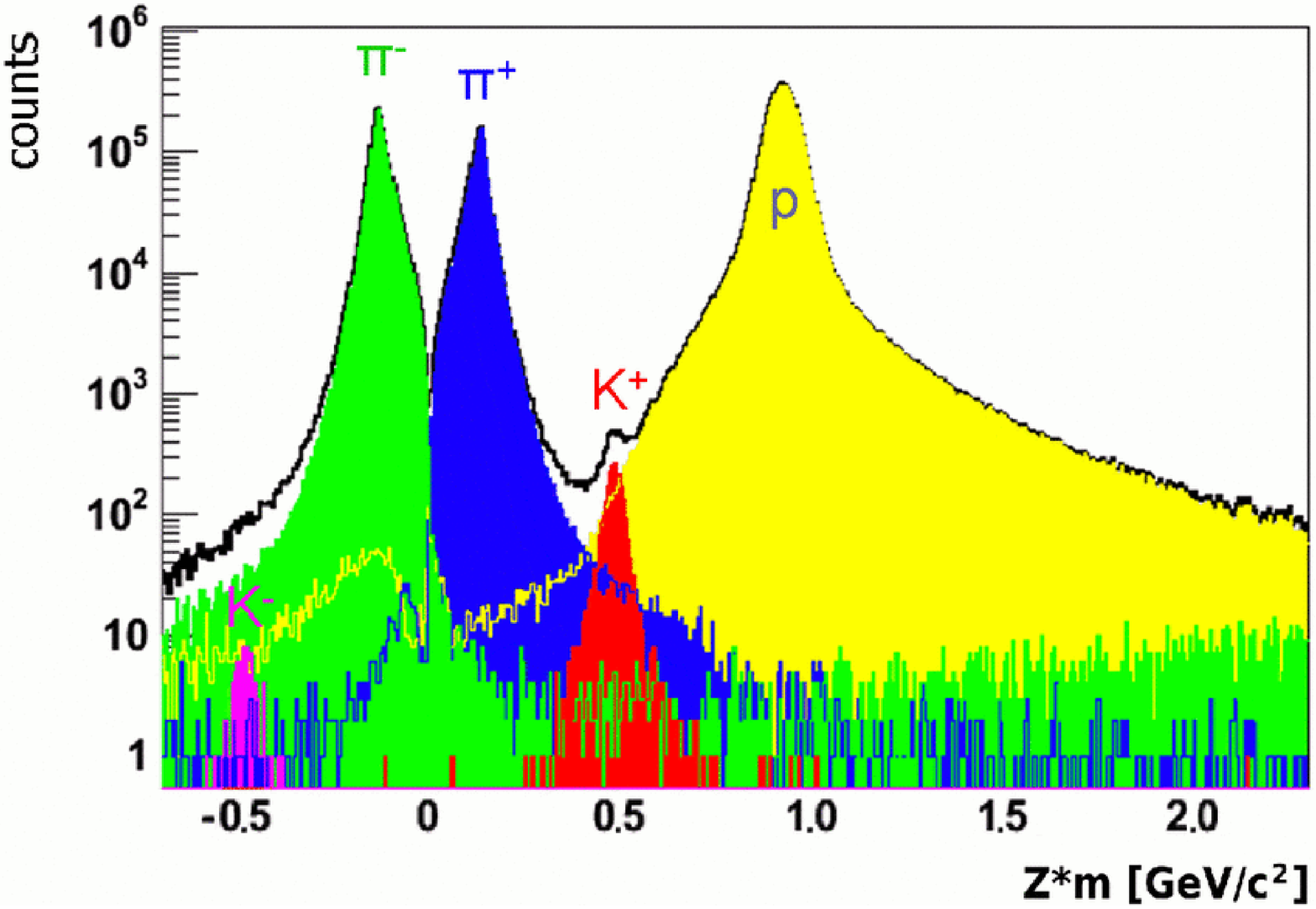}}
\subfigure{\includegraphics[width=0.49\textwidth]{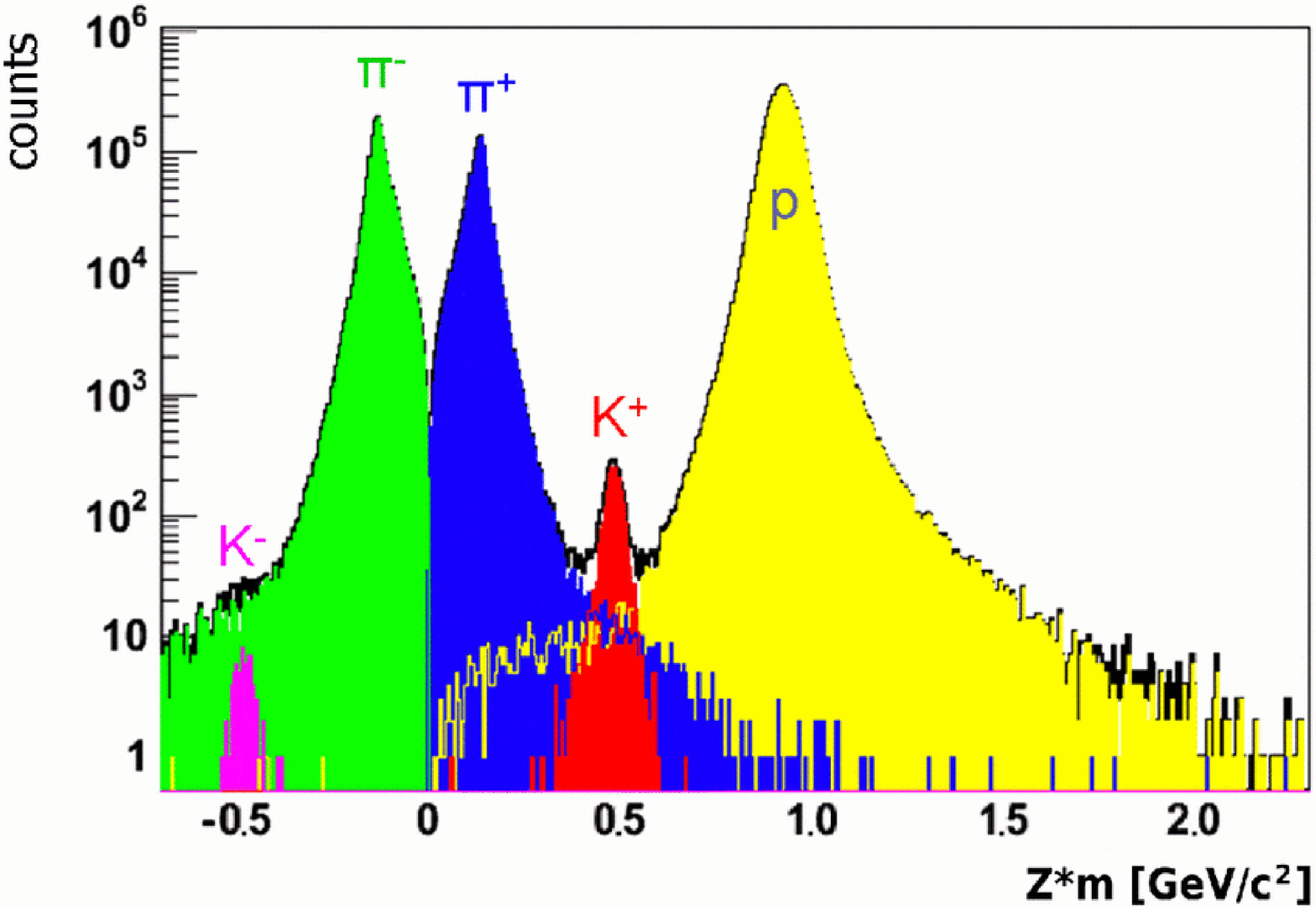}}}
 \caption{Left:
    Charge $Z$ times mass distribution using the \emph{Runge Kutta} method. No selection
    is made on the track quality. Right: Same distribution with a track quality cut
     $\chi^2_{RK} = 5$.
     These spectra are built out of tracks
    from $71\cdot10^6$ events obtained from full GEANT simulations of 2~AGeV C\,+\,C
    collisions in the UrQMD model. Different particle species
    are indicated by different colors. Details are given in~\protect\cite{SadPhD}.}
\label{nov02sim_cutsTQnoRK_meta1}
\end{figure}

Presently our Runge-Kutta method does neither take into account
the energy loss of the particle nor its multiple scattering.
This is however acceptable as the total material budget (MDCI to MDCIV)
stays typically below 0.5~\%
radiation length. Furthermore, the
procedure assumes presently that the error matrix attached to
any given MDC hit is diagonal, {\it i.e.} that the parameters
characterizing a given hit are all uncorrelated. A least-square
minimization procedure solves the linear equations with respect
to the track parameters.

As the result of \emph{Runge-Kutta} tracking, parameters of charged
particle trajectory are estimated, i.e. momentum and the initial
direction vector. In addition to that, a specific $\chi^2_{RK}$
value is provided which can be used either as a criterion for track
quality selection or as a method for particle identification
~\cite{SadPhD}. This is especially important for the
identification of rare tracks like $K^{+}$ and $K^{-}$ (see
fig.~\ref{nov02sim_cutsTQnoRK_meta1}), which shows mass times polarity distributions
of simulated tracks for C\,+\,C collisons at 2~AGeV with (right) and without (left)
track quality selection.

\subsubsection {Momentum resolution}\label{Chapter_tracking_performance_momentum}

\begin{figure}[htb]
\centering
 \mbox{
\subfigure{\includegraphics[width=0.48\linewidth,clip=true]{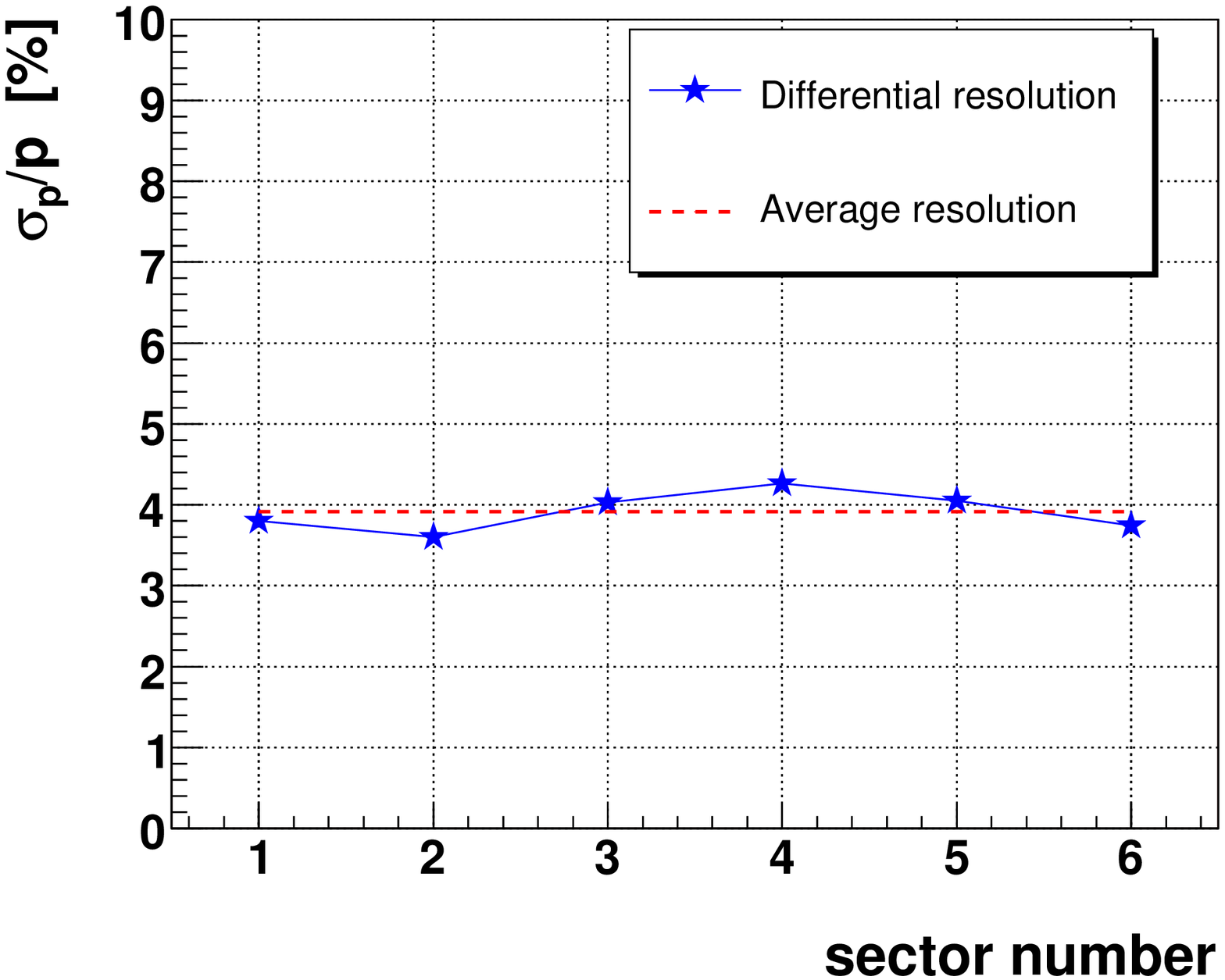}}
\subfigure{\includegraphics[width=0.48\linewidth,clip=true]{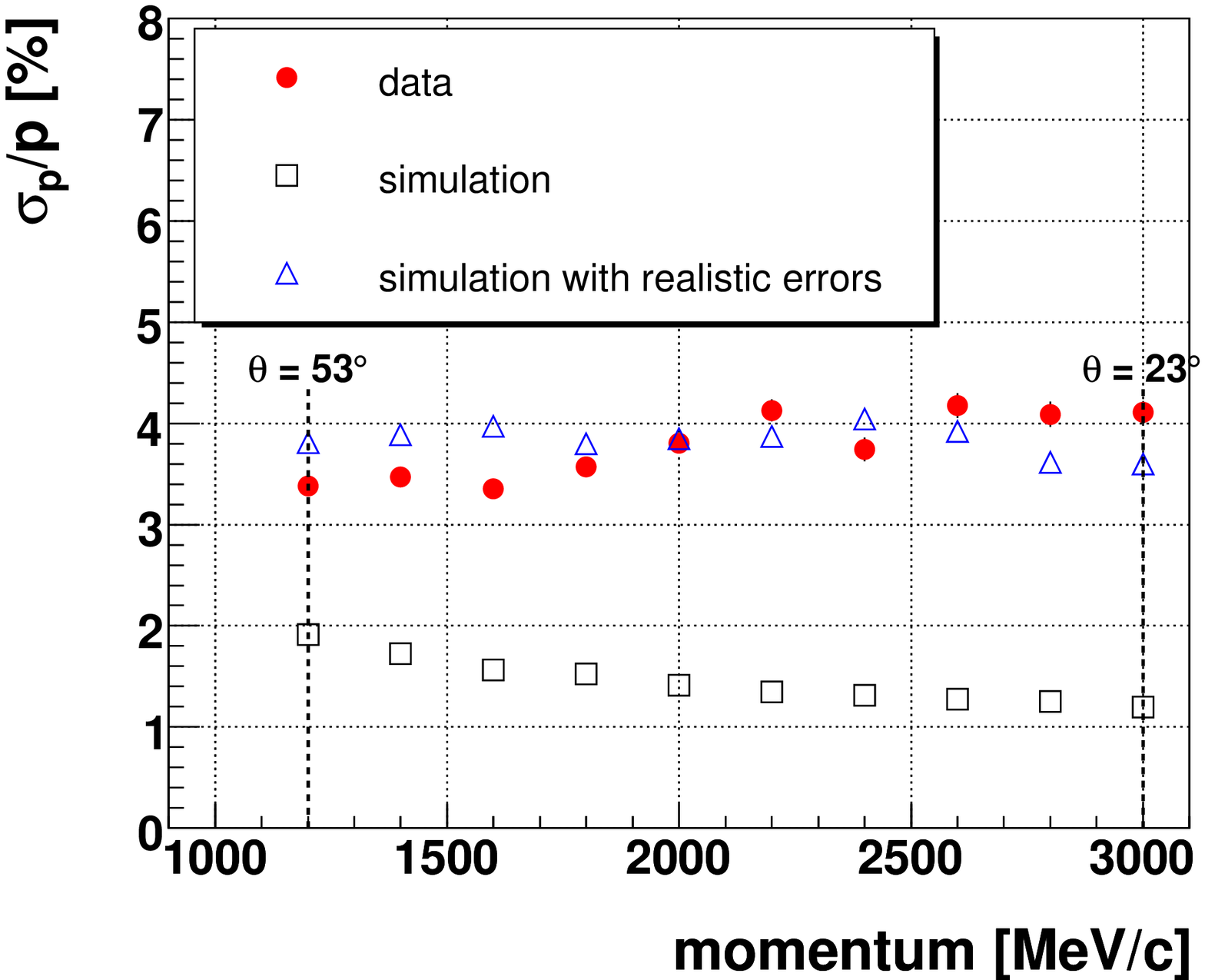}}}
 \caption{Left: Integrated momentum resolution for the 6 sectors
 of HADES. Right: Momentum resolution as a function of
 momentum. A full featured GEANT simulation assuming the drift cell
 resolution obtained from {\sc Garfield} simulations~\protect\cite{MarkertPHD}
 (black open squares) and a four times worse resolution to account
 for uncertainties originating from calibration and alignment (blue
 open triangles) is also shown. In both figures, 3.5~GeV proton-proton elastic
 events were used. Elastic scattering laboratory angles corresponding
 to the momentum range shown are indicated.}
\label{fig_mom_res}
\end{figure}

Data from proton-proton elastic collisions at 3.5~GeV have been used
to investigate the dependence of the momentum resolution on the
particle momentum. As the kinematics is a two-body one, and since
the angular resolution does not contribute significantly to the momentum
resolution, the momenta of the scattered protons can be calculated almost exactly
from the reconstructed polar angle.
From the residual of the momentum derived from
the reconstructed polar angle and the reconstructed {\emph Runge-Kutta}
momentum, integrated over all particle tracks, a resolution of
about 4~\% has been obtained (fig.~\ref{fig_mom_res} left).

\begin{figure}[htb]
  \centering
     \begin{minipage}[c]{0.40\linewidth}
 \centering
 \caption{Simulated momentum resolution for protons with realistic errors
({\sc Garfield} errors multiplied by 4) shown for the \emph{Runge-Kutta} method
(full symbols) and compared to the case without applying any detector
resolution ('Runge-Kutta ideal'; open symbols) for 2 different
polar angles.}\label{compResTheta}
  \end{minipage}
     \hspace{0.02\linewidth}
     \begin{minipage}[c]{0.52\linewidth}
     \includegraphics[width=\linewidth,clip=true]{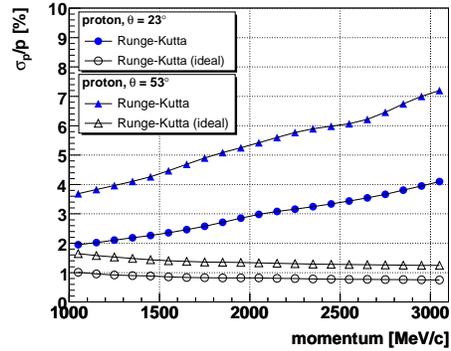}
    \end{minipage}
\end{figure}

Under the same conditions a full-featured GEANT simulation,
based on the time resolution from {\sc Garfield} calculations, would
give an average momentum resolution of 1.5~\%. Figure~\ref{compResTheta}
shows the simulated resolution for protons at two laboratory
polar angles. The 'Runge-Kutta ideal' curve corresponds to the case
where the detector resolution is not included, i.e. one sees
only the effect of the multiple scattering. As one can see spatial resolution
plays a dominant role for $p>$ 1000 MeV/c.

The 'Runge-Kutta' curves in fig.~\ref{fig_mom_res}, close to experimental data,
include the realistic detector resolutions, obtained by scaling-up
by a factor 4 the nominal {\sc Garfield} resolution values.
At first glance, the apparent disagreement with the {\sc Garfield}
calculations is in conflict with results shown on fig.~\ref{fig_selftrack}.
However, since the wires of the two 0 degree layers are parallel, the width
of the correlation patterns shown on fig.~\ref{fig_selftrack} is independent
of any offset (wire geometrical misalignment or miscalibration) of the two adjacent wires.
When reconstructing a hit on a chamber with 6 layers and over areas greater
than typically the cell size squared, one averages over many wires and then gets
sensitive to several individual and uncorrelated offsets. This leads to the observed
deterioration of the resolution. Disentangling the respective role of the different offsets is
still under investigation.

\begin{figure}[htb]
\centering
 \mbox{
 \subfigure{\includegraphics[width=0.46\linewidth,clip=true]{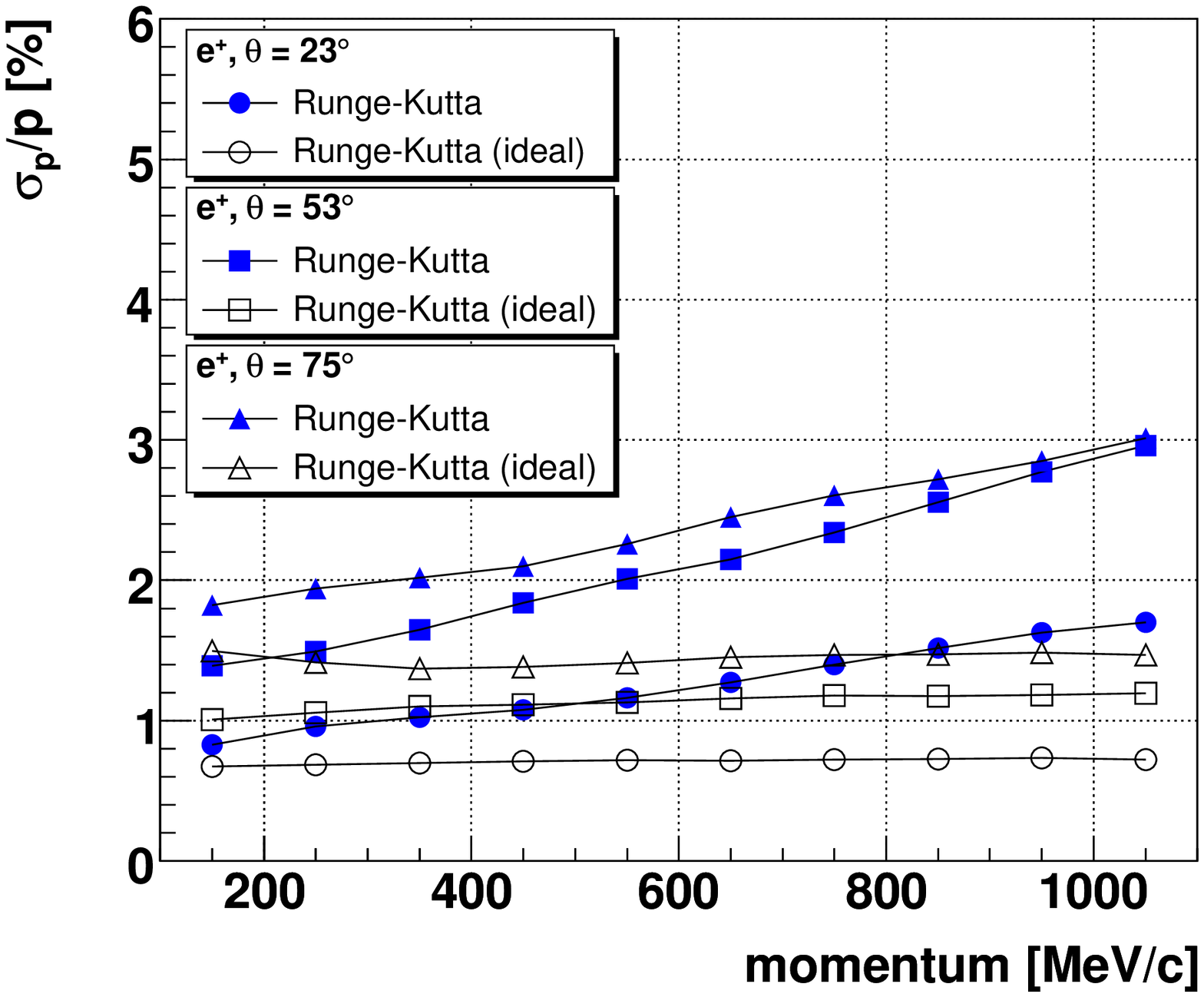}}
 \subfigure{\includegraphics[width=0.52\linewidth,clip=true]{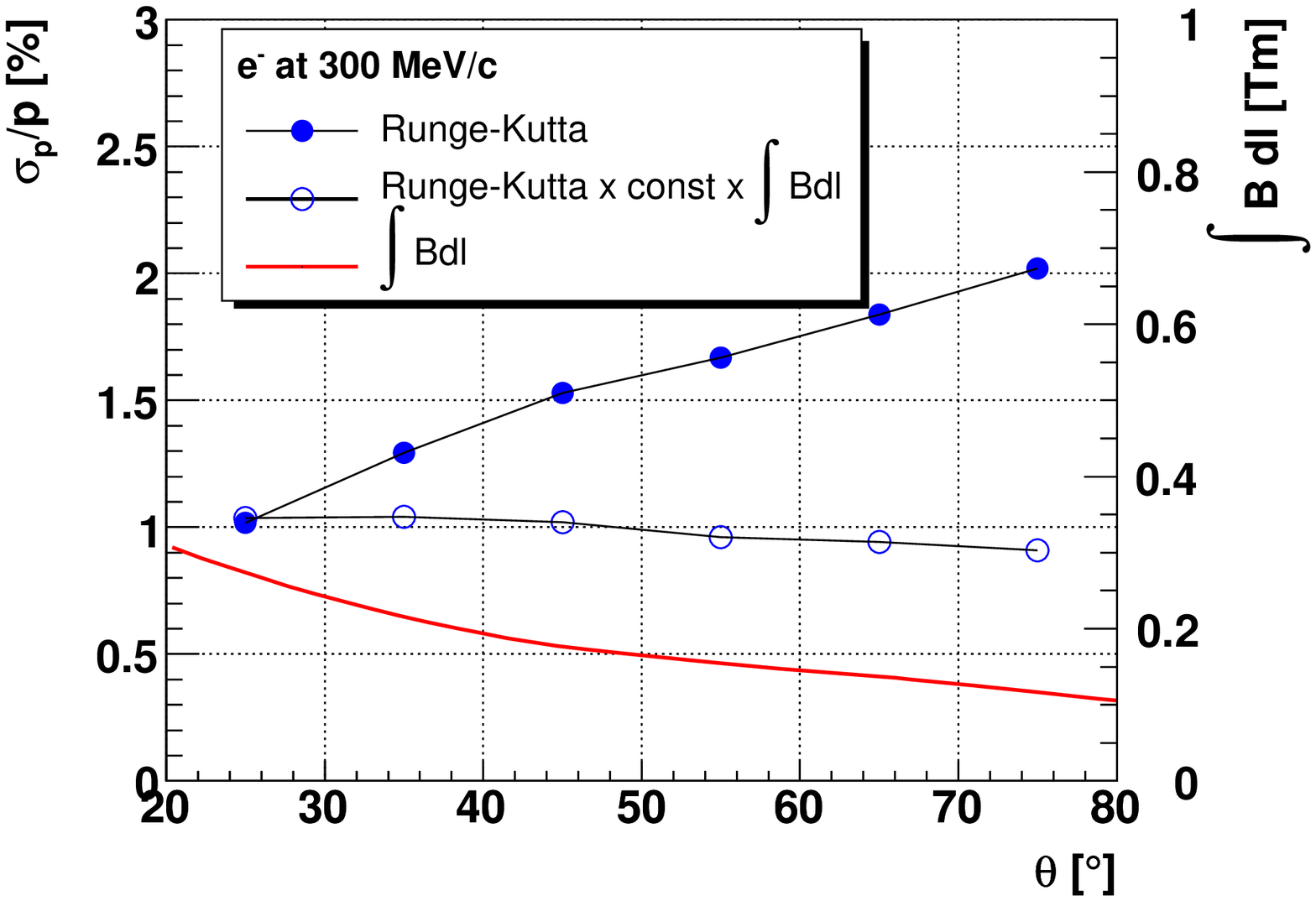}}}
  \caption{Left: Simulated momentum resolution for e$^{+}$ tracks
  with realistic errors shown for the 'Runge-Kutta' method
  (full symbols) and without applying any detector resolution
  ('Runge-Kutta ideal'; open symbols) for 3 polar angles. Right:
  Simulated resolution for 0.3~GeV/c e$^+$ (full symbols)
  against angle. The product of the resolution and field
  integral (open symbols) directly shows that the momentum dependence
  on the polar angle is due to the variation of the integrated
  magnetic field (solid line). The right-hand scale displays
  the integrated magnetic field separately.
}\label{fig_resol_lepton}
\end{figure}

After this
renormalisation, a satisfactory agreement between the observed
resolution and the simulation is obtained (fig.~\ref{fig_mom_res} right).
However, the large effect seen on the protons turns out to be
of much less importance for dielectron spectroscopy, which is
the goal HADES was built for. The average momentum of electrons
or positrons,
about 0.4\,-\,0.5~GeV/c from the $\omega$ decay, is much lower than
the proton average one and the effect of multiple scattering
is then dominant for electrons, as can be seen on the
fig.~\ref{fig_resol_lepton} left. Even with the realistic resolution,
{\it i.e.} detector resolution scaled up by a factor 4,
the relative worsening of the resolution stays below 40~\%.
Figure~\ref{fig_resol_lepton} right
displays the simulated resolution under realistic conditions
for 0.3~GeV/c electrons versus laboratory polar angle together with
the corresponding field integral. The product of the field integral and the
momentum resolution is nearly independent of the polar
angle, as expected. At a fixed momentum, this product
depends only on the relative distance between the detectors
and on their intrinsic resolution.

\begin{figure}[htb]
  \centering
     \begin{minipage}[c]{0.40\linewidth}
 \centering
 \caption{Simulated momentum resolution for e$^{+}$ tracks with
  realistic errors shown for different momentum reconstruction
  algorithms at a fixed polar angle of $53^{\circ}$.}
  \label{fig_compalg_lepton}
      \end{minipage}
     \hspace{0.02\linewidth}
     \begin{minipage}[c]{0.52\linewidth}
     \includegraphics[width=\linewidth,clip=true]{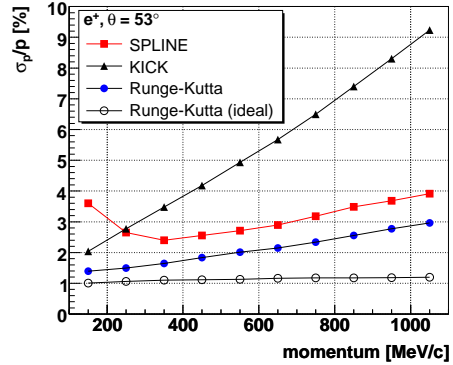}
    \end{minipage}
\end{figure}

A comparison of the different reconstruction
algorithms is presented on example of $e^+$ emitted at $53^{\circ}$
(fig.~\ref{fig_compalg_lepton} right). Whereas the \emph{kick plane} method
is entirely dominated by the resolution of the META detectors
(\mbox{$\sigma_{META}$= 6\,-\,13~mm} depending on polar angle), the curve
for the Spline is used to provide a reasonable start value
for the Runge-Kutta iterative fitting procedure.

\begin{figure}[htb]
\centering
\includegraphics[angle=-90,width=0.7\linewidth,clip=true]
{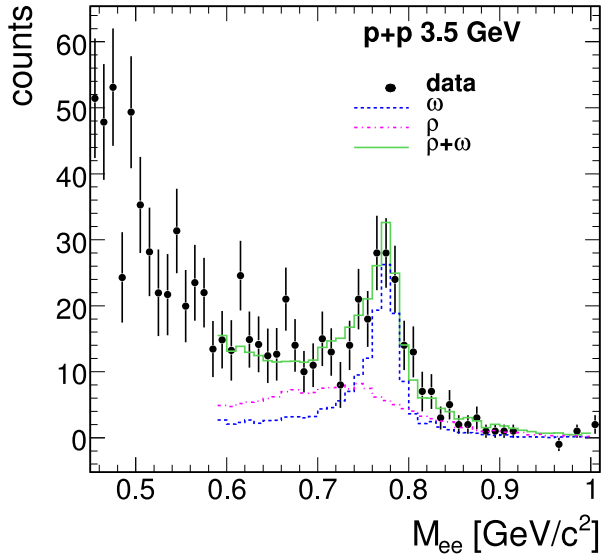}
\caption{e$^+$e$^-$ invariant mass spectrum in the $\omega$ region from 3.5~GeV
proton-proton reactions.
In this invariant mass region, the dominant contributions of the $\rho$
(dot-dashed histogram) and of the
$\omega$ (dashed histogram) are also indicated.}
\label{omega_peak}
\end{figure}

As it has been stated above, the relative
loss of resolution due to the deterioration of the overall chamber
resolution is moderate for the electrons from the $\rho/\omega$ decay.
The corresponding relative worsening of the e$^{\pm}$ invariant mass resolution
is of the order of $30\%$ only. This has been confirmed by a direct measurement of
the vector meson mass distribution. A preliminary
analysis of a recent experiment run on p\,+\,p at 3.5~GeV
(see fig.~\ref{omega_peak}) shows that the experimental peak width
($\sigma_{\omega}=21\pm2$  MeV/$c^2$) agrees quite well
with the simulated one (solid histogram), dominated in this invariant mass
region by the sum of 2 contributions,
coming from $\omega$ (dashed histogram) and $\rho$ (dot-dashed histogram)
two-body decays. One should also note that the peak position ($773\pm 3$) MeV/$c^2$
is shifted by $9$ MeV/$c^2$ as compared to the $\omega$ pole. This can be attributed to the electron
energy loss in the target and the detector materials as shown by the GEANT simulations (solid histrogram).

\subsection{Particle identification}\label{chapter_pid}

\subsubsection{Overview}

HADES is primarily designed as a dielectron spectrometer, hence a
main goal of the analysis is to achieve excellent electron-hadron
separation over the momentum range between 0.1 and 1~GeV/c. However,
for the normalization and interpretation of the dielectron data, a
simultaneous measurement of hadrons ($\pi$, K, p) is
mandatory as well.  Therefore, much attention has been paid to the
implementation of a universal Particle Identification (PID) method.
Two different approaches have been employed for performing
particle identification in HADES: (i) using a set of hard cuts on various
observables, namely momentum, velocity, energy loss in TOF and MDC, hit
patterns in RICH and Pre-Shower or (ii) applying a Bayesian method
to those observables \cite{barlow}. The basis of the latter one is a
statistical test of the hypothesis that the reconstructed track
properties are consistent with a given particle species.  In
practice, several observables from various sub-detectors are
combined together to Probability Density Functions (PDF). These are
determined for each observable and for all possible particle types
in detailed simulations. The particle identification probability is
then calculated based on the prior abundances of individual particle
types, as well as on the specific PDFs of measured variables.  Below
we describe in detail the Bayesian approach.

\subsubsection{Description of the Method}

\begin{figure}[ht]
  \vspace*{+.5cm}
  \center
  \includegraphics*[width=\textwidth]{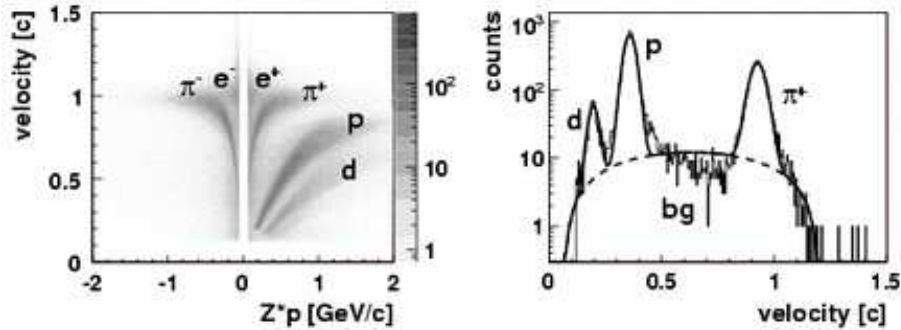}
  \vspace*{-0.1cm}
  \caption[]{Left: Velocity vs. $Zp$ of charged particles detected
  with HADES in C+C collisions at 2~AGeV (note the logarithmic z scale).
  Right: Projection of $\beta$ vs. $Zp$ onto the velocity axis for
  positive particles with a momentum of 0.35~GeV/c. Fits of PDFs are shown
  as well.}
  \label{pid1}
\end{figure}

The Bayesian PID method is illustrated for the velocity-vs.-momentum
relationship, referred to as the $\beta-p$ algorithm.
Particle velocity $\beta$ (in units of the speed of light) and momentum $p$ are indeed
the two observables that typically provide identification power, as
illustrated in fig.~\ref{pid1}. Well separated branches
corresponding to positive and negative pions, protons and deuterons are visible.

The first step in the Bayesian PID method consists in calculating
the PDFs of the chosen observables for each particle type.  For
example, in case of the $\beta-p$ algorithm, these are obtained
from the velocity distributions projected for various momentum and
polar angle bins.  The velocity distributions are fitted with a set
of Gauss functions and a second order polynomial to describe signal
and background, respectively (see fig.~\ref{pid1}).  The
background term accounts for incorrectly reconstructed tracks. The
fitted distributions are finally normalized to unity to be used as
PDFs.

In a similar way, PDF distributions are created for other
observables, {\emph e.g.} for the energy loss signals from the TOF and TOFINO
detectors, as well as from the MDC planes (see sect.~\ref{chapter_elos})
\cite{MarkertPHD,AlexSchmah2}. More specifically, for electron
identification in the RICH detector, various ring parameter
distributions are used: ring radius, the number of contributing
fired pads and their analog sum, and amplitude of the Pattern Matrix
(see sect.~\ref{Chapter_RICHIPU}). Likewise,
electron-hadron separation in the Pre-Shower detector is based on a
comparison of the integrated charge induced in consecutive layers of
the Pre-Shower detector (see sect.~\ref{Chapter_showerid}).

The second step of the method consists in merging the probabilities
from individual PID algorithms.  For each particle type, the products
of probabilities from all algorithms are calculated and the Bayes
formula \cite{barlow} is applied to take into account the relative particle
abundances.  With a set of $k$ independent experimental variables and
$h$ being a particular particle hypothesis with known probability
density function, the likelihood to observe the value of the
discriminating variables for this particle hypothesis is given by
\begin{equation}
L(\vec{x}|h) = \prod_{k}{f_k(x_k|h)},\label{eq:pid1}
\end{equation}
where $f_k(x_k|h)$ is the probability that a track with measurement $x_k$ is a
particle of species $h{\nobreakspace}={\nobreakspace}$e$^{\pm}$, $\pi^{\pm}$, K$^{\pm}$, p, d.

The probability $P(h|\vec{x})$ that a given track corresponds to the particle
type $h$ is then given by the Bayes theorem
\begin{equation}
P(h|\vec{x}) = \frac{L(\vec{x}|h)
\mathcal{P}(h)}{\sum_{h=e^{\pm},\pi^{\pm},K^{\pm},p,d}{L(\vec{x}|h)
\mathcal{P}(h)}},
\label{eq:pid2}
\end{equation}
where $\mathcal{P}(h)$ is the probability to find particle $h$ for a given
momentum $p$ and polar angle $\theta$ (relative abundance).  The sum
of $P(h|\vec{x})$ over all particle types $h$ is normalized to 1.
Knowing all $P(h|\vec{x})$ values for the track, the particle ID is
assigned by selecting the particle type $h$ having the largest
probability.

Two quality parameters, namely the PID efficiency and the PID purity, are evaluated to
test the performance of the method (see {\emph e.g.}~\cite{hommez} for details).
They can be studied systematically in realistic simulations using heavy-ion
collision events generated within the UrQMD transport model~\cite{Bass}.

\subsubsection{PID performance}

A detailed analysis of the $\beta-p$ PID method reveals that $\pi^+$
mesons can be separated from protons up to momenta of $p\simeq1.0$~GeV/c
with a purity better than 80~\% and with an efficiency of almost 100~\%.
At higher momenta, the efficiency drops rapidly for $\pi^+$ because of
the rather limited TOFINO time resolution~\cite{Agakichiev_et_al}.

Kaon identification has been successfully demonstrated ~\cite{kminphi} over a momentum 
range ($150<p<800$ MeV/c) and for polar angles $\theta>45^{0}$ (TOF region) using combination 
of the $\beta-p$ method and the energy loss measurements in TOF and MDCs. After exchange of the TOFINO by the RPC \cite{rpc} detector kaon identification in the full HADES acceptance will become possible.
\
\begin{figure}[ht]
  \vspace*{+.5cm}
  \center \includegraphics*[width=\textwidth]
  {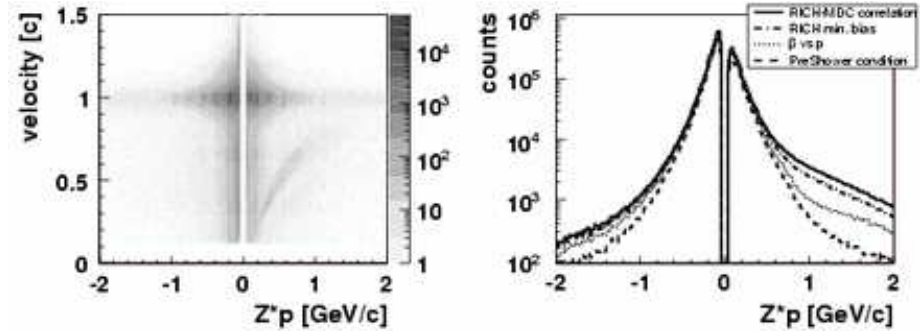}
  \vspace*{-0.1cm}
  \caption[]{Left: Velocity vs. $Zp$ scatter plot for charged particle
  tracks correlated with a RICH ring. Data are from 2~AGeV C\,+\,C collisions.
   Right: Projections on the $Zp$ axis, illustrating the effect of
   different electron identification conditions.}
  \label{pid2}
\end{figure}

\begin{figure}[ht]
  \vspace*{+.5cm}
  \center \includegraphics*[width=\textwidth]
  {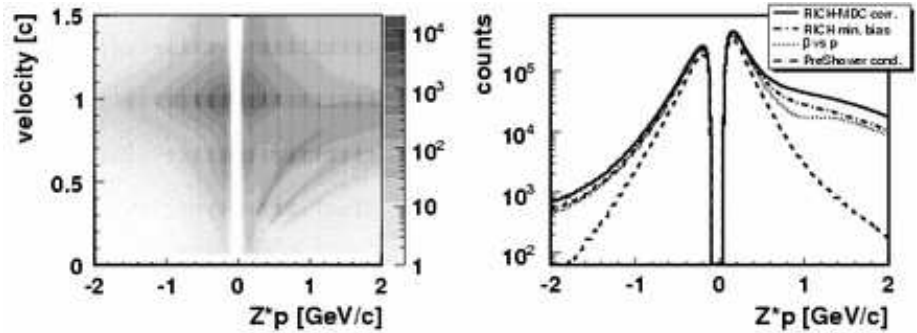}
  \vspace*{-0.1cm}
  \caption[]{Same as fig.~\ref{pid2}, but for Ar\,+\,KCl collisions
  at 1.76~AGeV.}
  \label{pid_ArKCl}
\end{figure}

Below we present in more details electron identification which is the main goal of the HADES detector.
Figure~\ref{pid2} left displays the same distribution as in
fig.~\ref{pid1}, but for LVL2-triggered events with an additional
condition applied on the spatial correlation between electron tracks
reconstructed in the RICH and MDC. Furthermore, only tracks
involving spatially correlated hits, within a $2\sigma$ window,
on the TOF are shown.

Figure~\ref{pid2} right shows the projection of the two-dimensional
$\beta$ vs. $Zp$ distribution and its reductions due to various electron
identification conditions successively applied on: i) aforementioned track correlation in RICH-MDC ii) ring quality like ring radius, number of contributing pads per ring (see sect.~\ref{chapter_richperf}) iii) $\beta-p$ correlation and iv) Pre-shower electron signature (see sect.~\ref{Chapter_showerid}).
A detailed investigation of
measured electron distributions and dedicated C\,+\,C Monte Carlo
simulations proved that the residual contamination of hadronic
background in the final electron sample is less than 3~\% and
contributes mostly at higher momenta ($p>0.6$~GeV/c).  The
efficiency of the PID method is close to 80~\% and drops to
70~\% with higher positron and electron momenta. These results depend
to some degree on the particular reaction investigated.
Figure~\ref{pid_ArKCl} shows the case of 1.76~AGeV Ar\,+\,KCl collisions,
with approximately 5 times larger track multiplicity.
A significantly larger intensity of positive charged tracks,
correlated with RICH rings as compared to negative ones, indicates
an increased contribution from misidentified protons.
However, using Pre-Shower electron conditions, a clean lepton identification
up to $p\,=\,1$~GeV/c could be achieved.

\subsection{Pair reconstruction} \label{pair_reco}
\subsubsection{Overview}
In contrast to Compton scattering of high energy photons and weak decays
of muons, all other reactions and decay processes create correlated
electron-positron pairs. The main aim of the HADES analysis is to
reconstruct with high accuracy the dielectron invariant mass from
the observed electron and positron three-momenta.  Hence, in the
next major step of the data analysis, identified leptons are
combined into unlike-sign (e$^+$e$^-$) and like-sign (e$^+$e$^+$,
e$^-$e$^-$) lepton pairs. In this way, all pair combinations are
reconstructed, the correlated true pairs as well as uncorrelated
pairs, constituting the so-called Combinatorial Background (CB).  At beam
energies available at the SIS, most of the CB
arises from photon conversion ($\sim$\,60~\%), either in the target or
in the RICH radiator gas, and from Dalitz decays of the ubiquitous
$\pi^0$ mesons ($\sim$\,25~\%).  Both sources lead to pairs with
predominantly small opening angles, the so-called close pairs
with typically $\alpha_{e^+e^-}<10^o$.

\begin{figure}[\htb]
\begin{center}
\includegraphics[viewport=20 110 555 395,width=0.8\textwidth]{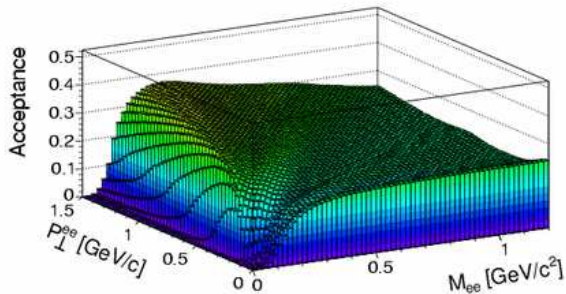}
\caption[Pair acceptance]{Geometrical acceptance for e$^+$e$^-$ pairs
with opening angle $\alpha_{e^+e^-}>9^{\circ}$ as a function of
pair mass and transverse momentum. The acceptance is averaged over
a distribution in rapidity in the interval [0,2] representative of the reactions at
SIS energies.} \label{pair_acceptance}
\end{center}
\end{figure}

The acceptance of the
spectrometer for pairs is shown as a function of the pair invariant
mass and of the transverse momentum in fig.~\ref{pair_acceptance}.  It
is determined by the detector geometry and the deflection
in the magnetic field for both e$^+$ and e$^-$ as a function of
the particle charge $Z$, momentum $p$ and angles $\theta$ and $\phi$.
The geometrical pair acceptance of HADES is obtained as a product of
the two single-electron acceptances (see fig.~\ref{pair_acceptance}),
which together with the momentum resolution function
constitute the HADES acceptance filter (available upon request from
the authors).

\vspace*{0.5cm}

\subsubsection{Rejection of close pairs}

Fortunately, close pairs with very small opening angle leading to
overlapping tracks in the tracking system can be rejected efficiently
by applying a condition on the $\chi^2$ of track segment fits reconstructed
in the analysis.  From simulations we observe that this cut rejects more
than 90~\% of close pairs with opening angles $\alpha_{e^+e^-}<3^{\circ}$,
while still achieving an efficiency of 95~\% for single electron tracks.
The number of uncorrelated pairs is further reduced by applying a cut on
the pair opening angle (typically $\alpha_{e^+e^-}>9^{\circ}$)
and by rejecting pairs which share common hits in the inner or the outer
detector parts.  In addition, both legs of pairs rejected by either one
of these two cuts are marked and are not allowed to contribute to the
pair sample retained for further physics analysis. Those pairs are
used to create pair observable distributions, like the pair invariant mass,
the pair transverse momentum and the pair rapidity.  Likewise, the
combinatorial background of uncorrelated pairs is determined from this
pair sample to which all further operations described below are applied.

\subsubsection{Combinatorial background}

\begin{figure}[hb]
\begin{center}
\includegraphics[width=0.7\textwidth]{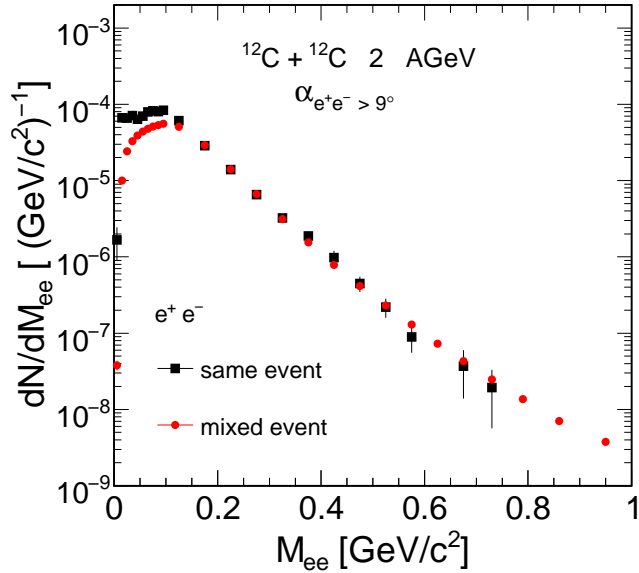}
\caption[CB comparison of methods]{Comparison of the like-sign pair technique
and the mixed-event one for the combinatorial background, properly
normalized to each other in the intermediate-mass region 0.15\,-\,0.5~GeV/c.}
\label{CB_comp_2_methods}
\end{center}
\end{figure}

The combinatorial background of remaining uncorrelated pairs
has to be modeled in order to be able to subtract it from the total
e$^+$e$^-$ yield. Commonly two strategies can be followed \cite{CB}
to build up the CB spectrum: (i) the like-sign pair technique and (ii) the
mixed-event technique.  In the first method, like-sign
e$^+$e$^+$ and e$^-$e$^-$ pairs are formed and subjected to the same
selection criteria as the unlike-sign pairs.  From the reconstructed
like-sign invariant-mass distributions ~dN$^{++}$/$dM_{ee}$ and
~dN$^{--}$/$dM_{ee}$, the respective CB distribution is obtained as
their geometric mean $dN_{CB}$/$dM_{ee}$ =
2 $\sqrt{dN^{++}/dM_{ee} \cdot dN^{--}/dM_{ee}}$.  The signal
distribution is calculated by subtracting  the CB from the total e$^+$e$^-$
yield, i.e. ~dN$^{+-}$/$dM_{ee}$ - dN$_{CB}$/$dM_{ee}$.  In
the same way, this method can be applied to other pair observables,
{\it e.g.} the pair rapidity or transverse momentum.

The advantage of the like-sign pair technique is that it naturally
includes residual correlations due to total energy and charge
conservation, as well as collective flow patterns, but its
statistical accuracy is limited and may not be sufficient in certain
kinematical regions.  Here, the mixed-event technique can help. A
pair distribution is then built by combining two tracks from
different events which are inherently uncorrelated.  This method
offers a large number of mixed combinations even for small event
samples. We also make sure that only tracks from the same event class
are combined, {\emph i.e.} (i) they originate from the same target
segment and (ii) they belong to the same track-multiplicity bin.
Finally, we build mixed-event e$^+$e$^-$ pair distributions and
normalize them to the integral of the corresponding like-sign
distributions within statistically significant and correlation-free
ranges of the like-sign distributions (see
fig.~\ref{CB_comp_2_methods}).

As one can see the mixed background agrees very well with the like-sign CB above $M_{ee}>150$ MeV/c$^2$. At lower masses both distributions substantially differs. It is due to already mentioned correlations between conversion pairs stemming from both photons originating from the $\pi^0\rightarrow \gamma \gamma$ decay process. For this reason at low masses we use like-sign background which,as it has been checked by simulations, properly describes spectral shape of the CB.

\begin{figure}[htb]
\centering
 \mbox{
\subfigure{\includegraphics[width=0.48\linewidth,clip=true]{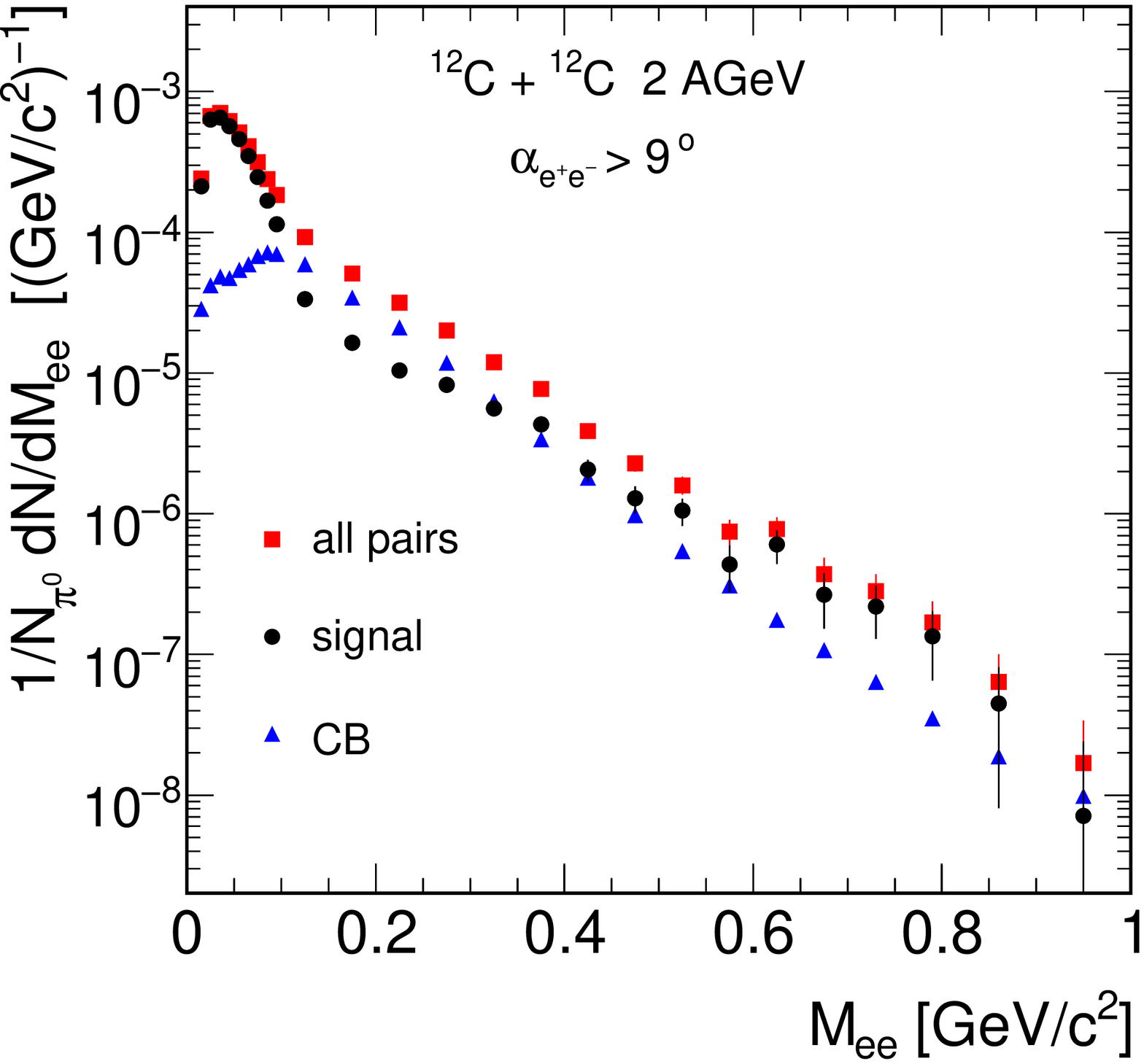}}
\subfigure{\includegraphics[width=0.48\linewidth,clip=true]{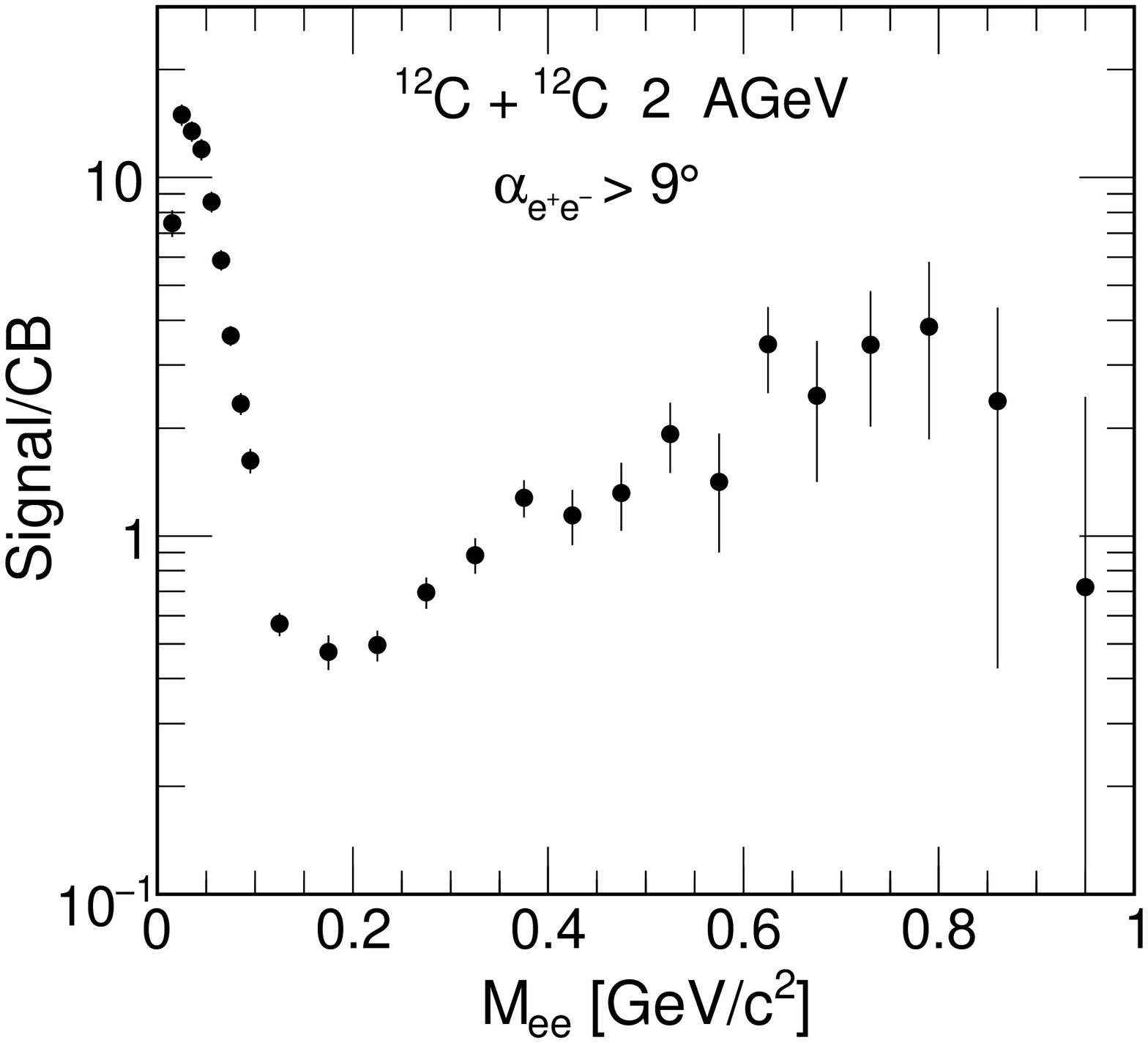}}}
 \caption{Left: e$^+$e$^-$ invariant mass
 spectrum for the C\,+\,C reaction at 2 AGeV. The signal (full circles) is obtained after subtraction of the
 combinatorial background (full triangles) from the all pairs signal (full squares) as explained in the text.
 Right: Signal to combinatorial background ratio as a function of the
 dilepton invariant mass in the hard-cut analysis of the same data sample.}
\label{SB_ratio}
\end{figure}
\begin{figure}[h]
\centering
 \mbox{
\subfigure{\includegraphics[width=0.5\textwidth]{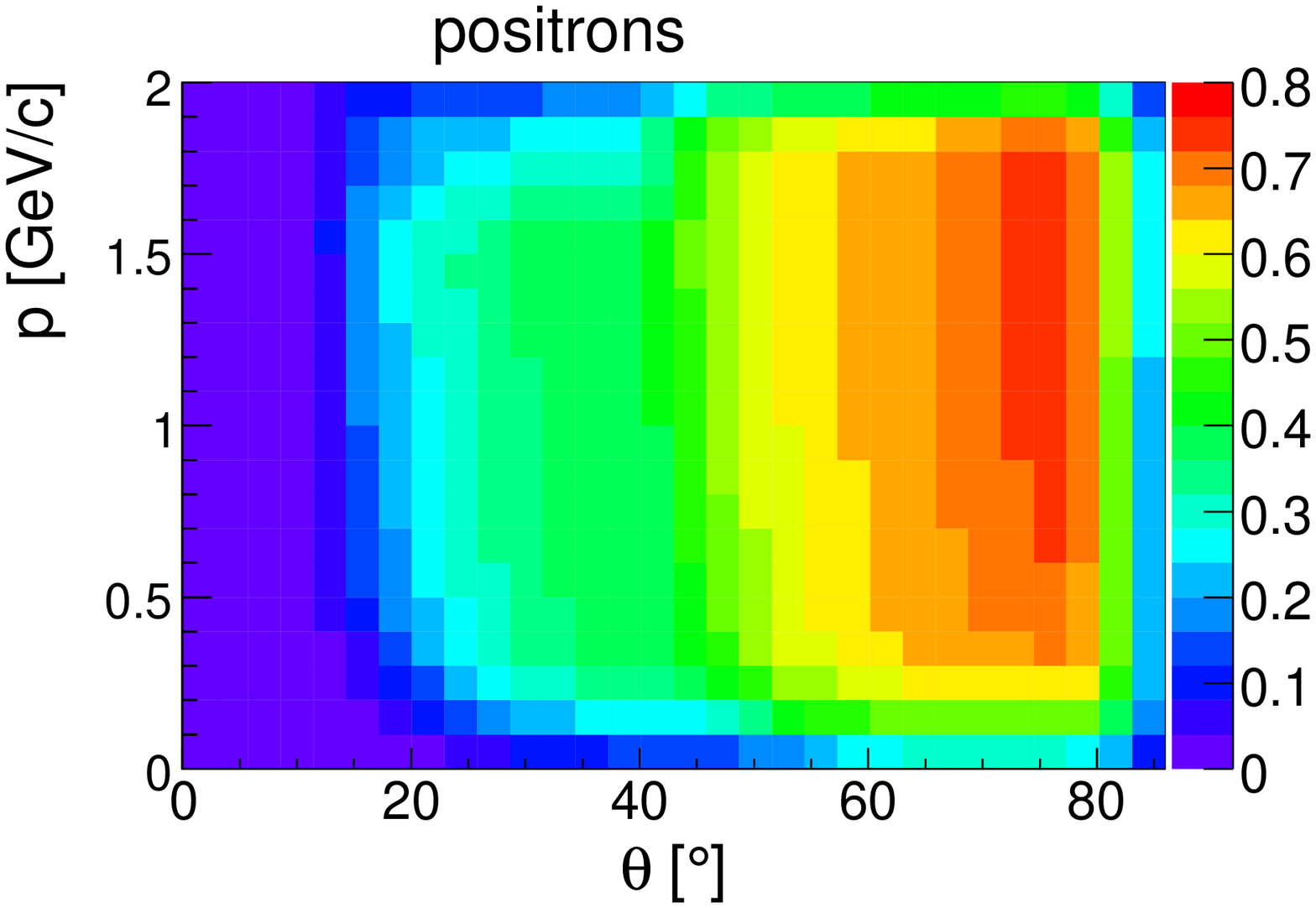}}
\subfigure{\includegraphics[width=0.5\textwidth]{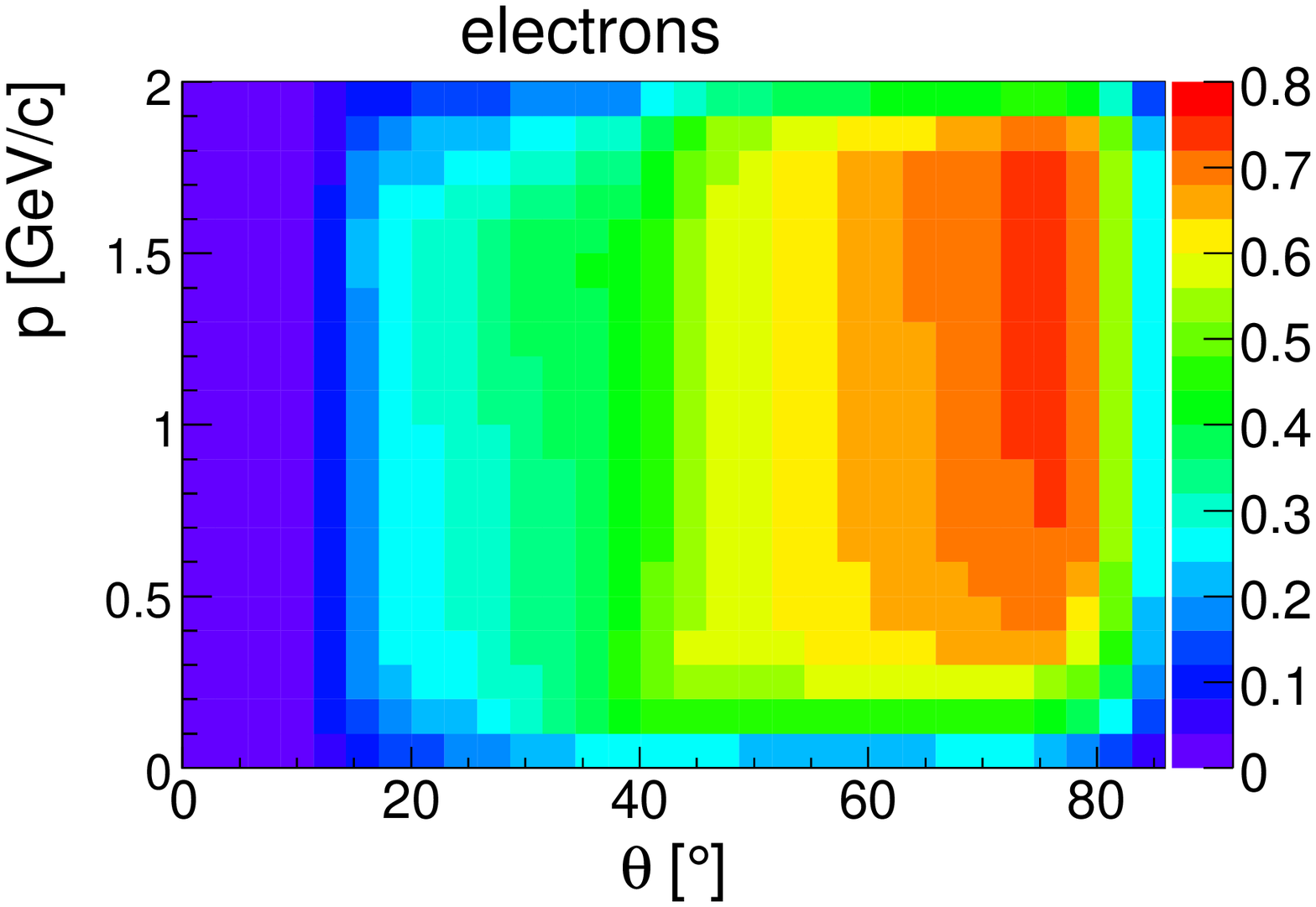}}}
\caption[reconstruction efficiency]{Left: Positron combined detection and
reconstruction efficiency, averaged over the azimuthal angle. Right: Same but for the electron.
The geometrical acceptance together with the bending and focusing
effects of the magnetic field explains the slight different behavior
between the electron and the positron efficiencies.}
\label{lepton_efficiency}
\end{figure}

Figure~\ref{SB_ratio} illustrates the signal-to-combinatorial
background ratio observed in C\,+\,C collisions at 2~AGeV. A good
ratio, about a factor two in the intermediate- and high-mass region,
directly reflects a small conversion contribution, being a
consequence of the concepts used for the RICH construction - short
path length and thin mirror - and for the tracking system detectors
which are made of low-Z material such as He-based counting gas and Al
wires.

\subsubsection{Efficiency corrections}

We correct the spectra for detector and reconstruction inefficiencies
by Monte-Carlo simulations embedding electron tracks with uniform
momentum and iso\-tropic angular distributions into A\,+\,A events, either
generated with the UrQMD transport model or taken from measured data.
The embedded tracks are further digitized and processed through the same
analysis chain as the measured data.
The single-electron efficiencies,
$\epsilon_-(z,p,\theta,\phi)$ and $\epsilon_+(z,p,\theta,\phi)$, are then deduced.
Values integrated over the azimuthal angle are shown in
fig.~\ref{lepton_efficiency} for both electrons and positrons.

The data are then corrected on a pair-by-pair basis with the weighting
factor $1/E_{+-}$, with $E_{+-}=\epsilon_+ \cdot \epsilon_-$ for given
electron momenta and emission angles, $E_{+-}$ ranging typically from
10~\% at $M_{ee}=0.1$ GeV/c$^2$ to 20~\% at $M_{ee}=1$ GeV/c$^2$.
The CB is treated likewise and subtracted, as described above, to
obtain the efficiency-corrected pair signal distribution.  This
prescription relies on the assumption that the single-leg efficiencies
are independent, as was carefully checked in our simulations and proven
to be valid within 15~\% for pairs with opening angles
$\alpha_{e+e-}>9^{\circ}$.


\clearpage
\section{Conclusion}

We have presented a description of the High-Acceptance Dielectron Spectrometer
HADES installed at
the SIS facility in GSI Helmholtzzentrum f\"ur Schwerionenforschung.
Given its hexagonal structure, HADES is nearly azimuthally symmetric
and covers a polar angular range from 18$^{\circ}$ to 85$^{\circ}$.
The dedicated trigger system permits the measurements of dielectron
spectra in reasonably short beam times.
Having this setup at our disposal, a rich physics
program ranging from hadron to heavy-ion physics is in progress with
unprecedented statistics and mass resolution.
Actually, HADES was used for data taking in p\,+\,p, d\,+\,p,
 p\,+\,Nb, C\,+\,C and Ar\,+\,KCl reactions at various beam energies.
First physics results on e$^+$e$^-$ invariant mass spectra
in C\,+\,C collisions at 1 and 2~AGeV have been published \cite{cc2,cc1}.

The detector components and their
respective performances have been described in detail.
Emphasis in electron identification is put on the Ring Imaging
Cherenkov counter. Additional electron selectivity is gained from
the Pre-Shower and TOF detectors which finally results in a
very good purity.
The tracking system, consisting of two layers of multi-wire drift chambers
in front of a superconducting magnet and two layers behind it, allows
for momentum determination of charged particles including also charged hadrons.
The performance for dilepton decays of the light vector mesons enables a
clear identification of $\rho$ and $\omega$ decays as well as the
investigation of predicted medium modifications of spectral properties
such as mass or width. Moreover, combination of measurements of the energy
loss in the MDCs and in the TOF detector, of the velocity and of the
momentum in the tracking system allows for pion, kaon and
proton separation, important for the reaction characterization and normalization.

There is a large variety of theoretical predictions of dielectron spectra
ranging from elementary p\,+\,p and p\,+\,n channels
~\cite{Shyam03}
to heavy-ion collisions
~\cite{Titov07}.
These predictions, differing in many details, can now be verified or falsified
by measurements with HADES. Indeed, theory-based interpretations
~\cite{Schmidt08}
of our spectra~\cite{cc2,cc1} show the strong interest in dielectron
spectroscopy as a tool for understanding hadronic reactions, \emph{per se} or
as a part of heavy-ion collisions, in the non-perturbative domain.
The motivation can be extended to the quest of medium-modified hadron properties
and the origin of the masses of the hadrons.
To accomplish these goals, the large geometrical acceptance of
the spectrometer has been designed which results in reduced statistical and systematic
errors and enables the investigation of exclusive 
elementary reactions channels induced by protons or pions. The
overall detector design has been optimized with respect to low
photon conversion probabilities in order to minimize the combinatorial
background.
The apparatus is ready for extending the employed projectiles to pions, aimed at
studying pion-induced dilectron emission off protons, deuterons and
heavy target nuclei. These reactions are helpful in disentangling the numerous
channels contributing to the complex spectra in heavy-ion collisions.
Interesting theoretical predictions  have been made
~\cite{Lutz03}
which await experimental verification.
While dielectron spectroscopy is the primary goal of HADES, the precise
tracking system allows also for the investigation of other rare probes in
particular those which contain strange quarks.

After having exploited the opportunities at SIS18, HADES is foreseen to
operate at SIS100 within FAIR, thus extending dilepton spectroscopy
in a hitherto unexplored region of beam energies.

\clearpage
\section{Acknowledgments}
The authors are grateful to all national funding agancies.
This work was supported in part by grants from BMBF (TM872 I TP2,
06MT190 TP2, 06F 140 and 06DR135), DFG, GSI (TM Koe1K, TM FR1 and TM KR2),
Czech Republic (MSMT LC07050 and LA316, GAASCR IAA100480803),
EU 6th Framework Program (RII3-CT-2004-506078 and RII3-CT-2005-515876), by MLL M\"{u}nchen,
by the Spanish Funding Agency (FPA2000-2041-C02-02), by INTAS (94-1233, 96-0468, 06-1000012-8861),
by 528/92/LN Stiftung f\"{u}r Deutsch-Polnische  Zusammenarbeit KBN 5P03B 140 20, by CNRS/IN2P3
and by PGIDT02PXIC20605PN, FPA2003-7581-C02-02.

\end{document}